\pdfoutput=1
\documentclass[11pt,runningheads]{llncs}
\usepackage[a4paper,hmargin=1.1in,vmargin=1.1in]{geometry}
\usepackage{combelow}

\renewcommand{\paragraph}{\subsubsection}

\newcommand{\app}[1]{Appendix~\ref{#1}}
\newcommand{\inaux}{}

\usepackage{tikz}
\usetikzlibrary{arrows,calc}

\usepackage{tikz}
\usetikzlibrary{cd}
\usepackage{pgfplots}
\pgfplotsset{compat=1.18}
\usepackage{xcolor}
\usepackage{colortbl}
\definecolor{maygreen}{RGB}{214,236,214}
\definecolor{mayred}{RGB}{248,220,220}

\tikzcdset{scale cd/.style={every label/.append style={scale=#1},
    cells={nodes={scale=#1}}}}

\definecolor{light-gray}{gray}{0.83}

\usetikzlibrary{backgrounds,calc,plotmarks, arrows, matrix,shapes,shapes.symbols, fit,tikzmark,
  decorations.pathmorphing, decorations.pathreplacing, positioning}

\usepackage{spverbatim}
\usepackage{verbatim}
\usepackage{fancyvrb}
\usepackage{multirow}
\usepackage{booktabs}
\usepackage{subfigure}
\usepackage{adjustbox}
\usepackage{graphics}
\usepackage[noend]{algpseudocode}
\usepackage{algorithm}
\usepackage{microtype}
\usepackage{amssymb}

\usepackage{url}

\usepackage{enumitem}

\usepackage{amsmath}
\usepackage{bbm}
\usepackage{epsfig}
\usepackage{amsfonts}
\usepackage{stmaryrd}

\usepackage{hyperref}
\usepackage{amssymb}
\usepackage{pythonhighlight}

\usepackage{environ}
\RenewEnviron{python}{}{}

\usepackage{xspace}

\renewcommand{\vec}[1]{\boldsymbol{#1}}

\newcommand{\abs}[1]{\left|#1\right|}

\newcommand{\ceil}[1]{\lceil#1 \rceil}
\newcommand{\near}[1]{\left\lfloor#1 \right\rceil}

\newcommand{\lwe}{\ensuremath{\mathsf{LWE}}\xspace}
\newcommand{\rlwe}{\ensuremath{\mathsf{RLWE}}\xspace}

\newcommand{\ckks}{\ensuremath{\mathsf{CKKS}}\xspace}
\newcommand{\ship}{\ensuremath{\mathsf{SHIP}}\xspace}

\newcommand{\paco}{\ensuremath{\mathsf{PaCo}}\xspace}
\newcommand{\spru}{\ensuremath{\mathsf{SPRU}}\xspace}
\newcommand{\tutti}{\ensuremath{\mathsf{TUTTI}}\xspace}
\newcommand{\accmul}{\ensuremath{\mathsf{AccMul}}\xspace}

\newcommand{\rns}{\ensuremath{\mathsf{RNS}}\xspace}

\newcommand{\stc}{\ensuremath{\mathsf{S2C}}}
\newcommand{\cts}{\ensuremath{\mathsf{C2S}}}

\newcommand{\pr}[2]{{\sf Pr}_{#1 \rightarrow #2}}
\newcommand{\tr}[2]{{\sf Tr}_{#1 \rightarrow #2}}
\newcommand{\ct}{{\sf ct}}
\newcommand{\pt}{{\sf pt}}
\newcommand{\sk}{{\sf sk}}
\newcommand{\pk}{{\sf pk}}

\newcommand{\nmax}{n_{\text{max}}}

\newcommand{\intv}[1]{\llbracket #1\rrbracket}

\renewcommand{\paragraph}{\subsubsection}

\hypersetup{pdftitle={TUTTI: Parallel CKKS Bootstrapping in under 3 ms on CPU},
      pdfauthor={Robin K\"ostler and Antonio Guimar\~aes},
      colorlinks=true,
      citecolor=black!70!green,
      linkcolor=black!70!red}

\begin{document}

\newcommand{\paperkeywords}{\keywords{Fully homomorphic encryption \and CKKS \and Bootstrapping \and Low latency \and Sparse secrets}}

\title{\tutti: Parallel CKKS Bootstrapping in under 3\,ms on CPU}
\titlerunning{\tutti: Parallel CKKS Bootstrapping in under 3\,ms on CPU}
\author{Robin K\"ostler\inst{1} \and Antonio Guimar\~aes\inst{2}}
\authorrunning{R. K\"ostler and A. Guimar\~aes}
\institute{University of Luxembourg, Luxembourg \and IMDEA Software Institute, Madrid, Spain}
\maketitle

\begin{abstract}
This paper introduces \tutti, a new \ckks bootstrapping procedure that outperforms previous low-latency solutions in multi-threaded latency for any number of messages and, most notably, has a lower latency than TFHE when bootstrapping a single one.
Our bootstrapping builds upon \spru (CK, TCHES 2027), improving it with new methods for exploiting parallelism and increasing the number of messages, as well as with a refined security analysis, of independent interest, of the block-binary secret keys of LMSS (AsiaCCS 2023).
\tutti bootstraps $4096$ real-number messages in $46$ to $83.4$\,ms on $192$ threads to restore $1$ to $9$ multiplicative levels, outperforming \ship (CHKS, Eurocrypt 2025) by a factor of up to $3$ on the same machine.
Unlike \ship, \tutti gains immediately from bootstrapping fewer messages, with a $256$-message bootstrapping taking just $15$ to $34$\,ms on $32$ threads.
For a single message, \tutti takes $4.6$\,ms on $32$ threads, whereas TFHE-rs needs $7.3$\,ms for a $4$-bit bootstrapping on the same machine and thread count.
We show that this advantage persists across different architectures: on an Apple M5 Pro laptop (ARM), \tutti takes $2.4$\,ms on $14$ threads, against $4.5$\,ms for TFHE-rs.
Differences between \ckks and lightweight schemes such as TFHE remain on several fronts, from functionality to key sizes and security assumptions, but our work shows that multi-threaded latency is no longer one of them, closing a long-standing gap in the practical FHE literature.
\paperkeywords
\end{abstract}

\section{Introduction}\label{s:intro}
Latency and throughput have long been the two axes that govern performance in fully homomorphic encryption (FHE). Both are set in practice by \textit{bootstrapping}~\cite{GEN09}, an operation that refreshes encryption noise and allows computations of unbounded depth. The idea of evaluating homomorphic operations in batches was first introduced by Smart and Vercauteren~\cite{SV10,SV14}, who brought the concept of Single-Instruction-Multiple-Data (SIMD)~\cite{Fly66,Fly72} to FHE. Batching was a breakthrough in throughput, paving the way for FHE with only polylogarithmic overhead~\cite{GHS12a} and for the first homomorphic evaluations of AES on many blocks at once~\cite{GHS12}. It came at the cost of latency, however, since the large parameters that SIMD schemes require made a bootstrapping take on the order of minutes for a thousand messages~\cite{HS15}.

Brakerski and Vaikuntanathan~\cite{BV14} took the opposite direction and studied bootstrapping with polynomially bounded parameters. Their work was followed by a family of lightweight schemes built on the bootstrapping of Alperin-Sheriff and Peikert (AP)~\cite{AP14}, such as FHEW~\cite{DM15} and TFHE~\cite{CGGI16,CGGI20}, which today bootstrap a single message in just a few milliseconds~\cite{TFHErsBench}. This created a fundamental divide in the practical FHE literature, and applications had to choose between low latency and high throughput, with few options in between. Several lines of research have since worked to close this divide. The AP bootstrapping framework was extended to batched settings with improved throughput~\cite{MS18,LW23a,LW23b,GPV23,DKMS24,GP25,PMHSY25}; scheme-switching techniques move data between schemes~\cite{BGGJ20,LHHMQ21,BCKPS23}; and, more recently, a series of works~\cite{BCKS24,SPRU25,SHIP25,PaCo25,SWIFT26} improved the latency of the Cheon--Kim--Kim--Song (\ckks) scheme~\cite{CKKS17}.

\ckks is a particularly promising scheme that excels in SIMD performance on fixed-point arithmetic and has gained significant traction thanks to its numerous applications to privacy-preserving machine learning~\cite{KSKLC18,LLLKKN22,EGR25,MYJK25}. Its fixed-point encoding has also been shown to enable bootstrapping designs that vastly outperform the digit-extraction-based bootstrapping~\cite{GHS12b,HS15,CH18,GV23,GIKV23,MHWW24} of schemes such as BGV~\cite{BGV12} and BFV~\cite{Bra12,FV12}, to the point that BFV ciphertexts are now bootstrapped faster through \ckks~\cite{KSS24,Kim25b}. Still, conventional \ckks bootstrapping takes about nine seconds on a single core to bootstrap $2^{15}$ real messages~\cite{BCKS24}, and it parallelizes poorly: on $32$ cores, it still takes $1.09$\,s for $2^{15}$ messages, and $624$\,ms for $2^{14}$~\cite{SHIP25}. This imposes significant limitations on latency-sensitive applications, such as private neural-network inference, where bootstrapping is either avoided through client interaction~\cite{JVC18,PZMZS24} or accounts for most of the inference time~\cite{AB24,MYJK25}.

To address this problem, Cheon, Hanrot, Kim and Stehl\'e introduced \ship~\cite{SHIP25}, a parallel bootstrapping for \ckks that requires a relatively low multiplicative depth and smaller parameters, thus enabling the bootstrapping of $4096$ real messages in $215$\,ms on a $32$-core CPU. Concurrently, Coron and K\"ostler introduced \spru~\cite{SPRU25}, a method that bootstraps significantly fewer messages in exchange for improved latency, refreshing a single one $5\times$ faster than conventional bootstrapping. Both are based on techniques that embed $\mathbb{Z}_q$ into the complex roots of unity, which removes the usual polynomial approximation of the modular reduction function and, with it, most of the multiplicative depth of the bootstrapping circuit. Two main limitations persist in this line of work.
\begin{enumerate}
  \item Schemes are still designed for a specific number of messages, and provide limited advantage away from it. For a polynomial ring of degree $N$ and a sparse secret of Hamming weight $h$, \ship performs $O(h\log(N/h))$ ring operations for any number of messages~\cite{SHIP25}, and only its linear transform becomes lighter when fewer messages are bootstrapped. Concretely, we estimate that reducing the number of messages improves its latency by at most $15\%$. \spru, in contrast, refreshes at most $N/(2h)$ messages per run, and its cost grows linearly with their number: it is overtaken by conventional bootstrapping from about $128$ messages on~\cite{SPRU25}.
  \item Even for a single message, the fastest existing CPU implementations remain more than an order of magnitude slower than state-of-the-art parallel TFHE bootstrapping. On $32$ threads of a $192$-core machine, our implementation of \ship takes at best $132$\,ms for $4096$ messages (Table~\ref{tab:ship-compare}), and bootstrapping a single message instead saves it at most $15\%$, which still leaves it more than $15\times$ above the $7.3$\,ms of a $4$-bit TFHE-rs bootstrapping on the same machine and thread count~\cite{TFHErs}.
\end{enumerate}

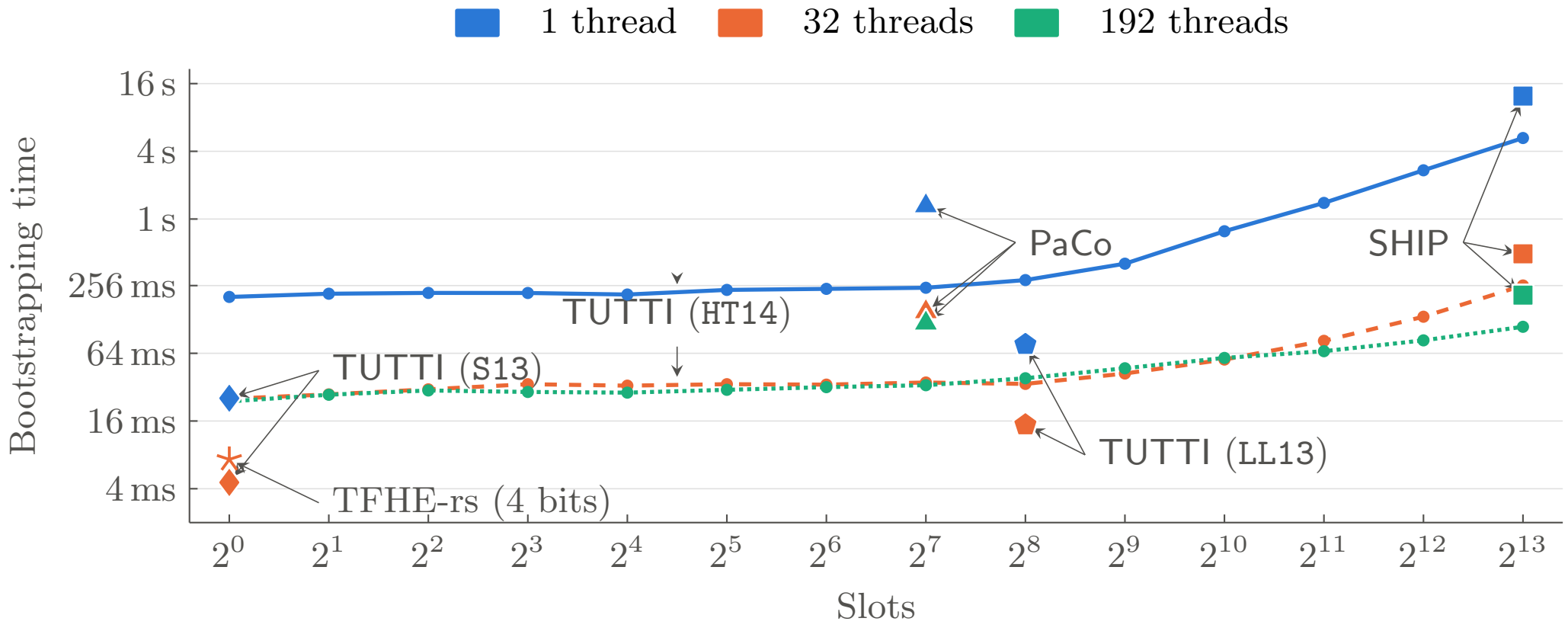
\begin{figure}[!ht]
\centering
\definecolor{slotsThr1}{HTML}{2A78D6}%
\definecolor{slotsThr32}{HTML}{EB6834}%
\definecolor{slotsThr192}{HTML}{1BAF7A}%
\definecolor{slotsInk}{HTML}{52514E}%
\begin{tikzpicture}
\begin{axis}[
  scale only axis, width=339.2pt, height=112pt,
  font=\small,
  xmin=-0.4, xmax=13.4,
  ymode=log, log basis y=2, ymin=2^1.0, ymax=2^14.4,
  xtick={0,...,13}, xticklabel={$2^{\pgfmathprintnumber{\tick}}$},
  ytick={4,16,64,256,1000,4000,16000},
  yticklabels={4\,ms,16\,ms,64\,ms,256\,ms,1\,s,4\,s,16\,s},
  xlabel={Slots}, ylabel={Bootstrapping time},
  axis lines*=left, axis line style={slotsInk, line width=0.4pt},
  tick style={slotsInk, line width=0.4pt}, major tick length=2pt,
  ymajorgrids, grid style={black!12, line width=0.3pt},
  tick label style={text=slotsInk}, label style={text=slotsInk},
  legend style={at={(0.5,1.03)}, anchor=south, draw=none, fill=none,
                legend columns=-1, column sep=8pt, font=\small},
  legend cell align=left,
  slotsSwatch/.style={legend image code/.code={\fill[##1] (0pt,-3pt) rectangle (10pt,3pt);}},
  every axis plot/.append style={line width=1pt, line join=round},
  clip=false,
]

\addplot[forget plot, slotsThr1, solid, mark=*, mark size=1pt] coordinates {
  (0,202.9) (1,216.7) (2,220.3) (3,220.0) (4,213.2) (5,234.3) (6,239.4)
  (7,244.8) (8,286.5) (9,400.7) (10,778.8) (11,1391.1) (12,2712.1) (13,5251.2)};

\addplot[forget plot, slotsThr32, dashed, mark=*, mark size=1pt, mark options={solid}] coordinates {
  (0,25.2) (1,27.6) (2,30.6) (3,34.0) (4,33.0) (5,34.0) (6,33.7)
  (7,35.2) (8,34.2) (9,42.3) (10,56.3) (11,82.6) (12,135.3) (13,256.5)};

\addplot[forget plot, slotsThr192, densely dotted, mark=*, mark size=1pt, mark options={solid}] coordinates {
  (0,23.9) (1,27.4) (2,29.9) (3,29.0) (4,28.6) (5,30.3) (6,32.0)
  (7,33.3) (8,38.4) (9,47.1) (10,58.2) (11,66.8) (12,83.4) (13,110.1)};

\addlegendimage{slotsSwatch, slotsThr1}\addlegendentry{1 thread}
\addlegendimage{slotsSwatch, slotsThr32}\addlegendentry{32 threads}
\addlegendimage{slotsSwatch, slotsThr192}\addlegendentry{192 threads}

\pgfplotsset{slotsRef/.style={only marks, forget plot, mark size=2.6pt,
                              mark options={draw=white, line width=0.5pt, fill=#1}}}
\addplot[slotsRef=slotsThr1,   mark=square*] coordinates {(13,12393.0)};
\addplot[slotsRef=slotsThr32,  mark=square*] coordinates {(13,490.5)};
\addplot[slotsRef=slotsThr192, mark=square*] coordinates {(13,210.6)};
\addplot[slotsRef=slotsThr1,   mark=triangle*, mark size=3.6pt] coordinates {(7,1306.9)};
\addplot[slotsRef=slotsThr32,  mark=triangle*, mark size=3.6pt] coordinates {(7,149.4)};
\addplot[slotsRef=slotsThr192, mark=triangle*, mark size=3.6pt] coordinates {(7,118.4)};
\addplot[slotsRef=slotsThr1,   mark=pentagon*, mark size=3.2pt] coordinates {(8,76.8)};
\addplot[slotsRef=slotsThr32,  mark=pentagon*, mark size=3.2pt] coordinates {(8,14.8)};
\addplot[slotsRef=slotsThr1,   mark=diamond*, mark size=3.8pt] coordinates {(0,25.51)};
\addplot[slotsRef=slotsThr32,  mark=diamond*, mark size=3.8pt] coordinates {(0,4.55)};
\addplot[only marks, forget plot, mark=star, mark size=3.2pt, slotsThr32, line width=0.7pt]
  coordinates {(0,7.29)};

\begin{scope}[every node/.style={text=slotsInk, font=\footnotesize},
              leader/.style={->, >=stealth, slotsInk, line width=0.3pt, shorten >=2pt}]
  \node (lola) at (axis cs:4.5,140) {\tutti\ ($\mathtt{HT14}$)};
  \draw[leader] (lola.north) -- (axis cs:4.5,223.5);
  \draw[leader] (lola.south) -- (axis cs:4.5,33.5);
  \node[anchor=east] (ship) at (axis cs:12.4,620) {\ship};
  \draw[leader] (ship.east) -- (axis cs:13,12393.0);
  \draw[leader] (ship.east) -- (axis cs:13,490.5);
  \draw[leader] (ship.east) -- (axis cs:13,210.6);
  \node[anchor=west] (paco) at (axis cs:7.9,620) {\paco};
  \draw[leader] (paco.west) -- (axis cs:7,1306.9);
  \draw[leader] (paco.west) -- (axis cs:7,149.4);
  \draw[leader] (paco.west) -- (axis cs:7,118.4);
  \node[anchor=west] (single) at (axis cs:0.9,45) {\tutti\ ($\mathtt{S13}$)};
  \draw[leader] (single.west) -- (axis cs:0,25.51);
  \draw[leader] (single.west) -- (axis cs:0,4.55);
  \node[anchor=west] (ll13) at (axis cs:8.6,8) {\tutti\ ($\mathtt{LL13}$)};
  \draw[leader] (ll13.west) -- (axis cs:8,76.8);
  \draw[leader] (ll13.west) -- (axis cs:8,14.8);
  \node[anchor=west] (tfhe) at (axis cs:0.9,3.0) {TFHE-rs (4 bits)};
  \draw[leader] (tfhe.west) -- (axis cs:0,7.29);
\end{scope}

\end{axis}
\end{tikzpicture}%
%
\caption{Bootstrapping latency by the number of bootstrapped messages (fixed-point real-number slots). \tutti on $\mathtt{S13}$ and $\mathtt{LL13}$ leaves $1$ multiplicative level to the application, \paco leaves $7$, and \tutti on $\mathtt{HT14}$ and \ship on $\mathtt{HT14}$ leave $9$. All have a similar precision.
The full parameter sets are given in Tables~\ref{tab:lola-params} and~\ref{tab:ship-params}.}
\label{fig:slots}
\end{figure}

In this paper, we build upon \spru to introduce \tutti, a new bootstrapping procedure that lifts both limitations and achieves state-of-the-art multi-threaded latency for any number of messages. Figure~\ref{fig:slots} summarizes the latency of \tutti for a varying number of messages and compares it with \ship~\cite{SHIP25}, \paco~\cite{PaCo25} and TFHE-rs~\cite{TFHErs}.

\subsection{Contributions and technical overview}

Let us start with a high-level overview of \spru, as introduced by Coron and K\"ostler~\cite{SPRU25}.
In a ring $\mathcal{R}_q=\mathbb{Z}_q[X]/(X^N+1)$, a \ckks ciphertext $(c_0,c_1)\in\mathcal{R}_q^2$ encrypts a message $m$ under a secret key $s$ if $m\approx c_0+c_1\cdot s \bmod q$, up to a small error.
For a binary $s$, each coefficient of $m$ is thus a sum of coefficients of $c_0$ and $c_1$ (up to sign) selected by the bits of $s$, followed by a reduction modulo $q$.
Conventional bootstrapping procedures evaluate that reduction with a high-degree polynomial approximation, which requires a large multiplicative depth.
\spru instead maps every value $x$ of the decryption equation to the complex root of unity $e^{2\pi\mathrm{I}x/q}$, on which the reduction modulo $q$ comes for free, since adding any multiple of $q$ to $x$ leaves the root unchanged (i.e., $e^{2\pi\mathrm{I}(x+kq)/q}=e^{2\pi\mathrm{I}x/q}$ for every integer $k$). In this way, it suffices to evaluate the decryption equation, without the reduction, in the exponent. Notice that sums turn into products, with one factor per nonzero bit of $s$, so \spru switches to a sparse secret with only $h$ nonzero bits~\cite{BTH22}. Further, following a block-binary structure~\cite{LMSS23}, the $N$ bits of the key are split into $h$ blocks, each holding exactly one nonzero bit.
Each factor is then a sum, over one block, of public roots weighted by the secret bits:
\begin{equation*}
  \underbrace{e^{2\pi\mathrm{I}\,m/q}}_{\text{root of the message}}
  \;=\;
  \underbrace{\prod_{b=1}^{h}}_{\text{depth }\log h}
  \Bigl(\,\sum_{j=1}^{N/h}\underbrace{s_{b,j}}_{\text{encrypted bit}}\cdot\underbrace{e^{2\pi\mathrm{I}\,c_{b,j}/q}}_{\text{public root}}\Bigr),
\end{equation*}
where the $c_{b,j}$ are coefficients of the ciphertext.
\spru evaluates this in three steps.
First, it computes the public roots in the clear and multiplies them by the encrypted secret bits, which form the bootstrapping key; these plaintext--ciphertext products are cheap and independent.
Second, it multiplies the $h$ factors, at a cost of only $\log h$ multiplicative levels.
Third, it takes the imaginary part, $\sin(2\pi m/q)\approx 2\pi m/q$ for a small message, which gives back $m$ up to a known factor.
All messages share the slots of one ciphertext, which is why one run of \spru covers at most $N/(2h)$ of them.

\begin{figure}[!ht]
\centering
\begin{tikzpicture}[x=0.5cm, y=0.5cm, font=\scriptsize,
  cell/.style={draw=black!50, minimum width=0.5cm, minimum height=0.5cm, inner sep=0pt, anchor=south west},
  arr/.style={->, >=stealth, black!70, line width=0.5pt}]
  \def\msgcol#1{\ifcase#1\or blue!22\or orange!35\or red!22\or yellow!40\or green!25\or cyan!22\or violet!22\or gray!25\fi}
  \node[anchor=south] at (4,4.6) {\spru: $2$ messages};
  \foreach \b in {1,...,4} {
    \node[cell, fill=\msgcol{1}] at ({2*(\b-1)},2.5) {$\b$};
    \node[cell, fill=\msgcol{2}] at ({2*(\b-1)+1},2.5) {$\b$};
  }
  \draw[arr] (4,2.3) -- node[right] {rotate and multiply} (4,-1.3);
  \foreach \i in {0,...,3} {
    \node[cell, fill=\msgcol{1}] at ({2*\i},-2.5) {$m_1$};
    \node[cell, fill=\msgcol{2}] at ({2*\i+1},-2.5) {$m_2$};
  }
  \draw[decorate, decoration={brace, mirror, amplitude=3pt}, black!60] (2,-2.6) -- node[below=2pt] {copies} (8,-2.6);
  \begin{scope}[xshift=6.5cm]
  \node[anchor=south] at (4,4.6) {\tutti: $8$ messages};
  \foreach \b in {1,...,4} {
    \foreach \a in {1,...,8} {
      \node[cell, fill=\msgcol{\a}] at ({\a-1},{4.5-\b}) {$\b$};
    }
  }
  \node[anchor=west, text=black!70] at (8.1,2.5) {$h$ accumulators};
  \draw[arr] (4,0.3) -- node[right] {multiply slot by slot} (4,-1.8);
  \foreach \a in {1,...,8} {
    \node[cell, fill=\msgcol{\a}] at ({\a-1},-2.5) {$m_{\a}$};
  }
  \end{scope}
\end{tikzpicture}
\caption{Where the factors of each message live, for $h=4$ blocks and $8$ slots.
A cell holds the factor of the block it is numbered with, and colours tell messages apart.
\spru (left) multiplies the factors within one ciphertext and keeps copies of each result; \tutti (right) stores them in $h$ accumulators and multiplies them slot by slot, so that every slot refreshes its own message.}
\label{fig:lola-overview}
\end{figure}
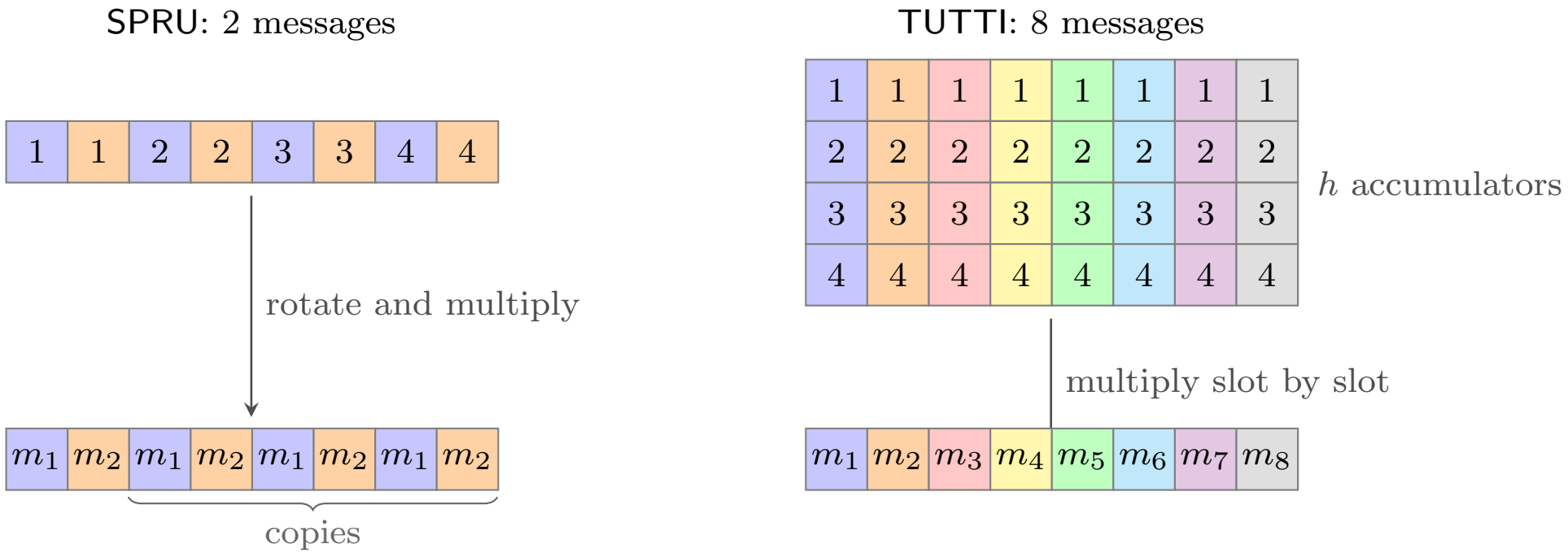

\paragraph{A parallel version of \spru.}
Recall that, in the equation above, the root of each message is a product of $h$ \emph{factors}, one per block of the secret: the factor of block $b$ is the term in parentheses, that is, the public root selected by the secret bit of block $b$.
The first step of \spru computes these factors, and the second multiplies them.
In \spru, both steps take place \emph{within} a single ciphertext: the first step writes the $h$ factors of a message into $h$ different slots, and the second multiplies them by rotating that ciphertext onto itself.
The resulting root ends up copied in all $h$ slots, so each message uses up $h$ slots, and these copies cannot serve other messages (Figure~\ref{fig:lola-overview}, left).
\tutti keeps the three steps and changes only where the first step writes the factors: the factor of block $b$ goes to the $b$-th of $h$ ciphertexts, which we call accumulators.
The second step then multiplies the accumulators slot by slot, without any rotation, and the third step is unchanged (Figure~\ref{fig:lola-overview}, right).
Every slot then serves a different message: with $h$ accumulators, a single run refreshes all $N/2$ slots, and in between, $2^k$ accumulators refresh $2^k$ times as many messages as \spru, always at the same depth $\log h$.
The only work that grows with the number of messages is the first step, which builds the accumulators: a batch of plaintext--ciphertext products that do not depend on each other, and hence spread over any number of cores.

\paragraph{Further optimizations.}
On top of this new layout, \tutti shortens each of the three steps.
First, the sparse secret is only needed during bootstrapping, so we can afford to restrict it further and take it in a subring, of the form $s(X^R)$: its blocks shrink by a factor $R$, and so does the number of products of the first step, while the key amounts to a secret of dimension $N/R$, which our security analysis takes into account (Section~\ref{s:dimension-reduction}).
Second, as in \ship, the products of the first step are rescaled by the auxiliary prime used for key switching, so they consume no multiplicative level (Section~\ref{s:extprod}).
Third, as in \ship, the linear transform that moves the messages between coefficients and slots (\stc) runs once on the incoming ciphertext, at the bottom of the modulus chain, instead of at the end of the bootstrapping, where it would act on a much larger modulus (Section~\ref{s:stc-placement}).
For a single message, we also remove most of the remaining rotations by computing more independent products in the first step, and add a second harmonic of the sine to the third step, which cancels its main approximation error; together, they bootstrap one message in $4.6$\,ms on $32$ threads, at $6.5$ bits of precision (Section~\ref{s:single-slot-impl}).
Finally, on machines with limited memory bandwidth, the full-slot bootstrapping can be split into several smaller ones that share a single key, almost $8$ times smaller in our experiments, at some cost in latency on many cores (Section~\ref{s:cosets}).

\paragraph{Refined security analysis of block-binary secrets.}
\spru, \ship and \tutti all rely on sparse secrets: the fewer their nonzero bits $h$, the shorter the product of the second step, but the easier the secret becomes to attack.
Their parameters therefore hinge on how sparse that secret may be for a given modulus, which in turn sets the ring degree $N$ and, with it, the cost of every operation.
Previous bootstrapping papers~\cite{LMSS23,SHIP25,SPRU25} estimated the combinatorial attacks on such secrets with a cost formula extrapolated from dense secrets, which, in the sparse regime, puts their cost at least $25$ bits below that of the actual attacks (Section~\ref{s:may}).
We instead cost each attack on the exact block structure: lattice attacks with the recent ring-aware hybrids of Hou, Jiang and Ogilvie~\cite{HouJO26}, and combinatorial attacks with Odlyzko's meet-in-the-middle and May's Meet-LWE, with and without guessing the error~\cite{Odlyzko90,May21,HKLS22,BLLLW26} (Section~\ref{s:sparse-keys}).
This gives, for each ring degree and Hamming weight, the largest base modulus that keeps $128$ bits of security, including for the subring secrets of \tutti.
As a consequence, a single message can be bootstrapped at $N=2^{13}$ instead of the $N=2^{15}$ used by \spru, and all our parameter sets push the sparse key to the limit that current attacks allow (Table~\ref{tab:param-security}).
Applied to \ship's secret, the same analysis keeps its base moduli within these limits, provided the switch into the sparse secret happens at the base modulus.

\paragraph{Implementation and experiments.}
We implement \tutti and \ship in the same library, with the same thread control, and compare them on the same machine, with \ship on its published chains and \tutti on sets of the same ring degree and levels (Section~\ref{s:implementation}).
\tutti trades total work for parallelism: on a single thread, \ship is faster on two of the four parameter sets, but from $48$ threads on, \tutti is faster on every set, by up to $3\times$ ($155$ against $467$\,ms on $\mathtt{HT14}$), at a precision on par or better.
Unlike \ship, \tutti also becomes faster as fewer messages are bootstrapped, down to $4.6$\,ms for a single one (Figure~\ref{fig:slots}).
This single-message latency is below that of TFHE: on the same $32$ threads, a $4$-bit bootstrapping of TFHE-rs~\cite{TFHErs} takes $7.3$\,ms, and on the $14$ threads of an Apple M5 Pro laptop, where an AArch64 port of \tutti takes $2.4$\,ms, TFHE-rs takes $4.5$\,ms, so the advantage persists across architectures.
The comparison with TFHE, however, needs some nuance. In particular, TFHE evaluates a lookup table on exact messages with bounded failure probability, whereas \tutti refreshes a fixed-point value at a measured precision; Section~\ref{s:implementation} details the comparison.
Our implementation and security estimates are open source.\footnote{\url{https://github.com/robinkoestler/tutti}}

\section{Preliminaries}
\label{s:prelims}

\paragraph{Notation.}
For $r\in\mathbb{R}$, we write $\near{r}$ for the nearest integer to $r$, with ties rounded up.
We also write $\intv{n}:=\{0,1,\ldots,n-1\}$ and let $\mathrm I$ be a fixed square root of $-1$.
All logarithms are in base $2$.
We represent ${\mathbb Z}_q$ by ${\mathbb Z}\cap(-q/2,q/2]$, and $z\bmod q$ denotes the centered reduction.
For a power-of-two $N$, we let ${\cal R}=\mathbb{Z}[X]/(X^N+1)$, ${\cal R}_q={\cal R}/q{\cal R}$ and ${\cal S}=\mathbb{R}[X]/(X^N+1)$.
We identify the elements of these rings with their coefficient vectors, written in plain type, and $\|a\|_\infty$ denotes the largest absolute value of a coefficient.
Slot vectors are written in bold, and $\vec{v}\odot\vec{v}'$ denotes the slotwise product in $\mathbb{C}^{N/2}$.
Whenever we count the slots, messages, values or coefficients that a bootstrapping refreshes, we count real numbers represented in fixed-point.

\subsection{The \ckks\ scheme}
\label{s:ckks}

\ckks\ \cite{CKKS17} is a leveled homomorphic encryption scheme for approximate arithmetic.
A ciphertext $\ct=(b,a)\in{\cal R}_{q_\ell}^2$ at level $\ell$ encrypts a plaintext $m\in{\cal R}$ under the secret key $\sk=(1,s)$ if
\[
b+a\cdot s=m+e\pmod{q_\ell}
\]
for a small error $e$.
Decryption therefore recovers $m$ only approximately.
Unless stated otherwise, the secret $s$ is dense and ternary, with coefficients uniform in $\{-1,0,1\}$.
We write ${\sf Enc}_{\pk}$ for public-key encryption.

We use the \rns\ variant of the scheme \cite{CHKKS18SAC}, in which $q_\ell=q_0\,p_1\cdots p_\ell$ is a product of word-sized primes over the base prime $q_0$.
An additional auxiliary prime $P$ is used for key switching.
The primes need not have the same bit length, so the number of levels measures the consumption of the modulus less precisely than the number of bits does.

\paragraph{Encoding.}
A plaintext holds a vector of $N/2$ complex numbers, called its slots.
More precisely, ${\cal S}$ is identified with $\mathbb{C}^{N/2}$ through the map
\[
{\sf DFT}(p)=\bigl(p(\zeta^{5^j})\bigr)_{j\in\intv{N/2}},\qquad \zeta=\exp(\mathrm I\pi/N).
\]
For a scaling factor $\Delta$, a vector $\vec{z}$ is encoded as ${\sf Ecd}(\vec{z})=\near{\Delta\cdot{\sf iDFT}(\vec{z})}$ and decoded by $\Delta^{-1}\cdot{\sf DFT}$ \cite{CKKS17,CHKKS18}.
Both maps are computed in \rns\ arithmetic at the current modulus \cite{BEHZ16}.

Shorter vectors are encoded by repetition.
For a power-of-two $n\mid N/2$, sparse packing encodes $\vec{z}\in\mathbb{C}^n$ as ${\sf Ecd}$ of $N/(2n)$ concatenated copies of $\vec{z}$, which is a polynomial in $\mathbb{Z}[X^{N/(2n)}]/(X^N+1)$.
For $n=1$ and a real $z$, this is the constant $\near{\Delta z}$.

\paragraph{Operations.}
The operation ${\sf Add}$ adds two ciphertexts, and therefore their slots.
The operation ${\sf ExtMult}(\ct,\pt)=(b\cdot\pt,\,a\cdot\pt)\bmod q_\ell$ multiplies a ciphertext by a plaintext $\pt\in{\cal R}$, and ${\sf Mult}$ multiplies two ciphertexts and relinearizes the result.
The product of two encodings carries an extra factor $\Delta$, since
\[
{\sf Ecd}(\vec{z}_1\odot\vec{z}_2)\simeq\Delta^{-1}\cdot{\sf Ecd}(\vec{z}_1)\cdot{\sf Ecd}(\vec{z}_2).
\]
A product is therefore followed by the rescaling ${\sf RS}(\ct)=\near{\ct/p_\ell}\bmod q_{\ell-1}$, which consumes one level.
We write ${\sf MultR}={\sf RS}\circ{\sf Mult}$ and ${\sf ExtMultR}={\sf RS}\circ{\sf ExtMult}$.
Relinearization and the automorphisms of Section~\ref{s:rot-conj} change the secret key, and a key switch restores it.

\paragraph{Key switching.}
A key switch turns an encryption under a secret $\bar s$ into an encryption of the same message under $s$.
It uses a public switching key, made of encryptions of multiples of $\bar s$ under $s$, and adds a small error.
We use two variants, detailed in \app{app:ks}.
The GHS switch~\cite{GHS12} works at the modulus $Pq_\ell$, extended by the auxiliary prime $P$: the ciphertext is multiplied with the switching key at that modulus, and a division by $P$, which we call ModDown, removes the factor $P$ together with most of the switching error.
In the \rns variant, the ciphertext is first split into dnum digits, each a group of primes of $q_\ell$, and the key holds one sample per digit~\cite{CHKKS18SAC}.
The BV switch~\cite{BV11} needs no auxiliary prime: it splits the ciphertext into $D$ digits in base $2^\omega$, its key holds $D$ samples modulo $q_\ell$, and it costs $2D$ multiplications in ${\cal R}_{q_\ell}$.
For security, the difference that matters is the modulus of the published samples, $Pq_\ell$ for GHS against $q_\ell$ for BV.
We use GHS for every switch under the dense key, and BV for the switch into the sparse key (Section~\ref{s:sparse-keys}).

\subsection{Slot operators}
\label{s:rot-conj}

The automorphism $\Psi_r$, defined by $\Psi_r(m)(X)=m(X^{5^r})$, rotates the slots of $m$ by $r$ positions to the left:
\[
{\sf DFT}(\Psi_r(m))_j={\sf DFT}(m)_{j+r\bmod N/2}.
\]
The automorphism ${\sf conj}(m)(X)=m(X^{-1})$ conjugates the slots.
Homomorphically, ${\sf Rot}(\ct,r)$ and ${\sf Conj}(\ct)$ apply these maps to both components of $\ct$ and then key-switch.
Neither consumes a level.
The operator ${\sf Im2}(m)=-X^{N/2}(m-{\sf conj}(m))$ returns twice the imaginary part of each slot.
We evaluate it as ${\sf ExtMult}({\sf Conj}(\ct)-\ct,X^{N/2})$, without rescaling.

\paragraph{Trace and product.}
The next two operators combine the slots of a single ciphertext.
For powers of two $d'\mid d\mid N/2$, the trace~\cite{AP13,CHKKS18} and the product operator~\cite{SPRU25} are the compositions
\begin{align*}
\tr{d}{d'}&=({\sf id}+\Psi_{d'})\circ({\sf id}+\Psi_{2d'})\circ\cdots\circ({\sf id}+\Psi_{d/2}),\\
\pr{d}{d'}&=({\sf id}\cdot\Psi_{d'})\circ({\sf id}\cdot\Psi_{2d'})\circ\cdots\circ({\sf id}\cdot\Psi_{d/2}),
\end{align*}
where $({\sf id}\cdot\Psi_r)(m)=m\cdot\Psi_r(m)$.
Given the sparse packing of $\vec{z}\in\mathbb{C}^d$, they return the sparse packing of the $d'$ partial sums and partial products
\[
t_k=\sum_{i\in\intv{d/d'}} z_{k+id'}
\qquad\text{and}\qquad
t_k=\prod_{i\in\intv{d/d'}} z_{k+id'},
\qquad k\in\intv{d'},
\]
respectively.
For example, $\pr{4}{1}=({\sf id}\cdot\Psi_1)\circ({\sf id}\cdot\Psi_2)$ acts on four slots as
\[
(z_0,z_1,z_2,z_3)\;\mapsto\;(z_0z_2,\,z_1z_3,\,z_0z_2,\,z_1z_3)\;\mapsto\;(p,p,p,p),\qquad p=z_0z_1z_2z_3.
\]
Notice that each layer writes every partial product into two slots, which Section~\ref{s:double-slots} will exploit.

Homomorphically, ${\sf id}+\Psi_r$ is ${\sf Add}(\ct,{\sf Rot}(\ct,r))$ and ${\sf id}\cdot\Psi_r$ is ${\sf MultR}(\ct,{\sf Rot}(\ct,r))$.
The trace therefore costs $\log(d/d')$ automorphisms and no level, while the product operator costs $\log(d/d')$ automorphisms and $\log(d/d')$ levels.
When the factors of a product lie in distinct ciphertexts, no automorphism is needed, and $\bigodot_{i\in I}\ct_i$ denotes the binary tree of ${\sf MultR}$, of depth $\lceil\log|I|\rceil$.

\paragraph{Slot-to-coefficient.}
\label{s:slottocoeff}
Let $\vec{t}\in\mathbb{R}^n$ be sparsely packed as $v={\sf Ecd}(\vec{t})\in\mathbb{Z}[X^{N/(2n)}]/(X^N+1)$.
The linear map $\stc_n$ moves the slot values into the coefficients:
\begin{equation}
\label{eq:mXStC}
\stc_n(v)=\sum_{j\in\intv{n}}\near{\Delta\cdot t_j}\cdot X^{jN/n}\in\mathbb{Z}[X^{N/n}]/(X^N+1).
\end{equation}
There are two standard ways to evaluate it.
The first is a matrix--vector product by the diagonal method \cite{HS14,CHKKS18}.
With the baby-step giant-step technique, it takes ${\cal O}(\sqrt n)$ rotations and ${\cal O}(n)$ plaintext products, in one level.
The second is a homomorphic FFT, which decomposes the matrix into sparse factors, with ${\cal O}(\log n)$ operations but one level per factor \cite{CHH18,CCS19}.
The bootstraps of this paper are shallow, and their product takes only $\log h$ levels, so we do not spend several more levels on this map.
We always use the matrix--vector product, as does \ship, so $\stc_n$ takes one level.
We denote the inverse map by \cts.
For real $\vec t$, refreshing $n$ slots and refreshing the $n$ coefficients of~\eqref{eq:mXStC} are thus the same task.
A complex vector of $n$ slots amounts to $2n$ coefficients, its real and imaginary parts.

\subsection{\spru\ bootstrapping}
\label{s:spru-boot}

Classical \ckks\ bootstrapping \cite{CHKKS18} starts by raising the modulus of the ciphertext, which adds an unknown integer polynomial to the plaintext.
To remove this polynomial, the bootstrapping needs an approximate reduction modulo~$1$.
This reduction has to act on each coefficient, whereas homomorphic operations act on the slots, so the circuit moves the coefficients to the slots by \cts, evaluates the reduction slotwise, and returns to the coefficients by \stc.

\spru\ \cite{SPRU25} and \ship\ \cite{SHIP25} avoid both \cts\ and the reduction.
They evaluate the decryption equation directly on the unit circle, using a sparse secret at $q_0$ that is reached by the sparse-secret encapsulation of Section~\ref{s:sparse-keys}.
We recall \spru\ in three steps: we write decryption as a matrix--vector product, map it to the unit circle, and pack the result into the slots.

\paragraph{Decryption matrix.}
Let $h\mid N$ be a power of two and $B=N/h$.
The \spru\ secret is block-binary (Section~\ref{s:windowed-secrets}): $s\in\{0,1\}^N$ has exactly one $1$ in each of the $h$ blocks of length $B$, and $s_0=1$.
Let $n$ be a power of two with $1\le n\le\nmax:=B/2$, and let $\ct=(c_0,c_1)$ encrypt $m=\sum_{a\in\intv{n}}m_aX^{aN/n}$ modulo $q$.
For $i\in\intv{N}$ and $a\in\intv{n}$, let $C_{i,a}$ be the coefficient of $X^{aN/n}$ in $c_1\cdot X^i$, to which we add the coefficient of $X^{aN/n}$ in $c_0$ when $i=0$.
The matrix $\mathbf{C}=(C_{i,a})\leftarrow{\sf DecMat}(\ct)$ then satisfies, error included,
\begin{equation}
\label{eq:spru-decmat}
  m_a=\sum_{i\in\intv{N}} s_i\, C_{i,a}\pmod q.
\end{equation}
We fix $s_0=1$ so that $c_0$ can be merged into row $i=0$.
Decryption is then a sum over the $h$ blocks, and the product in~\eqref{eq:spru-circle} below needs no additional factor for $c_0$.
This reveals the first block of the secret, which we take into account in the security analysis (the anchored instance of Section~\ref{s:sparse-budget}).

\paragraph{Roots of unity.}
The map $x\mapsto\exp(2\pi\mathrm{I}\,x/q)$ turns the sum in~\eqref{eq:spru-decmat} into a product.
Since each block of $s$ contains a single $1$, it selects exactly one root:
\begin{equation}
\label{eq:spru-circle}
  \exp\!\left(2\pi\mathrm{I}\,\frac{m_a}{q}\right)
  =\prod_{b\in\intv{h}}\;
    \sum_{j\in\intv{B}}
      s_{bB+j}\cdot\exp\!\left(2\pi\mathrm{I}\,\frac{C_{bB+j,a}}{q}\right).
\end{equation}
To recover $m_a$, we scale each root by $\delta=(q/(4\pi\Delta))^{1/h}$.
Twice the imaginary part of the product is then $\frac{q}{2\pi\Delta}\sin(2\pi m_a/q)\simeq m_a/\Delta$ for $|m_a|\ll q$, and ${\sf Im2}$ extracts it.
The quality of this approximation depends on the ratio $\gamma=q/\Delta$ of the modulus to the scale of the message: for $|m_a|/\Delta\le1$, the angle $2\pi m_a/q$ is at most $2\pi/\gamma$.

\paragraph{Packing.}
The slots must hold the candidate roots of~\eqref{eq:spru-circle}.
There are $B$ of them for each of the $n$ coefficients and $h$ blocks, so $nN$ values in total, and we spread them over $2n$ encodings of $N/2$ slots.
Encoding $u\in\intv{2n}$ takes the sub-block of offsets $j(u,k)=u\cdot B/(2n)+k$, $k\in\intv{B/(2n)}$, of every block.
The map $\iota(k,b,a)=k\cdot hn+b\cdot n+a$ is a bijection from $\intv{B/(2n)}\times\intv{h}\times\intv{n}$ onto $\intv{N/2}$, and slot $\iota(k,b,a)$ holds offset $j(u,k)$ of block $b$ for coefficient $a$.
The secret and the public packing share this layout:
\begin{align}
\label{eq:spru-index}
  \tilde{s}^{u}_{\iota(k,b,a)}=s_{bB+j(u,k)},\qquad
  \rho^{u}_{\iota(k,b,a)}=\delta\cdot\exp\!\left(2\pi\mathrm{I}\,\frac{C_{bB+j(u,k),a}}{q}\right),
\end{align}
with $S_u={\sf Ecd}(\tilde{\vec s}^{u})$ and $E_u={\sf Ecd}(\vec\rho^{u})$.
For example, for $N=8$, $h=2$, $B=4$ and $n=1$, the two secret packings are the following.
\begin{center}
\small
\begin{tabular}{lcccc}
\toprule
Slot $\iota(k,b,0)$ & $0$ & $1$ & $2$ & $3$\\
$(k,b)$ & $(0,0)$ & $(0,1)$ & $(1,0)$ & $(1,1)$\\
\midrule
$\tilde{\vec s}^{0}$ & $s_0$ & $s_4$ & $s_1$ & $s_5$\\
$\tilde{\vec s}^{1}$ & $s_2$ & $s_6$ & $s_3$ & $s_7$\\
\bottomrule
\end{tabular}
\end{center}

We first sum the slotwise products $\tilde{\vec s}^u\odot\vec\rho^u$ over $u$, and then over $k$ with $\tr{N/2}{hn}$.
Only the offset of the single $1$ of each block survives, so slot $bn+a$ now holds the scaled root that block $b$ selects for coefficient $a$.
In the example, the trace adds slots $0$ and $2$, which cover the first block, and slots $1$ and $3$, which cover the second.
Next, $\pr{hn}{n}$ multiplies the $h$ blocks, and ${\sf Im2}$ yields $m_a/\Delta$, repeated $N/(2n)$ times.
This is the input of $\stc_n$.
For $n=1$, the input is the constant $m_0/\Delta$ in every slot, whose encoding is the constant polynomial $m_0$, so $\stc_1$ is the identity.

\paragraph{Homomorphic evaluation.}
The bootstrapping key consists of the $2n$ ciphertexts ${\sf cs}_u\leftarrow{\sf Enc}_{\pk}(S_u)$, encrypted at a modulus high enough for the circuit (Algorithm~\ref{alg:spru-keygen}\inaux).
Algorithm~\ref{alg:spru-boot} builds each $E_u$ in the clear and evaluates the steps above, computing the slotwise products as ${\sf ExtMultR}({\sf cs}_u,E_u)$.
The plaintext--ciphertext products consume one level, which Section~\ref{s:extprod} moves to the auxiliary prime.
The product operator consumes $\log h$ levels, the trace and ${\sf Im2}$ consume none, and $\stc_n$ consumes one more, or none for $n=1$, where it is the identity.
The output encrypts $\sum_a m_a X^{aN/n}$ at the remaining modulus, which is larger than $q$.

\begin{algorithm}[h]
\caption{\spru\ bootstrapping \cite{SPRU25}}
\label{alg:spru-boot}
\begin{algorithmic}[1]
  \Require $q$, key $({\sf cs}_u)_{u<2n}$, ciphertext $\ct$, $\delta=(q/(4\pi\Delta))^{1/h}$
  \Ensure Refreshed ciphertext $\ct'$
  \State $\mathbf{C}\leftarrow{\sf DecMat}(\ct)$, ${\sf acc}\leftarrow 0$
  \For{$u=0$ to $2n-1$}
    \ForAll{$0\le k<B/(2n)$, $0\le b<h$, $0\le a<n$}
      \State $\rho^{u}_{\iota(k,b,a)}\leftarrow\delta\cdot\exp(2\pi\mathrm{I}\cdot C_{bB+j(u,k),a}/q)$
    \EndFor
    \State $E_u\leftarrow{\sf Ecd}(\vec\rho^{u})$
    \State ${\sf acc}\leftarrow{\sf Add}({\sf acc},{\sf ExtMultR}({\sf cs}_u,E_u))$
  \EndFor
  \State \Return $\ct'\leftarrow\stc_n\bigl({\sf Im2}\bigl(\pr{hn}{n}(\tr{N/2}{hn}({\sf acc}))\bigr)\bigr)$
\end{algorithmic}
\end{algorithm}

\paragraph{The operator $\accmul$.}
\label{s:accmul}
A \emph{single} $\accmul$ turns one public packing $\vec\rho^{u}$ into its contribution to the accumulator of Algorithm~\ref{alg:spru-boot}.
It encodes the packing by ${\sf iDFT}$, applies an $\mathsf{NTT}$ to reach the \rns\ representation, computes the plaintext--ciphertext product with ${\sf cs}_u$, and adds the result to the accumulator.
Single $\accmul$s are independent of one another, so they are the units that we distribute over the cores.
The \emph{full} $\accmul$ load is the set of all single $\accmul$s of one bootstrapping, which is $2n$ here and more in Section~\ref{s:more-slots}.
Computing the roots that fill $\vec\rho^{u}$ is not part of $\accmul$ (see Section~\ref{s:exponentiation-cost}).

\subsection{\ship\ bootstrapping}
\label{s:ship-boot}

\ship\ \cite{SHIP25} evaluates the same unit-circle identity with the other side of the product encrypted.
Let $\ct=(c_0,c_1)$ hold the messages in its coefficients $m_t$ modulo $q_0$, under a sparse key $s$ of weight $h$, and let $\omega=\exp(2\pi\mathrm{I}/q_0)$.
Writing $(\cdot)_t$ for the coefficient of $X^t$, decryption gives
\begin{equation}
\label{eq:ship-circle}
  \omega^{m_t}=\omega^{(c_0)_t}\prod_{j\,:\,s_j\ne0}\omega^{(s_j\,X^j c_1)_t}.
\end{equation}
For each nonzero $s_j$, the factor over all $t$ is a vector of roots of unity read off $c_1$, shifted by the position $j$ and signed by $s_j$.
\spru\ builds such a factor by multiplying public roots with encrypted bits of the secret.
\ship\ instead starts from public vectors and rotates them by the encrypted position $j$, which is a \emph{blind rotation}.
Its secret is windowed ternary (Section~\ref{s:windowed-secrets}): each nonzero coefficient lies in a public window, so only its position within the window has to stay encrypted.

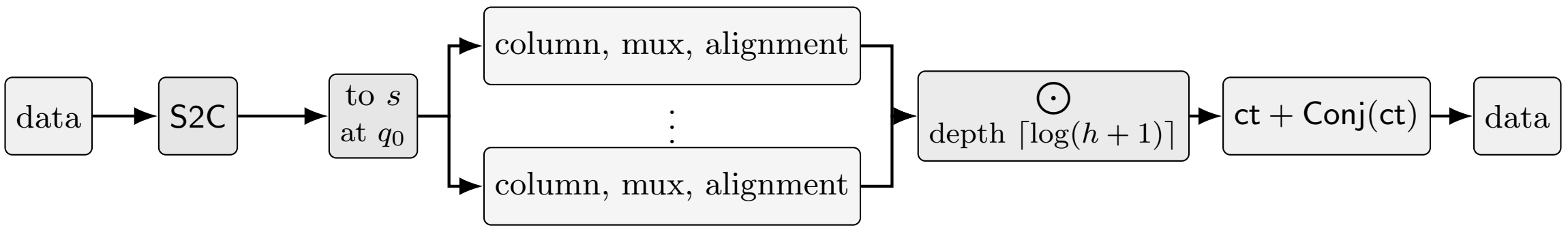
\begin{figure}[ht]
\centering
\begin{adjustbox}{max width=\textwidth}
\begin{tikzpicture}[
  stage/.style={draw,rounded corners=2pt,align=center,font=\small,inner sep=3pt,minimum height=0.75cm},
  arr/.style={-{Latex},thick}
]
  \node[stage,fill=black!6] (data1) at (0,0) {data};
  \node[stage,fill=black!10] (s2c) at (1.45,0) {$\stc$};
  \node[stage,fill=black!10] (ks) at (3.15,0) {to $s$\\[-1pt]{\scriptsize at $q_0$}};
  \node[stage,fill=black!4] (br0) at (6.05,0.68) {column, mux, alignment};
  \node[stage,fill=black!4] (br1) at (6.05,-0.68) {column, mux, alignment};
  \node[font=\small] at (6.05,0) {$\vdots$};
  \node[stage,fill=black!8] (odot) at (9.75,0)
    {$\bigodot$\\[-1pt]{\scriptsize depth $\lceil\log(h+1)\rceil$}};
  \node[stage,fill=black!6] (conj) at (12.4,0) {$\ct+{\sf Conj}(\ct)$};
  \node[stage,fill=black!6] (data2) at (14.25,0) {data};
  \draw[arr] (data1) -- (s2c);
  \draw[arr] (s2c) -- (ks);
  \draw[arr] (ks.east) -- ++(0.3,0) |- (br0.west);
  \draw[arr] (ks.east) -- ++(0.3,0) |- (br1.west);
  \draw[arr] (br0.east) -- ++(0.3,0) |- (odot.west);
  \draw[arr] (br1.east) -- ++(0.3,0) |- (odot.west);
  \draw[arr] (odot) -- (conj);
  \draw[arr] (conj) -- (data2);
\end{tikzpicture}
\end{adjustbox}
\caption{\ship\ bootstrapping.
After \stc\ and the switch to the sparse key $s$ at $q_0$, the $h$ blind rotations run independently and return their factors at the top of the modulus chain, where the product tree and the conjugation follow.}
\label{fig:ship-boot}
\end{figure}

Figure~\ref{fig:ship-boot} shows the circuit.
At the bottom of the chain, \ship\ applies \stc\ and switches to the sparse key at $q_0$, as \spru\ does in Section~\ref{s:spru-boot}.
Each of the $h$ factors is then computed by a blind rotation that combines two methods \cite[Section~4.3]{SHIP25}, with one step $\theta$ between them, and it starts with the column method, which needs public inputs.
\begin{itemize}
  \item \emph{Column.} The low part of the position, modulo $\theta$, is resolved by $4\theta$ plaintext--ciphertext products between public vectors and encrypted masks of the key. They use the auxiliary modulus and no key switch. Its cost grows linearly with $\theta$. This is the order of their implementation; their Algorithm~4 exchanges the two parts of the position.
  \item \emph{Multiplexer.} The high part is resolved one digit at a time. Section~4.3 of \cite{SHIP25} uses binary digits, and their Section~5.1 generalizes this to a base $\nu$. For each digit, an encrypted multiplexer selects among the corresponding rotations, with one key switch. The window of half-width $w$ requires $\lceil\log_\nu\lceil 2w/\theta\rceil\rceil$ digits.
  \item \emph{Alignment.} The start of the window is public, so an ordinary rotation moves the result there. In the implementation that we use, this takes one key switch.
\end{itemize}
The public term $\omega^{(c_0)_t}$, scaled by $\gamma/(4\pi\mathrm{I})$, and the $h$ factors are then multiplied by $\bigodot$, of depth $\lceil\log(h+1)\rceil$, and $\ct+{\sf Conj}(\ct)$ extracts twice the real part, which is again $\frac{\gamma}{2\pi}\sin(2\pi m_t/q_0)$.

The published sets take $h=31$, so the product has depth $5$.
Their experiments favour $\nu=4$, and all four sets take it, with $\theta$ between $6$ and $17$ and a multiplexer range of $64=4^3$ \cite[Table~2]{SHIP25}.
This gives $3$ digits, hence four key switches in sequence per factor in the implementation that we use (Table~\ref{tab:ship-params}).
The $h$ blind rotations are independent, but the chain of key switches within each of them bounds the latency even with unlimited cores.
Only \stc\ depends on the number of slots.
Table~\ref{tab:lola-vs-ship} in Section~\ref{s:comparison} compares a factor of \ship\ with one of \tutti.

\section{Sparsely structured secret keys}
\label{s:sparse-keys}

\spru and \ship evaluate the decryption equation at the base prime \(q_0\) homomorphically, and their cost grows with the Hamming weight \(h\) of the secret in that equation: \spru multiplies \(h\) block roots at depth \(\log h\), and \ship performs \(h\) blind rotations.
A sparse secret is weaker than a dense one at the same modulus, and the ciphertext chain needs a large modulus.
We therefore use sparse-secret encapsulation \cite{BTH22}, which relies on two keys.
The data lives under a dense ternary key, which must resist attacks at the full modulus \(PQ\).
The decryption equation that bootstrapping evaluates uses a sparse key, which is exposed only at \(q_0\).
A key switch at \(q_0\) moves a ciphertext from the dense to the sparse key, and the bootstrapping key, an encryption of the sparse key under the dense one, moves it back.
The dense key bounds \(\log PQ\) (Table~\ref{tab:dense-key}), and the sparse key bounds \(\log q_0\) (Sections~\ref{s:sparse-budget} and~\ref{s:may}).
Section~\ref{s:param-security} checks our parameter sets against both bounds.
Our parameter sets ($\mathtt{S13}$, $\mathtt{S13m2}$, $\mathtt{LL13}$, $\mathtt{LL14}$, $\mathtt{HT14}$, $\mathtt{HT15}$) are listed in Table~\ref{tab:lola-params}, and \ship's ($\mathtt{LL13}$, $\mathtt{LL14}$, $\mathtt{HT14}$, $\mathtt{HT15}$) in Table~\ref{tab:ship-params}.

\paragraph{Assumptions and cost model.}
Every published key except the switching key into the sparse key is an \rlwe sample under the dense key.
The switching key consists of \rlwe samples under the sparse key, so it must live at a modulus the sparse key tolerates: the BV switch keeps it at \(q_0\), whereas a GHS switch would place it at \(Pq_0\) (Section~\ref{s:ckks}).
Our implementations of \tutti and \ship both use the BV switch.
As every bootstrappable scheme does, we assume circular security, and we target IND-CPA security; IND-CPA\(^{\rm D}\) \cite{LM21} is out of scope, as for the schemes we compare with.
The error is discrete Gaussian with standard deviation \(\sigma=3.2\), and costs are classical.
A parameter set is accepted when the cheapest attack is estimated to cost at least \(128\) bits in word operations, the unit of the lattice estimator \cite{APS15,LatticeEstimator} at commit $\mathtt{53da598}$; Section~\ref{s:may} converts the combinatorial costs into that unit.

\paragraph{Block-binary and windowed secrets.}
\label{s:windowed-secrets}
The block-binary secret of \spru\ (Section~\ref{s:spru-boot}) has exactly one \(1\) in each of the \(h\) blocks of length \(B=N/h\).
A windowed secret, as in \ship \cite{SHIP25}, is ternary of weight \(h\): its \(k\)-th nonzero coefficient lies in the window of half-width \(w\) centered at \(kB\), and \ship takes \(w=\max(175,\,B/2)\).

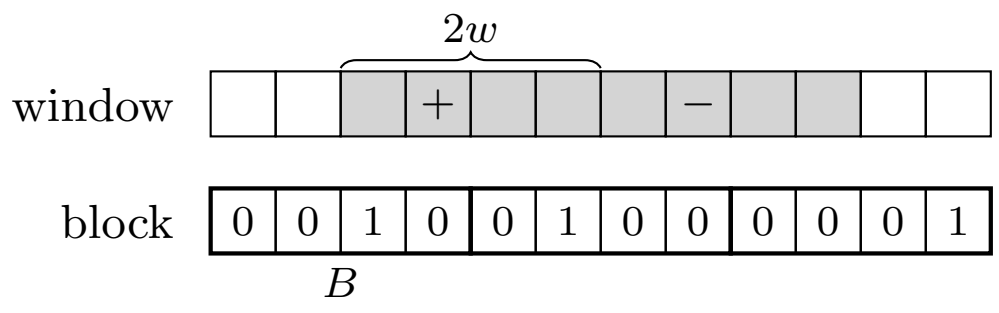
\begin{center}
\begin{tikzpicture}[x=0.43cm,y=0.43cm]
  \node[anchor=east,font=\small] at (-0.25,0.5) {block};
  \foreach \i/\v in {0/0,1/0,2/1,3/0,4/0,5/1,6/0,7/0,8/0,9/0,10/0,11/1}{
    \draw (\i,0) rectangle ++(1,1);
    \node[font=\scriptsize] at (\i+0.5,0.5) {\v};
  }
  \draw[thick] (0,0) rectangle (4,1);
  \draw[thick] (4,0) rectangle (8,1);
  \draw[thick] (8,0) rectangle (12,1);
  \node[font=\scriptsize] at (2,-0.45) {$B$};
  \node[anchor=east,font=\small] at (-0.25,2.3) {window};
  \fill[light-gray] (2,1.8) rectangle (6,2.8);
  \fill[light-gray] (6,1.8) rectangle (10,2.8);
  \foreach \i in {0,...,11}{
    \draw (\i,1.8) rectangle ++(1,1);
  }
  \node[font=\scriptsize] at (3.5,2.3) {$+$};
  \node[font=\scriptsize] at (7.5,2.3) {$-$};
  \draw[decorate,decoration={brace,amplitude=3pt}] (2,2.85) -- (6,2.85)
    node[midway,above=1pt,font=\scriptsize] {$2w$};
\end{tikzpicture}\\[2pt]
{\small Schematic, \(w=B/2\): one \(1\) per block, one \(\pm 1\) per gray window around a block boundary.}
\end{center}

\paragraph{Subring secrets.}
Some of our parameter sets restrict the block-binary key further, to the form \(s(X)=s'(X^R)\) for a divisor \(R\) of \(B\); Section~\ref{s:dimension-reduction} explains what this saves.
The restriction leaves an \rlwe problem of degree \(N/R\):

\begin{proposition}
\label{prop:subring-secret}
Let $N$ be a power of two and let $R\mid N$.
Let $s'\in\mathbb{Z}[Y]/(Y^{N/R}+1)$ and set $s(X)=s'(X^R)$.
For any modulus $q$ and any distribution $\chi$ on $\mathbb{Z}$, the search and decision coefficient-form \rlwe\ problems for one sample in dimension $N$ with secret $s$ and coefficient error $\chi$ are equivalent to the same problems for $R$ samples in dimension $N/R$ with secret $s'$ and coefficient error $\chi$.
\end{proposition}

The equivalence is standard: ring switching \cite{GHPS12} and subring secret encapsulation \cite{MHWW25} rest on the same decomposition.
We include the proof, a rearrangement of coefficients, for completeness.

\begin{proof}
Write $\mathcal{R}=\mathbb{Z}[X]/(X^N+1)$ and $\mathcal{R}'=\mathbb{Z}[Y]/(Y^{N/R}+1)$, and $\mathcal{R}_q$, $\mathcal{R}'_q$ for their reductions modulo $q$.
Coefficient-form \rlwe\ is the variant of the search and decision problems of~\cite[Definitions~3.2 and~3.3]{LPR13} with coefficient-wise independent errors; we do not assume an arbitrary error distribution in the canonical embedding.

The substitution $\phi\colon\mathcal{R}'\to\mathcal{R}$, $Y\mapsto X^R$, is a ring homomorphism, well defined because $(X^R)^{N/R}=X^N=-1$, and it reduces modulo $q$.
For $a\in\mathcal{R}_q$ and $0\le j<R$, let $a^{(j)}\in\mathcal{R}'_q$ collect the coefficients of $a$ in the degrees congruent to $j$ modulo $R$:
\begin{equation}
\label{eq:subring-split}
  a^{(j)}(Y)=\sum_{k=0}^{N/R-1} a_{j+kR}\,Y^k,
  \qquad
  a=\sum_{j=0}^{R-1} X^j\,\phi\!\left(a^{(j)}\right).
\end{equation}
The map $a\mapsto(a^{(0)},\ldots,a^{(R-1)})$ only reorders coefficients.
It is therefore a bijection $\mathcal{R}_q\to(\mathcal{R}'_q)^R$ of $\mathbb{Z}_q$-modules, it maps a uniform $a$ to $R$ independent uniform $a^{(j)}$, and it maps an error with independent coefficients from $\chi$ to $R$ independent errors of the same kind.

Multiplication by $s=\phi(s')$ preserves the residue classes: $X^j\phi(a^{(j)})\cdot\phi(s')=X^j\phi(a^{(j)}\cdot s')$, so the decomposition of $a\cdot s$ is $(a^{(j)}\cdot s')_{j<R}$.
A sample $b=a\cdot s+e$ thus decomposes into the $R$ samples $b^{(j)}=a^{(j)}\cdot s'+e^{(j)}$ of dimension $N/R$ under $s'$, and the inverse map rebuilds one sample under $s$ from $R$ such samples.
The same bijection of $\mathcal{R}_q^2$ identifies the uniform distribution on pairs with $R$ independent uniform pairs.
An algorithm for the search or the decision problem therefore transfers in either direction, converting a recovered secret between $s'$ and $\phi(s')$ in the search case, with $k$ samples in dimension $N$ corresponding to $kR$ samples in dimension $N/R$.
Success probabilities and advantages are identical, and the conversion costs $O(kN)$.
For a random secret, independence is conditional on $s'$, and the correspondence holds for any distribution of $s'$, pushed forward to $s=\phi(s')$.
\end{proof}

In our setting, the coefficients of \(s'\) are \((s_0,s_R,s_{2R},\ldots)\).
Since \(R\mid B\), the unique \(1\) in each block of \(s\) lies on the \(R\)-grid, and those grid positions form a contiguous block of length \(B/R\) in \(s'\).
Thus \(s'\) is a block-binary secret of degree \(N/R\) and weight \(h\), and each published sample under \(s\) gives \(R\) samples under \(s'\).
The attacks below need fewer samples than one sample already provides (Section~\ref{s:sparse-budget}), so the sparse key is estimated at degree \(N/R\).

\paragraph{Dense key.}
\label{s:dense-budget}
Table~\ref{tab:dense-key} gives the largest integer bit length \(\log q\) at which the cheapest classical attack of the lattice estimator still costs at least \(128\) bits.
The estimates at these caps lie between \(128.0\) and \(129.1\) bits, whether the attack gets the \(N\) \lwe samples of one \rlwe sample or unlimited samples, as from the many published keys.
On \(\mathtt{LL13}\), \tutti\ and \ship\ both use a dense ternary key of Hamming weight \(512\), with as many entries \(-1\) as \(+1\) (Section~\ref{s:params}).
With the hybrid attacks included, it costs \(128.3\) bits at \(212\) and \(127.7\) at \(213\) for \(N=2^{13}\), and May's Meet-LWE \cite{May21} costs over \(2^{1000}\) on it at that degree.

\begin{table}[t]
\centering
\caption{Largest \(\log q\) at which the dense key costs at least \(128\) bits.}
\label{tab:dense-key}
\setlength{\tabcolsep}{6pt}
\small
\begin{tabular}{lccccc}
\toprule
\(N\) & \(2^{12}\) & \(2^{13}\) & \(2^{14}\) & \(2^{15}\) & \(2^{16}\) \\
\midrule
Uniform ternary & \(106\) & \(214\) & \(430\) & \(868\) & \(1746\) \\
Hamming weight \(512\) & --- & \(212\) & \(428\) & \(862\) & --- \\
\bottomrule
\end{tabular}
\end{table}

\subsection{Lattice attacks on the sparse key}
\label{s:sparse-budget}

Hybrid lattice attacks have long exploited the sparsity of such secrets \cite{Alb17,SC19,CP19}.
The most recent ones also exploit the ring structure of sparse-secret \rlwe in power-of-two cyclotomic rings: Ogilvie's rotated primal hybrid and Hou and Jiang's enhanced hybrid decoding attack, merged into one paper \cite{HouJO26}.
Table~\ref{tab:sparse-block} applies both to \spru's block-binary secret.

Since \spru fixes \(s_0=1\) (Section~\ref{s:spru-boot}), the first block of the secret is known.
An attacker therefore faces what we call the \emph{anchored} instance: the remaining \(h'=h-1\) blocks, of total dimension \(N'=N-B\), each still with exactly one \(1\).
For each pair \((N,h)\), Table~\ref{tab:sparse-block} gives the largest integer bit length \(\log q_0\) at which the rotated primal hybrid still costs at least \(128\) bits on this instance.
The enhanced hybrid was more expensive on every row.
Both attacks need fewer equations than a single \rlwe sample provides: even on the row \((2^{11},64)\), which has the fewest, they use \(765\) of its \(N'=2016\).
The additional samples that a subring secret gives (Proposition~\ref{prop:subring-secret}) are therefore of no use to them, and indeed, rerunning both attacks with these samples, for \(R\le 32\) and degrees up to \(2^{16}\), left the caps unchanged.

\begin{table}[t]
\centering
\caption{Largest \(\log q_0\) at which the rotated primal hybrid \cite{HouJO26} costs at least \(128\) bits on the block-binary secret anchored by \(s_0=1\), of dimension \(N'=N-B\) and weight \(h'=h-1\).
A key \(s'(X^R)\) is checked at degree \(N/R\).}
\label{tab:sparse-block}
\setlength{\tabcolsep}{6pt}
\begin{tabular}{ccccc}
\toprule
\(N\) & \(h\) & \(N'\) & \(h'\) & \(\log q_0\) \\
\midrule
\(2^{11}\) & \(64\) & \(2016\)  & \(63\) & \(27\)  \\
\(2^{12}\) & \(64\) & \(4032\)  & \(63\) & \(51\)  \\
\(2^{13}\) & \(64\) & \(8064\)  & \(63\) & \(102\) \\
\(2^{14}\) & \(64\) & \(16128\) & \(63\) & \(209\) \\
\midrule
\(2^{13}\) & \(32\) & \(7936\)  & \(31\) & \(27\)  \\
\(2^{14}\) & \(32\) & \(15872\) & \(31\) & \(59\)  \\
\bottomrule
\end{tabular}
\end{table}

This modulus caps the sparse key, not the ciphertext chain, which stays under the dense key.
At the same degree the sparse cap lies well below the dense one, \(102\) bits against \(214\) at \(N=2^{13}\) and \(h=64\), and halving the weight cuts it further, to \(27\) bits.
The hybrid's meet-in-the-middle stage ignores the blocks (\app{app:may-out}).

\ship's windowed secret has weight \(h=31\), the anchored weight \(h'\) of the rows with \(h=32\).
\ship fixes no coefficient, so its instance keeps the full dimension \(N\) instead of \(N'=N-B\), and a cap of Table~\ref{tab:sparse-block} can only be conservative for it.
Its base primes lie within these caps: \(25\) bits on \(\mathtt{LL13}\) against \(27\), \(48\) and \(28\) bits on \(\mathtt{LL14}\) and \(\mathtt{HT14}\) against \(59\), and \(50\) bits on \(\mathtt{HT15}\), whose degree \(2^{15}\) is larger than that of the \(59\)-bit row.
\ship's own lattice check used the GHS modulus \(P\cdot q_0\), at \(55\), \(100\), and \(105\) bits \cite{SHIP25}, above these caps, so our implementation of \ship also switches at \(q_0\) by BV (Section~\ref{s:bv-ks}).

\subsection{Revisiting May's attack}
\label{s:may}

A sparse key has few possible values, so it can also be attacked combinatorially, by searching for the secret directly instead of reducing a lattice.
We cost three such attacks on the anchored instance of Section~\ref{s:sparse-budget}, whose search space has size \(S=B^{h'}\), at the modulus of the same row of Table~\ref{tab:sparse-block}.
Each of them reads one \rlwe sample as \(N'\) uniform \lwe equations modulo \(q_0\) and does not use the ring.
We describe each attack and then its cost on our rows, which Table~\ref{tab:may-spru} collects.
\app{app:may} gives the details, and our script reproduces May's table for NTRU-HPS to within \(1\) to \(5\) bits.
Our main finding comes last: the closed exponents that earlier bootstrapping papers derived from May's attack underestimate these costs by at least \(25\) bits.
Costing the attacks instead admits a smaller ring degree for the same security, and with it a lower latency: a single message can be bootstrapped at \(N=2^{13}\) instead of \(N=2^{15}\).

\paragraph{Odlyzko's meet-in-the-middle.}
The unknowns are the \(h'\) remaining blocks, each with a single \(1\) at an unknown offset in \(\intv{B}\).
This attack \cite{Odlyzko90,HSW03} splits the secret as \(s=s_1+s_2\), where \(s_1\) holds the first \(\lceil h'/2\rceil\) blocks and \(s_2\) the others, so that each part has about \(B^{h'/2}\) candidates.
It builds two lists, of \(a\cdot s_1\) and of \(b-a\cdot s_2\) over all candidates.
Since \(b=a\cdot s+e\) with a small error \(e\), the correct pair agrees up to \(e\), and sorting both lists on the top bits of a few coordinates finds it, so the attack does not have to guess the error.
Its larger list has \(T_{\mathrm{Odl}}=B^{\lceil h'/2\rceil}\) elements, which is \(S^{1/2}\) up to the rounding of an odd \(h'\).
This is the column \(T_{\mathrm{Odl}}\) of Table~\ref{tab:may-spru}, against which we compare the two other attacks.

\paragraph{May's Meet-LWE.}
May's attack \cite{May21} makes these lists smaller with \emph{representations}.
It writes \(s=s_1+s_2\) in many ways, also with cancellations \(0=1+(-1)\) on zero coordinates, so that the secret has many representations as a pair, and finding any one of them suffices.
The lists, that is, the candidate sets for \(s_1\) and \(s_2\), can therefore be filtered: a candidate is kept only if \(a\cdot s_1\) takes a fixed value on a few coordinates of the \lwe\ equation, which only a fraction of the representations satisfy.
Each of \(s_1\) and \(s_2\) is found in the same way, by splitting it again, so the attack is a tree of lists with one level per split, and the number of levels is its \emph{depth} (\app{app:may-rep1}).
Fixing an exact value on these coordinates requires guessing the error there.
For a Gaussian error, this guess costs more than for May's ternary one, and it grows as \(q_0\) shrinks.
As a result, Meet-LWE with error guessing beats Odlyzko's split on three rows only, by at most \(8\) bits, and we list it only in Table~\ref{tab:may-params} of the appendix.

\paragraph{Guessing-free Meet-LWE.}
This variant \cite{KM21,HKLS22,BLLLW26} drops the guess.
Its filter keeps a candidate when \(a\cdot s_1\) lies in a box of the error's size on the selected coordinates, instead of taking an exact value, and two partial sums match when they agree up to that box.
We instantiate it at depth \(2\) on the representations of \app{app:may-rep1}, with the cost model of \cite[Section~3.3]{BLLLW26}, which gives the column \(T_{\mathrm{free}}\) of Table~\ref{tab:may-spru}.
On the rows with \(h=64\) a deeper tree may be cheaper, so these entries are upper bounds.
A deeper tree would have to gain at least \(30\) bits to reach \(128\), whereas with error guessing, where \app{app:may} searches depths up to \(4\), the best tree never lies more than \(8\) bits below Odlyzko's split.

\begin{table}[t]
\centering
\caption{Rounded bit costs of combinatorial attacks (green) on the anchored block-binary secret at the moduli of Table~\ref{tab:sparse-block}: Odlyzko's split \(T_{\mathrm{Odl}}\) and the depth-\(2\) guessing-free Meet-LWE \(T_{\mathrm{free}}\), in list elements, and \(T_{\mathrm{free}}\) in word operations, \(T_{\mathrm{ops}}\).
The closed powers \(T_{\mathrm{BBK}}=S^{0.28}\) \cite{LMSS23} and \(T_{\mathrm{SHIP}}=S^{0.4}\) \cite{SHIP25} of \(S=B^{h'}\) (red) are not attacks.
Bold: the cheapest attack, in list elements and, for \(N=2^{13}\), \(h=32\), in word operations.}
\label{tab:may-spru}
\setlength{\tabcolsep}{3.5pt}
\small
\begin{tabular}{ccc|>{\columncolor{maygreen}}c>{\columncolor{maygreen}}c>{\columncolor{maygreen}}c>{\columncolor{mayred}}c>{\columncolor{mayred}}c}
\toprule
\(N\) & \(h\) & \(\log q_0\) & \(T_{\mathrm{Odl}}\) & \(T_{\mathrm{free}}\) & \(T_{\mathrm{ops}}\) & \(T_{\mathrm{BBK}}\) & \(T_{\mathrm{SHIP}}\) \\
\midrule
\(2^{11}\) & \(64\) & \(27\)  & \(160\) & \(\le\mathbf{158}\) & \(\le 165\) & \(88\)  & \(126\) \\
\(2^{12}\) & \(64\) & \(51\)  & \(192\) & \(\le\mathbf{186}\) & \(\le 194\) & \(106\) & \(151\) \\
\(2^{13}\) & \(64\) & \(102\) & \(224\) & \(\le\mathbf{216}\) & \(\le 224\) & \(123\) & \(176\) \\
\(2^{14}\) & \(64\) & \(209\) & \(256\) & \(\le\mathbf{246}\) & \(\le 254\) & \(141\) & \(202\) \\
\midrule
\(2^{13}\) & \(32\) & \(27\)  & \(128\) & \(\mathbf{125}\) & \(\mathbf{132}\) & \(69\)  & \(99\)  \\
\(2^{14}\) & \(32\) & \(59\)  & \(144\) & \(\mathbf{139}\) & \(146\) & \(78\)  & \(112\) \\
\bottomrule
\end{tabular}
\end{table}

\paragraph{The row \(N=2^{13}\), \(h=32\).}
This is the only row within \(10\) bits of \(128\), so it is the only one where the unit of the cost matters.
Its guessing-free top list has \(L=2^{125.1}\) elements.
May counts list elements and omits the polynomial factors of sorting and hashing \cite[Section~2.3]{May21}, whereas our criterion counts word operations.
Hashing one element on about \(\log L\) coordinates costs one word operation per coordinate, which adds \(\log\log L\approx 7\) bits.
In word operations, the row therefore costs \(132\) bits and Odlyzko's split \(135\), the figures of Table~\ref{tab:param-security}.
Bi et al.'s hash repetitions and the memory of the lists only raise these costs (\app{app:may-free}).
Our set \(\mathtt{S13}\) uses this row at \(\log q_0=24<27\), which leaves the list sizes, and with them these costs, unchanged.

\paragraph{Closed exponents underestimate the cost of these attacks.}
May's Meet-LWE \cite{May21} is the reference attack when bootstrapping papers choose parameters for sparse secrets, and three of them quote it as a closed power of the key-space size.
In the sparse regime, these powers lie at least \(25\) bits below the attacks instantiated above, so they are far too conservative.
Lee, Min, Seo, and Song \cite{LMSS23} estimate a block-binary key by \(T=S^{0.28}\) with \(S=(B+1)^{h}\), allowing a block to be entirely zero.
\ship quotes \(T\approx(4w)^{0.4h}\) scalar products, the factor \(4\) accounting for ternary signs, and requires \((4w)^{0.4\cdot 31}>2^{115}\) for an estimated \(128\) bits at \(N\ge 2^{13}\) \cite{SHIP25}.
Coron and K\"ostler use \(T=S^{0.25}\) with \(S=(N/h)^{h-1}\) \cite[Section~4.5]{SPRU25}.
On \(S=B^{h'}\), these closed powers claim \(69\) and \(99\) bits on the row \(N=2^{13}\), \(h=32\), where the cheapest attack costs \(125\) bits in list elements and \(132\) in word operations (Table~\ref{tab:may-spru}).
The reason is the regime: May's exponents \(0.28\) to \(0.3\) come from the dense ternary secrets of NTRU and BLISS \cite{HPS98,DDLL13}, of relative weight \(0.3\) to \(0.667\), and he claims the form \(T\approx S^{1/4}\) for none of the smaller weights \cite{May21}.
Here \(h/N\in\{2^{-9},\ldots,2^{-5}\}\), and the saving from representations shrinks relative to \(\log S\) as \(1/\log(N/h)\).
This is what lets a single message be bootstrapped at \(N=2^{13}\), where the estimate of \spru\ forced \(N=2^{15}\), and our parameter sets rely on the instantiated attacks (Section~\ref{s:param-security}).

\paragraph{\ship's windowed secret.}
The same attacks apply to \ship's secret, where each window has \(4w\) signed positions.
Odlyzko's split over the windows is the cheapest of them on every published set, at no less than \(142\) bits at the base prime (Table~\ref{tab:may-ship} in \app{app:may-ship}).
These costs leave room to narrow the windows, down to the smallest half-width at least \(N/(2h)\), at no less than \(136\) bits.
On \(\mathtt{LL13}\), this takes the half-width from \(175\) to \(133\), and on the other sets the published half-width is already at this limit.
Even on \(\mathtt{LL13}\), a narrower window hardly affects the latency of \ship.
The number of multiplexer digits stays \(3\), and with it the sequential key switches of a blind rotation, so at most the column step \(\theta\) could shrink, from \(6\) to \(5\).
We therefore keep the published windows.

\paragraph{Hybrid attacks.}
Hybrids of the guessing-free lists with lattice reduction gain up to \(10\) bits \cite{HKLS22} and \(18\) bits \cite{BLLW22,BLLLW26} on unstructured homomorphic-encryption sets.
Whether these gains carry over to block and window constraints, and combine with the ring isometries of \cite{HouJO26}, is open.
Like the bootstrapping papers we compare with, our estimates cover the attacks instantiated above.
As for any set at the \(128\)-bit line, such an improvement would call for smaller base primes or larger rings.

\subsection{Security of our parameter sets}
\label{s:param-security}

Table~\ref{tab:param-security} checks each parameter set of Table~\ref{tab:lola-params} against the bounds of this section.
The dense key is checked at \(\log PQ\) against Table~\ref{tab:dense-key} at degree \(N\).
The sparse key has degree \(N/R\) (Proposition~\ref{prop:subring-secret}) and is checked at \(q_0\) against the row \((N/R,h)\) of Tables~\ref{tab:sparse-block} and~\ref{tab:may-spru}.
Every set uses a base prime at or below the cap of its row.
No attack of this section becomes cheaper as the modulus decreases: the lattice attacks become more expensive, Odlyzko's split and the lists of Meet-LWE do not depend on \(q_0\), and the error guess grows.
The combinatorial costs at the cap therefore carry over to the smaller \(q_0\).

\begin{table}[t]
\centering
\caption{Security of the parameter sets of Table~\ref{tab:lola-params} against the caps of Tables~\ref{tab:dense-key} and~\ref{tab:sparse-block}; the last column gives the costs of Table~\ref{tab:may-spru} in word operations at the cap of each row.
The cap of $\mathtt{LL13}$ is the one of its dense key of Hamming weight $512$.}
\label{tab:param-security}
\setlength{\tabcolsep}{4pt}
\small
\begin{tabular}{lcccccc}
\toprule
Set & \(N\) & \(R\) & \(\log PQ\) (cap) & \((N/R,h)\) & \(\log q_0\) (cap) & Odlyzko / guessing-free \\
\midrule
\(\mathtt{S13}\)  & \(2^{13}\) & \(1\) & \(208.4\) (\(214\)) & \((2^{13},32)\) & \(24\) (\(27\)) & \(135\) / \(132\) \\
\(\mathtt{S13m2}\) & \(2^{13}\) & \(1\) & \(213.6\) (\(214\)) & \((2^{13},32)\) & \(23\) (\(27\)) & \(135\) / \(132\) \\
\(\mathtt{LL13}\) & \(2^{13}\) & \(4\) & \(212.0\) (\(212\)) & \((2^{11},64)\) & \(24\) (\(27\)) & \(167\) / \(\le 165\) \\
\(\mathtt{LL14}\) & \(2^{14}\) & \(4\) & \(430.0\) (\(430\)) & \((2^{12},64)\) & \(48\) (\(51\)) & \(200\) / \(\le 194\) \\
\(\mathtt{HT14}\) & \(2^{14}\) & \(8\) & \(429.5\) (\(430\)) & \((2^{11},64)\) & \(27\) (\(27\)) & \(167\) / \(\le 165\) \\
\(\mathtt{HT15}\) & \(2^{15}\) & \(8\) & \(769.0\) (\(868\)) & \((2^{12},64)\) & \(50\) (\(51\)) & \(200\) / \(\le 194\) \\
\bottomrule
\end{tabular}
\end{table}

\section{The \texorpdfstring{\tutti}{TUTTI} bootstrapping}
\label{s:more-slots}

The original \spru\ bootstrapping \cite{SPRU25} (Algorithm~\ref{alg:spru-boot} in Section~\ref{s:spru-boot}) refreshes at most \(\nmax=B/2\) coefficients.
\spru\ already refreshes further coefficients by monomial shifts \cite[Algorithm~13]{SPRU25}: a shift adjusts the decryption matrix \({\sf DecMat}(\ct)\) of Equation~\eqref{eq:spru-decmat}, and the shifted bootstrappings share one key, run independently and are added, but every extra run takes cores from each bootstrapping.

\tutti\ instead refreshes all coefficients within one run, so that every core works on a single bootstrapping.
It exploits that the product operator of Section~\ref{s:rot-conj} wastes slots: each of its layers \({\sf id}\cdot\Psi_r\) multiplies the two halves of a period and writes the result into both.
\tutti\ fills one of the two copies with a further group of coefficients, layer by layer, up to all \(N/2\) coefficients at the same depth \(\log h\) (Section~\ref{s:double-slots}).
We write \(\tutti_n\) for our bootstrapping of \(n\) coefficients; for \(n\le\nmax\), it is \spru\ with the optimizations of this section and of Section~\ref{s:techniques}.

Compared with \spru\ on \(\nmax\) coefficients, \(\tutti_n\) with \(t=n/\nmax\) accumulators differs in two ways.
Its \accmul\ load is \(t\) times larger, and a tree of \(t-1\) products of distinct accumulators replaces the first \(\log t\) layers of the product operator at the same depth, which removes their rotations.
With many cores, the main challenge is therefore to share the \accmul\ load.
A subring secret shrinks this load (Section~\ref{s:dimension-reduction}), and many cores share it (Section~\ref{s:parallelizability}); \stc\ moves to the bottom of the chain (Section~\ref{s:stc-placement}), and cosets trade cores for a smaller key (Section~\ref{s:cosets}).
In the full-slot case the circuit takes the shape of \ship's, a product of \(h\) ciphertexts without rotations; Section~\ref{s:comparison} compares the two.

\subsection{From \texorpdfstring{$2\nmax$}{2 nmax} coefficients to all \texorpdfstring{$N/2$}{N/2} slots}
\label{s:double-slots}

We first show how \tutti\ refreshes twice as many coefficients as \spru\ by filling the slots that one layer of the product operator wastes, and then fill those of the further layers, up to all \(N/2\) coefficients.
Like \spru, the construction of this section ends with \(\stc_n\); Section~\ref{s:stc-placement} moves it to the bottom of the chain.

\paragraph{Bootstrapping twice as many coefficients.}
\spru\ refreshes \(2\nmax\) coefficients by two shifted runs.
Each run accumulates its own slot vector in Algorithm~\ref{alg:spru-boot}: \(\vec a\in\mathbb{C}^{N/2}\) for the first group of \(\nmax\) coefficients and \(\vec b\) for the second.
For \(n=\nmax\) the trace \(\tr{N/2}{hn}\) is the identity, since \(hn=N/2\), so each run applies the product operator directly and returns \(\vec p=\pr{N/2}{\nmax}(\vec a)\) and \(\vec p\,'=\pr{N/2}{\nmax}(\vec b)\), respectively.
The first layer of each product operator, \({\sf id}\cdot\Psi_{N/4}\), multiplies the two halves of its vector and writes every product twice, so each run wastes half of its slots in that layer.

\tutti\ only changes the packing.
Instead of accumulating \(\vec a\) and \(\vec b\) separately, the packings of the key and of the public roots place their entries interleaved in two accumulators,
\[
\vec v = (a_0, b_0, a_1, b_1, \ldots, a_{N/4-1}, b_{N/4-1}),
\quad
\vec w = (a_{N/4}, b_{N/4}, \ldots, a_{N/2-1}, b_{N/2-1}).
\]
This costs nothing, since it is a choice of packing.
As slot \(bn+a\) of \(\vec a\) and \(\vec b\) holds block \(b\), \(\vec v\) carries the blocks \(b<h/2\) of both groups and \(\vec w\) the others, which is the split by blocks that the generalization below uses.

A single Hadamard product \(\vec v\odot\vec w\) now performs the first layer of both product operators at once.
Its even slots hold the output of that layer on \(\vec a\) and its odd slots its output on \(\vec b\), so no slot is wasted.
The remaining layers, \(\pr{N/2}{2\nmax}\), then give \(N/(4\nmax)\) copies of
\[
\vec o = (p_0, p'_0, p_1, p'_1, \ldots, p_{\nmax-1}, p'_{\nmax-1}) \in \mathbb{C}^{2\nmax},
\]
which holds the \(2\nmax\) values of both runs in interleaved order, at the unchanged depth \(\log h\).
The packing also realizes any slot permutation for free, and one such permutation reorders \(\vec o\) as \((p_0, p_1, \ldots, p_{\nmax-1}, p'_0, p'_1, \ldots, p'_{\nmax-1})\).
This is a \(2\nmax\)-slot encoding, and \(\stc_{2\nmax}\) completes the bootstrapping of \(2\nmax\) coefficients.

\paragraph{Generalization to more coefficients.}
The case above is $t=2$, with the two accumulators $\vec v$ and $\vec w$.
For $1\le k\le\log h$ and $t=2^k$, the same construction uses $t$ accumulators $\vec v^{(g)}$, and $\vec v^{(g)}$ carries only the $h_t=h/t$ blocks $\{gh_t,\ldots,(g+1)h_t-1\}$, for all $n=t\cdot\nmax$ coefficients.
The Hadamard product $\bigodot_{g\in\intv{t}}\vec v^{(g)}$ multiplies across the $t$ groups of blocks, and $\pr{N/2}{n}$ multiplies the $h_t$ blocks within a group.
At $k=\log h$ one has $n=N/2$, the product operator is the identity, and a product tree of depth $\log h$ replaces it, with no rotations.
These vectors already fill every slot, so the trace after $\accmul$ is the identity.
${\sf Im2}$ and $\stc$ are unchanged from ordinary $\spru$.
Like \ship, $\tutti_{N/2}$ refreshes real values, so a complex vector of $N/2$ slots takes two refreshes (Section~\ref{s:slottocoeff}).
The proposition records the slot contents, including the $\accmul$ encodings; it uses the sequential slot order, which differs from the interleaved order of the example by a slot permutation.

\begin{proposition}[$2^k$-vector product]
\label{prop:2k-vector}
Let $1\le k\le\log h$, $t=2^k$, $h_t=h/t$, and $n=t\cdot\nmax$.
Let $\ct$ encrypt $m$ under the block-binary secret $s$, with coefficients $m_a$ of $X^{aN/n}$ for $a\in\intv{n}$, and let $\mathbf{C}\leftarrow{\sf DecMat}(\ct)\in\mathbb{Z}_q^{N\times n}$, so that $m_a=\sum_{i\in\intv{N}} s_i\,C_{i,a}\pmod q$ as in~\eqref{eq:spru-decmat}.
For $g\in\intv{t}$ and $u\in\intv{B}$, define slot vectors in $\mathbb{C}^{N/2}$ by
\begin{equation}
\label{eq:2k-index}
  \tilde{s}^{u,(g)}_{rn+a}=s_{(gh_t+r)B+u},
  \qquad
  \rho^{u,(g)}_{rn+a}=\delta\cdot\exp\!\Bigl(2\pi\mathrm{I}\,
      C_{(gh_t+r)B+u,\,a}/q\Bigr)
\end{equation}
for $r\in\intv{h_t}$ and $a\in\intv{n}$, and set
\begin{equation}
\label{eq:2k-accmul}
  \vec v^{(g)}=\sum_{u\in\intv{B}} \tilde{\vec s}^{u,(g)}\odot\vec\rho^{u,(g)}.
\end{equation}
Then every slot $rn+a$ of
\begin{equation}
\label{eq:2k-product}
  \pr{N/2}{n}\!\Bigl(\bigodot_{g\in\intv{t}}\vec v^{(g)}\Bigr)
\end{equation}
holds
\begin{equation}
\label{eq:2k-output}
  \delta^h\cdot\exp\!\bigl(2\pi\mathrm{I}\, m_a/q\bigr),
\end{equation}
and the multiplicative depth of~\eqref{eq:2k-product} is $k+\log h_t=\log h$.
\end{proposition}

\begin{proof}
Fix a slot $rn+a$ and a group $g$, and let $b=gh_t+r$.
The secret has a single $1$ in block $b$, at offset $j^\star_b$, so exactly one summand of~\eqref{eq:2k-accmul} survives in that slot:
$v^{(g)}_{rn+a}=\delta\cdot\exp(2\pi\mathrm{I}\,C_{bB+j^\star_b,\,a}/q)$, the root that block $b$ selects for coefficient $a$.
The Hadamard product multiplies these roots over $g\in\intv{t}$, at fixed $r$.
The operator $\pr{N/2}{n}$ multiplies the slots congruent modulo $n$, that is, over $r\in\intv{h_t}$ at fixed $a$, and writes the result to each of them.
Every block $b=gh_t+r\in\intv{h}$ thus contributes exactly once, and by~\eqref{eq:spru-decmat} the product is
\begin{align*}
  \prod_{b\in\intv{h}}\delta\cdot\exp\!\Bigl(2\pi\mathrm{I}\,\frac{C_{bB+j^\star_b,\,a}}{q}\Bigr)
  &=\delta^h\cdot\exp\!\Bigl(\frac{2\pi\mathrm{I}}{q}\sum_{i\in\intv{N}}s_iC_{i,a}\Bigr)\\
  &=\delta^h\cdot\exp\!\Bigl(2\pi\mathrm{I}\,\frac{m_a}{q}\Bigr).
\end{align*}
For the depth, the Hadamard product of $t=2^k$ vectors costs $k$ levels, and $\pr{N/2}{n}$ costs $\log h_t$ more, since $(N/2)/n=h_t$.
\end{proof}

At $k=\log h$ one has $h_t=1$ and $n=N/2$, so $\pr{N/2}{n}$ is the identity and~\eqref{eq:2k-product} is $\bigodot_{g\in\intv{h}}\vec v^{(g)}$.
For $n<N/2$ the output holds the $n$ values $h_t$ times, the sparse packing of Section~\ref{s:ckks}.
The proposition holds for an arbitrary $m$, since ${\sf DecMat}$ reads only the coefficients of $X^{aN/n}$; the cosets of Section~\ref{s:cosets} rely on this.
Homomorphically, $\vec v^{(g)}$ is the accumulator of the encodings $S_u^{(g)}={\sf Ecd}(\tilde{\vec s}^{u,(g)})$, encrypted in the key, and $E_u^{(g)}={\sf Ecd}(\vec\rho^{u,(g)})$, built in the clear; Algorithm~\ref{alg:lola-boot} records $\tutti_n$, and Algorithm~\ref{alg:lola-keygen}\inaux{} its key generation.

\begin{algorithm}[t]
\caption{$\tutti_n$ bootstrapping}
\label{alg:lola-boot}
\begin{algorithmic}[1]
  \Require $q$, key $({\sf cs}_u^{(g)})$, ciphertext $\ct$ encrypting $n$ coefficients
  \Require $\delta=(q/(4\pi\Delta))^{1/h}$, and $t,h_t,\nmax,n$ as above
  \Ensure Refreshed ciphertext $\ct'$
  \State $\mathbf{C}\leftarrow{\sf DecMat}(\ct)$
  \For{$g=0$ to $t-1$}
    \State ${\sf acc}^{(g)}\leftarrow 0$
    \For{$u=0$ to $B-1$}
      \ForAll{$r\in\intv{h_t}$, $a\in\intv{n}$}
        \State $\rho^{u,(g)}_{rn+a}\leftarrow\delta\cdot\exp(2\pi\mathrm{I}\cdot C_{(gh_t+r)B+u,\,a}/q)$
      \EndFor
      \State $E_u^{(g)}\leftarrow{\sf Ecd}(\vec\rho^{u,(g)})$
      \State ${\sf acc}^{(g)}\leftarrow{\sf Add}({\sf acc}^{(g)},{\sf ExtMult}({\sf cs}_u^{(g)},E_u^{(g)}))$
    \EndFor
    \State ${\sf acc}^{(g)}\leftarrow{\sf RS}({\sf acc}^{(g)})$\Comment{one division, by $P$ (Section~\ref{s:extprod})}
  \EndFor
  \State ${\sf acc}\leftarrow\bigodot_{g\in\intv{t}}{\sf acc}^{(g)}$
  \State ${\sf acc}\leftarrow\pr{N/2}{n}({\sf acc})$
  \State \Return $\ct'\leftarrow\stc_n({\sf Im2}({\sf acc}))$
\end{algorithmic}
\end{algorithm}

\subsection{Dimension reduction}
\label{s:dimension-reduction}

Algorithm~\ref{alg:lola-boot} builds the vectors $\vec v^{(g)}$ from $t\cdot B$ single $\accmul$s.
The load parallelizes, as low latency requires, but the absolute work is still large: for $t=h$, it takes $N$ single $\accmul$s.
We therefore restrict the block-binary secret to the shape $s(X)=s'(X^R)$ for a divisor $R\mid B$, independently of the packing of Section~\ref{s:double-slots}.
The unique $1$ in each block then lies on a multiple of $R$, that is, $s_i=0$ whenever $R\nmid i$.
Algorithms~\ref{alg:lola-keygen} and~\ref{alg:lola-boot} let $u$ run only over the grid $0,R,2R,\ldots,B-R$, so the key $({\sf cs}_u^{(g)})_{g\in\intv{t},\,u\in R\cdot\intv{B/R}}$ has $t\cdot B/R$ ciphertexts instead of $t\cdot B$.
The product depth stays $\log h$, and the Hadamard product, the operator $\pr{N/2}{n}$ and $\stc_n\circ{\sf Im2}$ are unchanged.
The \accmul\ load $t\cdot B/R=t\cdot(N/R)/h$ thus depends on the degree $N/R$ that security requires, not on the ring degree $N$.

The constraint has a price in security.
By Proposition~\ref{prop:subring-secret}, it leaves a block-binary key of degree $N/R$, which Section~\ref{s:sparse-keys} estimates at the row $(N/R,h)$ of Table~\ref{tab:sparse-block}.
For instance, for $N=2^{14}$ and $h=64$, the unconstrained key allows $\log q_0\le 209$, whereas $R=8$, of degree $N/R=2^{11}$, allows $\log q_0\le 27$.
The largest usable $R$ is the one whose row still admits the base prime, and Section~\ref{s:params} gives this choice for each of our parameter sets.

Secrets in the subring $\mathbb{Z}[X^R]/(X^N+1)$ are not new.
Ring switching moves ciphertexts into such a subring \cite{GHPS12}, and subring secret encapsulation switches to a dense secret of the subring before bootstrapping \cite{MHWW25}.
Here the subring is applied to the sparse key alone, and its purpose is the smaller \accmul\ load.

\subsection{Parallelization}
\label{s:parallelizability}
After dimension reduction, Algorithm~\ref{alg:lola-boot} runs $t\cdot B/R$ independent single $\accmul$s (Section~\ref{s:accmul}), which may be assigned to distinct cores.
The encoding is arithmetic-bound and takes the larger share of a single $\accmul$ on one core; the plaintext--ciphertext product takes most of the rest and reads its key ciphertext once, without reuse.
The full-slot key has $t\cdot B/R$ ciphertexts at the top modulus, several gigabytes for our full-slot sets (Section~\ref{s:params}).
On a machine with limited memory bandwidth, the key traffic, rather than the arithmetic, then bounds how far more cores help, and Section~\ref{s:cosets} shrinks the key for such machines.
Otherwise, the arithmetic of \accmul\ keeps scaling, so on many cores it stops being the largest stage, and \stc\ and the product operator set the latency (Section~\ref{s:where-time}).

\subsection{Placement of \texorpdfstring{\stc}{S2C}}
\label{s:stc-placement}
Without its final $\stc_n$, Algorithm~\ref{alg:lola-boot} maps the coefficients of its input to the slots of its output.
As in \ship, we therefore apply $\stc$ to the input instead of the output, and $\tutti_n$ stops after ${\sf Im2}$.
The bootstrapping then becomes the identity on slots, and the transform becomes cheaper: an opening $\stc_{N/2}$ runs once, at the two lowest primes, whereas a finishing one runs on every prime of the output modulus.
We evaluate the opening $\stc$ by baby-step giant-step, as in \ship\ \cite{HS14,SHIP25}; the baby steps are independent, so the transform parallelizes across the diagonals.
The transform is the half-width one of \cite{SCORE25}, $\mathbf{U}'\ct$ followed by the addition of its conjugate, whose result has slots $2\,\mathrm{Re}(\mathbf{U}'\vec z)$.
A fresh encryption has real slots, but a computed ciphertext does not, since its noise also has an imaginary part.
A complex $\mathbf{U}'$ would fold that imaginary part into the refreshed values, so we use $\mathrm{Re}(\mathbf{U}')$ instead.
The result on real slots is unchanged, the imaginary part cancels in the addition, and no operation is added.

Speed-ups of a finishing \stc\ for medium numbers of slots \cite{SCORE25,Zheng26} rely on the repeated layout of \spru's output, which an opening \stc\ does not see, and they would also have to make up for the larger modulus.
A finishing \stc\ at output level $\ell$ works modulo $q_\ell$, with $\ell+1$ primes, whereas an opening one works at level $1$, with two.
A plaintext product costs one multiplication per prime, so $\ell+1$ against $2$.
A key switch, with one digit per prime, extends each of its $\ell+1$ digits to the $\ell+2$ primes of $Pq_\ell$ and multiplies it with the key there, so it costs about $(\ell+1)(\ell+2)$ operations on single primes, against $2\cdot 3=6$.
A finishing transform is thus about $(\ell+1)/2$ times more expensive in its plaintext products and $(\ell+1)(\ell+2)/6$ times in its key switches, and a speed-up of a finishing \stc\ would have to exceed these factors to compete with an opening one.

\subsection{Sharing keys across cosets}
\label{s:cosets}

On a machine whose memory bandwidth limits the plaintext--ciphertext products (Section~\ref{s:parallelizability}), the monomial shifts of \spru\ can shrink the key.
For $1\le k<\log h$, $t=2^k$, $n=t\cdot\nmax$ and $h_t=h/t$, one $\tutti_n$ refreshes the coefficients of $X^{aN/n}$ with a key of $t\cdot B/R$ ciphertexts, $h_t$ times smaller than the full-slot key.
The $h_t$ cosets $\ct\cdot X^{-2j}$, $j\in\intv{h_t}$, share that key and refresh the coefficients of $X^{2\ell}$ with $\ell\equiv j\pmod{h_t}$, hence all $N/2$ real coefficients, in independent pipelines.
Algorithm~\ref{alg:lola-stc-bottom} shows the idea, and one $\stc_{N/2}$ at the bottom serves all cosets.
The output of a coset holds its $n$ values $h_t$ times (Proposition~\ref{prop:2k-vector}), so a $0$--$1$ mask $\vec\mu^{(j)}$ keeps the copy in the slot block $\{jn,\ldots,jn+n-1\}$, and the masked outputs are added, at the cost of one level.
Slot $jn+a$ then holds the coefficient of index
\[
  \pi(jn+a)=j+a\,h_t.
\]
To return the slots in their original order, the transform is preceded by the slot permutation that moves slot $s$ to slot $\pi(s)$, which merges into its diagonals at no cost.

\begin{algorithm}[t]
\caption{$\tutti_n$ with $h_t$ cosets}
\label{alg:lola-stc-bottom}
\begin{algorithmic}[1]
  \Require Ciphertext $\ct$ with $N/2$ real slots; shared $\tutti_n$ key; masks $\vec\mu^{(j)}$
  \Ensure Refreshed ciphertext $\ct'$, with the same slot order
  \State $\ct\leftarrow\stc_{N/2}(\pi(\ct))$\Comment{one merged linear map at the two lowest primes}
  \For{$j=0$ to $h_t-1$}\Comment{parallel cosets, shared key}
    \State $\ct_j\leftarrow{\sf ExtMult}(\ct,X^{-2j})$
    \State $\ct_j\leftarrow\tutti_n(\ct_j)$ through ${\sf Im2}$
    \State $\ct_j\leftarrow{\sf ExtMultR}(\ct_j,\vec\mu^{(j)})$
  \EndFor
  \State \Return $\ct'\leftarrow\sum_{j\in\intv{h_t}}\ct_j$
\end{algorithmic}
\end{algorithm}

The cosets interpolate between \spru, whose $h$ monomial shifts are the cosets of $k=0$, and $\tutti_{N/2}$, the single coset of $k=\log h$.

\paragraph{Advantages and drawbacks.}
Cosets divide the key traffic of \accmul\ by $h_t$, since one pass over the key serves every coset, which helps where memory bandwidth limits \accmul; the key's memory footprint shrinks by $h_t$ as well.
In exchange, the mask costs one level, cores spent on cosets are lost inside each $\tutti_n$, and every coset repeats its product operator: $h_t$ trees of $t-1$ multiplications and $\log h_t$ rotations, against one tree of $h-1$ multiplications.
On our machine, this pays off only on few threads (Section~\ref{s:coset-results}): with many threads, \accmul\ stays arithmetic-bound and the repeated product operators dominate.

\section{Optimizations}
\label{s:techniques}
This section describes the remaining optimizations of our implementation.
Section~\ref{s:opt-all} covers those that apply to all our parameter sets (Table~\ref{tab:lola-params}): they reduce the number of exponentiations, set the digit size of the sparse-secret encapsulation, and save a prime of the modulus chain.
Section~\ref{s:opt-single} covers the single-slot sets \(\mathtt{S13}\) and \(\mathtt{S13m2}\), for which we remove sequential automorphisms and improve the approximation of the sine.

\subsection{Optimizations for all parameter sets}
\label{s:opt-all}

\paragraph{One exponentiation per coefficient.}
\label{s:exponentiation-cost}
Algorithm~\ref{alg:spru-boot} builds \(2n\) encodings \(E_u\) with \(N/2\) roots each, so a direct implementation computes \(nN\) roots, and Algorithm~\ref{alg:lola-boot} builds \(t\cdot B/R\) encodings.
Most of these exponentiations are redundant.
Each root is \(\delta\cdot\exp(2\pi\mathrm{I}\,C_{i,a}/q)\) for an entry of \({\sf DecMat}(\ct)\).
Since \(c_1\cdot X^i\) is a negacyclic shift of \(c_1\), every entry is, up to sign, one of the \(N\) coefficients of \(c_1\), with a coefficient of \(c_0\) added in row \(i=0\).
A change of sign conjugates the root, so we compute the cosine and the sine once per distinct value and apply the sign to the sine.
This reduces the number of exponentiations from \(nN\) to \(N+n\), or to \(2N\) with the second harmonic of Section~\ref{s:second-harmonic}.

\paragraph{Digit size in the sparse-secret encapsulation.}
\label{s:bv-ks}
We switch to the sparse key at \(q_0\) with a BV decomposition \cite{BV11} (Section~\ref{s:sparse-keys}) into \(D=\ceil{\log q_0/\omega}\) digits in base \(2^\omega\) (\app{app:ks}).
The choice of \(\omega\) balances noise against running time.
Smaller digits give a smaller switching error (Equation~\eqref{eq:bv-err} in \app{app:ks}), but the switch costs \(2D\) multiplications and \(2D-1\) additions in \({\cal R}_{q_0}\).
For the full-slot sets, this cost is negligible compared with the rest of the bootstrapping, since \({\cal R}_{q_0}\) is a ring of a single prime, so we take \(\omega=1\).
The single-slot sets bootstrap in a few milliseconds, so the switch is no longer negligible.
On \(\mathtt{S13}\), we use \(\omega=3\), which gives \(D=\lceil 24/3\rceil=8\), that is, \(16\) multiplications and \(15\) additions.
This gains \(0.6\) to \(1\) bit of precision over \(\omega=4\), because the other error terms of the single-slot circuit are small (Section~\ref{s:opt-single}).

\paragraph{Plaintext--ciphertext products at the auxiliary modulus.}
\label{s:extprod}
As written, Algorithm~\ref{alg:spru-boot} spends a prime of the chain on the rescaling after its plaintext--ciphertext products.
Like \ship, which handles these products in its auxiliary modulus, we use the auxiliary prime \(P\) for this rescaling instead.
The only products at the top of the bootstrapping are the products \({\sf ExtMult}({\sf cs}_u^{(g)},E_u^{(g)})\) of \accmul.
We supply the key ciphertexts at the top modulus extended by \(P\), and we encode \(E_u^{(g)}\) at a scale of about \(P\).
The products are summed modulo \(PQ\), and one division by \(P\), the ModDown that ends every key switch, rescales the sum (Algorithm~\ref{alg:lola-boot}).
After \accmul, the ciphertext is therefore at the top of the ordinary modulus chain.
In Algorithm~\ref{alg:lola-boot}, and in Algorithm~\ref{alg:single-boot} of \app{app:single-slot}, the rescaling after \accmul\ denotes this division by \(P\).

\subsection{Optimizations for the single-slot sets}
\label{s:opt-single}

\paragraph{Fewer sequential steps.}
\label{s:single-slot-impl}
On \(\mathtt{S13}\) and \(\mathtt{S13m2}\), the trace \(\tr{N/2}{h}\) of Algorithm~\ref{alg:spru-boot} adds up the \(B\) offsets of each block with \(\log(N/(2h))\) sequential automorphisms at the top level, \(7\) on \(\mathtt{S13}\).
These automorphisms run one after the other, whereas \accmul{}s are independent, so we replace the trace by more \accmul{}s.
The key now holds one ciphertext \({\sf cs}_j\) per offset \(j\in\intv{B}\), which we call a \emph{key row}.
Slot \(\ell\) of \({\sf cs}_j\) encrypts \(s_{bB+j}\) with \(b=\ell\bmod h\), and slot \(\ell\) of \(E_j\) holds the matching root \(\delta\cdot\exp(2\pi\mathrm{I}\,C_{bB+j,0}/q)\).
As each block of \(s\) has a single \(1\), the sum \(\sum_j{\sf ExtMult}({\sf cs}_j,E_j)\) holds in slot \(\ell\) the root selected by block \(b\), ready for \(\pr{h}{1}\) and \({\sf Im2}\).

In this layout, each single \accmul\ is cheap.
The encoding \(E_j\) has period \(h\) in the slots, hence only \(2h\) nonzero coefficients, so its encoding and its \rns\ representation take transforms of \(2h\) instead of \(N\) points, \(128\) times fewer on \(\mathtt{S13}\).
The drawback is memory traffic rather than memory size: every bootstrapping reads the whole key, which is \(B\) rows or \(128\)\,MiB on \(\mathtt{S13}\), and at a latency of a few milliseconds this means tens of GB per second.
The idea is the same as in Section~\ref{s:double-slots}, applied to a different operator.
There, we split the blocks over several accumulators to save rotations of the product operator; here, we split the offsets over several key rows to save the rotations of the trace.

A key row can also hold several offsets, which reduces this traffic and interpolates between this circuit and Algorithm~\ref{alg:spru-boot}.
For a power of two \(G\mid B\), let slot \(\ell\) in group \(g=\lfloor\ell/h\rfloor\bmod G\) take offset \(g+Gj\) of its block.
This leaves \(B/G\) rows, and \(\tr{Gh}{h}\), with \(\log G\) sequential automorphisms, adds the groups before the product.
The case \(G=1\) is the circuit above, and \(G=B/2\) is Algorithm~\ref{alg:spru-boot}.
A machine with more memory bandwidth per core favours a smaller \(G\), since the extra rows are then cheap to read.
\(\mathtt{S13}\) and \(\mathtt{S13m2}\) use \(G=2\), with one automorphism and a \(64\)\,MiB key on \(\mathtt{S13}\).
Our implementation of \(\tutti_n\) for \(2\le n\le\nmax\) (Section~\ref{s:more-slots}) uses the same layout with \(G=1\): one accumulator of \(B/R\) rows, in which slot \(\ell\) serves block \(\lfloor\ell/n\rfloor\bmod h\) of coefficient \(\ell\bmod n\), followed by \(\pr{hn}{n}\).

\paragraph{A second harmonic.}
\label{s:second-harmonic}
The bootstrapping relies on the approximation \(\sin\varphi\approx\varphi\), whose error is significant at the small ratio \(\gamma=2^{4.5}\) of \(\mathtt{S13}\).
For inputs \(m_0/\Delta\in[-1,1]\), the angle \(\varphi=2\pi m_0/q\) satisfies \(|\varphi|\le2\pi/\gamma\) (Section~\ref{s:spru-boot}).
The output \(\frac{q}{2\pi\Delta}\sin\varphi\) of \({\sf Im2}\) then differs from \(m_0/\Delta=\frac{q}{2\pi\Delta}\varphi\) by about \(\frac{q}{2\pi\Delta}\cdot\varphi^3/6\), which is largest at the ends of the range.
A second harmonic of the sine cancels this term.
Indeed,
\[
\alpha\sin\varphi+\beta\sin 2\varphi=(\alpha+2\beta)\,\varphi-(\alpha+8\beta)\,\frac{\varphi^3}{6}+\cdots,
\]
so the weights \(\alpha=4/3\) and \(\beta=-1/6\) cancel the cubic term and keep the slope \(\alpha+2\beta=1\).
We therefore return
\[
\frac{q}{2\pi\Delta}\Bigl(\frac43\sin\varphi-\frac16\sin 2\varphi\Bigr)=\frac{q}{2\pi\Delta}\Bigl(\varphi-\frac{\varphi^5}{30}+\cdots\Bigr),
\]
and since the slope at \(\varphi=0\) is still \(1\), an error in the input reaches the output at the same size.
This is the five-point central difference \cite{BF11} applied to \(e^{\mathrm{I}\varphi}\), and the two-term sine series of Jutla and Manohar \cite{JM22}.

The circuit above changes in four places.
\begin{itemize}
  \item \emph{Public roots.} Each coefficient \(C_{i,0}\) also gives the root of the doubled angle, \(\exp(4\pi\mathrm{I}\,C_{i,0}/q)\). The matrix \({\sf DecMat}(\ct)\) itself is unchanged.
  \item \emph{Layout.} The even and the odd slots carry the first and the second harmonic of block \(\lfloor\ell/2\rfloor\bmod h\), which doubles every slot period above. Since the product multiplies \(h\) factors, the key rows carry the \(h\)-th roots of the weights, \((4/3)^{1/h}\) and \((1/6)^{1/h}e^{\mathrm{I}\pi/h}\), where \(e^{\mathrm{I}\pi/h}\) supplies the sign.
  \item \emph{Trace and product.} The partial trace becomes \(\tr{2Gh}{2h}\) and the product \(\pr{2h}{2}\), which leaves \(\frac43\delta^h e^{\mathrm{I}\varphi}\) and \(-\frac16\delta^h e^{2\mathrm{I}\varphi}\) in neighbouring slots.
  \item \emph{Final sum.} A rotation by one slot adds these two values before \({\sf Im2}\).
\end{itemize}
The second harmonic costs one sequential automorphism, but no level and no key row.
Together with the \(3\)-bit BV digits of Section~\ref{s:bv-ks}, it raises the precision on \(\mathtt{S13}\) from \(5.21\) bits with Algorithm~\ref{alg:spru-boot} to \(6.53\) bits (Table~\ref{tab:ship-compare}).

The technique would equally apply to \(\tutti_n\) with \(n\le N/(4h)\), whose slots hold at least \(2h\) copies of each value, \(h\) of them spare for the second harmonic.
We did not implement this variant.

\begin{figure}[t]
\centering
\begin{adjustbox}{max width=\textwidth}
\begin{tikzpicture}[
  slot/.style={draw,minimum width=0.62cm,minimum height=0.62cm,inner sep=0pt,font=\scriptsize},
  stage/.style={draw,rounded corners=2pt,align=center,font=\small,inner sep=3pt,minimum height=0.8cm},
  arr/.style={-{Latex},thick}
]
  \foreach \l in {0,...,15} {
    \pgfmathtruncatemacro{\etaval}{mod(\l,2)}
    \pgfmathtruncatemacro{\b}{mod(floor(\l/2),4)}
    \pgfmathsetmacro{\shade}{\etaval==0 ? 8 : 30}
    \ifcase\b \def\col{blue}\or \def\col{orange}\or \def\col{teal}\or \def\col{red}\fi
    \node[slot,fill=\col!\shade] (s\l) at (0.62*\l,0) {$b{=}\b$};
    \node[font=\tiny,above=0pt of s\l] {$\ell{=}\l$};
    \node[font=\tiny,below=0pt of s\l] {$\eta{=}\etaval$};
  }
  \draw[decorate,decoration={brace,mirror,amplitude=4pt}] ($(s0.south west)+(0,-0.38)$) -- node[below=4pt,font=\scriptsize] {group $g=0$: offsets $0,2,4,\dots$} ($(s7.south east)+(0,-0.38)$);
  \draw[decorate,decoration={brace,mirror,amplitude=4pt}] ($(s8.south west)+(0,-0.38)$) -- node[below=4pt,font=\scriptsize] {group $g=1$: offsets $1,3,5,\dots$} ($(s15.south east)+(0,-0.38)$);
  \node[font=\scriptsize,align=left,anchor=west] at ($(s15.east)+(0.25,0)$) {slot $\ell$ of row $j$:\\ key $a_\eta\, s_{bB+g+Gj}$\\ root $r_{bB+g+Gj,\eta}$};

  \begin{scope}[yshift=-2.5cm]
    \node[stage,fill=black!6] (acc) at (0,0) {$\sum_j{\sf ExtMult}({\sf cs}_j,E_j)$\\[-1pt]{\scriptsize $B/G$ rows}};
    \node[stage,fill=black!10] (tr) at (3.6,0) {$\tr{4h}{2h}$\\[-1pt]{\scriptsize adds the groups}};
    \node[stage,fill=black!10] (pr) at (6.6,0) {$\pr{2h}{2}$\\[-1pt]{\scriptsize multiplies the blocks}};
    \node[stage,fill=black!10] (rot) at (9.6,0) {$+\,{\sf Rot}(\cdot,1)$};
    \node[stage,fill=black!6] (im) at (12.1,0) {${\sf Im2}$};
    \draw[arr] (acc) -- (tr);
    \draw[arr] (tr) -- (pr);
    \node[font=\scriptsize,align=center,below=3pt of acc] {root if its offset\\is $g\bmod G$, else $0$};
    \draw[arr] (pr) -- (rot);
    \draw[arr] (rot) -- (im);
    \node[font=\scriptsize,align=center,below=3pt of tr] {root of block $b$,\\harmonic $\eta+1$};
    \node[font=\scriptsize,align=center,below=3pt of pr] {even: $\frac43\delta^he^{\mathrm{I}\varphi}$\\odd: $-\frac16\delta^he^{2\mathrm{I}\varphi}$};
    \node[font=\scriptsize,align=center,below=3pt of rot] {$\delta^h\bigl(\frac43e^{\mathrm{I}\varphi}$\\$-\frac16e^{2\mathrm{I}\varphi}\bigr)$};
    \node[font=\scriptsize,align=center,below=3pt of im] {$\frac{q}{2\pi\Delta}\cdot{}$\\$\frac{8\sin\varphi-\sin 2\varphi}{6}$};
  \end{scope}
\end{tikzpicture}
\end{adjustbox}
\caption{Single-slot bootstrapping with the second harmonic and \(G=2\) groups, drawn for \(h=4\).
Top: one period of \(2Gh=16\) slots; the colour is the block \(b\), light and dark the first and the second harmonic, and the braces the groups.
Bottom: the steps after \accmul, with the slot contents after each step.
With \(G=1\), the partial trace is dropped.}
\label{fig:single-slot}
\end{figure}
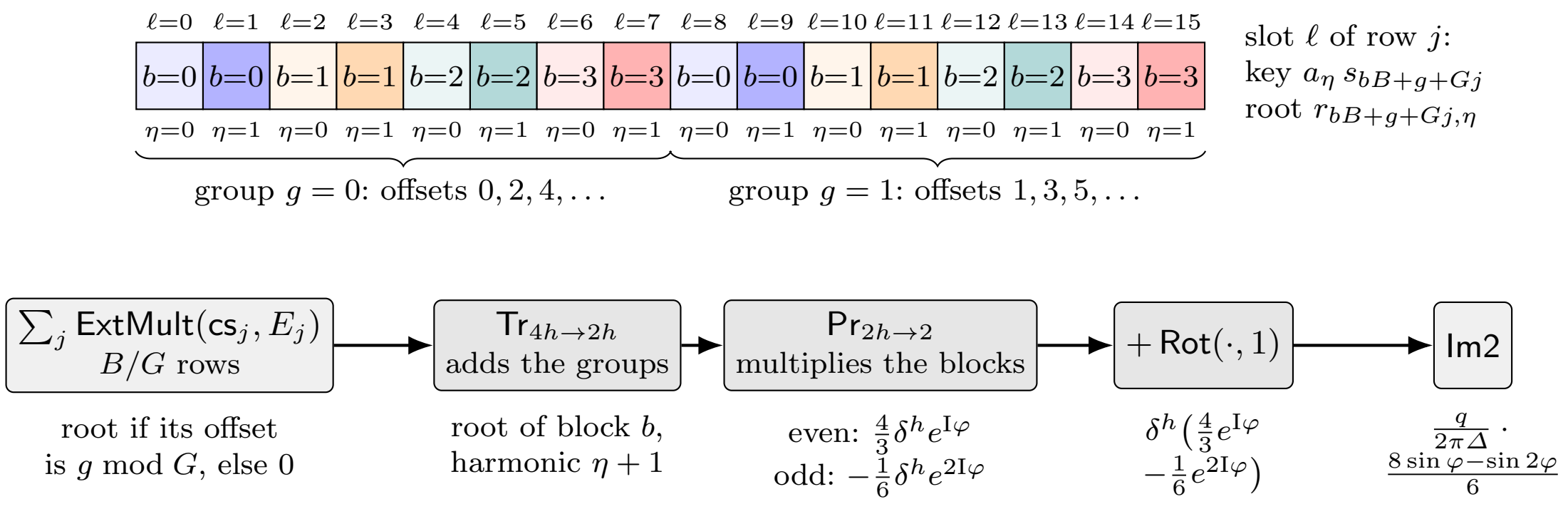

\paragraph{The complete circuit.}
Figure~\ref{fig:single-slot} shows the single-slot circuit with \(G\) groups and the second harmonic, as used by \(\mathtt{S13}\) with \(G=2\): one period of the slots and the steps after \accmul.
The pseudocode is in \app{app:single-slot}.
For a single slot, \(C_i\) denotes the constant coefficient of \(c_1\cdot X^i\), to which that of \(c_0\) is added for \(i=0\), so that \(m_0=\sum_i s_iC_i\bmod q\) as in~\eqref{eq:spru-decmat}.
Slot \(\ell\in\intv{N/2}\) of key row \(j\) serves harmonic \(\eta(\ell)+1\) of the secret coefficient with index \(i_j(\ell)\), where
\[
\eta(\ell)=\ell\bmod 2,\qquad i_j(\ell)=bB+g+Gj,
\]
with the block \(b=\lfloor\ell/2\rfloor\bmod h\) and the group \(g=\lfloor\ell/(2h)\rfloor\bmod G\) of slot \(\ell\), so every encoding has period \(2Gh\) in the slots.
In the figure, \(a_\eta\) is the amplitude of harmonic \(\eta+1\), that is \(a_0=(4/3)^{1/h}\) or \(a_1=(1/6)^{1/h}e^{\mathrm{I}\pi/h}\), and \(r_{i,\eta}=\delta\cdot\exp(2\pi\mathrm{I}\,(\eta+1)\,C_i/q)\) is the corresponding root.
Without the second harmonic, \(\eta(\ell)=0\), \(b=\ell\bmod h\), \(g=\lfloor\ell/h\rfloor\bmod G\), the amplitude is \(1\), the trace and the product become \(\tr{Gh}{h}\) and \(\pr{h}{1}\), and the final rotation is dropped.

\section{Experimental Results}
\label{s:implementation}
We implemented \tutti, \paco and \ship in  the \texttt{vfhe} library~\cite{vfhe}, so that all three share the same arithmetic and the same thread control.
We run all experiments on an \texttt{r8i.metal-96xl} instance on AWS (2x Intel Xeon 6975P-C processors), at $1$, $8$, $32$, $48$, $192$ and $384$ threads.
Latency is the mean over at least three runs after a warm-up.
Memory is the peak resident set of the process, keys included.
Precision follows the metric of \ship~\cite{SHIP25}: for real inputs $z_i=2i/n-1\in[-1,1]$ and the real parts $z'_i$ of the output slots, it is the median over runs of $-\log\max_i\abs{z'_i-z_i}$.
The worst slot is stricter than an average over the slots, and the inputs include the end of the range, $-1$.
For a single message, a run is $4096$ bootstraps of uniform $z\in[-1,1]$, with the maximum taken over them, and the median is over three runs.
The output of \tutti is at the scale from which one multiplication returns to the input scale, so that bootstrapped ciphertexts can be multiplied without further rescaling.

\subsection{Parameter selection}
\label{s:params}

Table~\ref{tab:lola-params} lists the parameter sets of our experiments.
$\mathtt{S13}$ and $\mathtt{S13m2}$ are the single-slot sets described below; all others refresh all $N/2$ real slots with $\tutti_{N/2}$.
$\mathtt{LL13}$, $\mathtt{LL14}$, $\mathtt{HT14}$ and $\mathtt{HT15}$ are compared with the \ship sets of the same name \cite{SHIP25}; $\mathtt{LL13}$ is a low-precision full-slot set with one multiplicative level.
On \tutti's $\mathtt{LL13}$, $R=4$ needs $h=64$, which spreads the budget over ten primes of about $21$ bits.
At such primes the rounding noise of rescaling and ModDown dominates, and its variance grows with $\|s\|^2$ of the dense key.
Both \tutti and \ship therefore use, on $\mathtt{LL13}$, a dense ternary key of Hamming weight $512$ instead of a uniform one ($\|s\|^2\approx 2N/3$): its noise is $1.7$ bits lower, for a budget of $212$ instead of $214$ bits (Table~\ref{tab:dense-key}), and it raises the precision on $\mathtt{LL13}$ from $3.9$ to $4.9$ bits for \tutti and, in our implementation, from $4.7$ to $5.0$ bits for \ship.
The ratio $\gamma$ trades the error of the sine, which falls as $\gamma^{-2}$, against the noise, which the output carries multiplied by $\gamma/2\pi$; with the lower noise, \tutti on $\mathtt{LL13}$ takes $\gamma=12$.
We run \ship\ on its published chains, although each exceeds the dense-key cap of Table~\ref{tab:dense-key} at its degree, by $4.5$ to $14$ bits, which puts it at $125.5$ to $126.3$ bits of security in our model (column Sec.\ of Table~\ref{tab:ship-compare}).

The modulus chain is listed from the bottom: the base prime $q_0$, one prime consumed by the opening $\stc$, the multiplicative levels left to the application, and $\log h$ primes consumed by the product of the $h$ \accmul outputs.
A single auxiliary prime $P$ serves both the ModDown after \accmul and every GHS key switch (rotations of $\stc$, automorphisms of the product and of ${\sf Im2}$, relinearization), with one digit per prime of the current level.
The sparse-secret encapsulation is the BV key switch of Section~\ref{s:bv-ks}, at $q_0$ only, with $D=\ceil{\log q_0/\omega}$ digits in base $2^\omega$.
The sparse key has degree $N/R$ (Proposition~\ref{prop:subring-secret}), so each pair $(N/R,h)$ with $\log q_0$ is a row in Table~\ref{tab:sparse-block}.
In $\mathtt{S13}$, $\mathtt{S13m2}$, $\mathtt{LL13}$ and $\mathtt{HT14}$, every prime, including $P$, lies below $2^{30}$, so our implementation stores each residue in a $32$-bit word, which is faster than the $64$-bit words of $\mathtt{LL14}$ and $\mathtt{HT15}$.
A key ciphertext holds $2N$ words per prime of the chain, $P$ included, so the bootstrapping key takes $1.25$, $10$, $4.5$ and $36$\,GiB on $\mathtt{LL13}$, $\mathtt{LL14}$, $\mathtt{HT14}$ and $\mathtt{HT15}$, and $64$ and $72$\,MiB on $\mathtt{S13}$ and $\mathtt{S13m2}$, whose rows have a smaller period (Section~\ref{s:single-slot-impl}).

\paragraph{How a set is chosen.}
Two caps determine a set: the dense-key budget of Table~\ref{tab:dense-key} bounds $\log PQ$ at degree $N$, and the row $(N/R,h)$ of Table~\ref{tab:sparse-block} bounds $\log q_0$.
All our full-slot sets take $h=64$ and the largest $R$ whose row admits the base prime, which is $R=8$ for the $27$-bit prime of $\mathtt{HT14}$ and the $50$-bit prime of $\mathtt{HT15}$, and $R=4$ for $\mathtt{LL13}$ and $\mathtt{LL14}$.
The levels left to the application are those of the \ship set of the same name, and the remaining budget goes to the other primes.
The \stc\ prime only scales the encoded transform, since $\gamma$ fixes the input scale $q_0/\gamma$, so it is the first to shrink when the budget is tight, to $18$ bits on $\mathtt{HT14}$.
The Boot primes take what is left, and they matter: on $\mathtt{HT14}$, raising two of them from $25$ to $26$ bits raises the precision from $4.73$ to $5.15$ bits at the same latency.

The ratio $\gamma$ is set last.
The error of the sine alone limits the precision to $7.3$ bits at $\gamma=2^5$ and to $19.3$ bits at $\gamma=2^{11}$.
The high-precision sets $\mathtt{LL14}$ and $\mathtt{HT15}$ therefore take $\gamma=2^{11}$, and their precision is then limited by the noise.
On $\mathtt{LL13}$, every bit of the budget counts, so the primes are requested at fractional widths: each width $b$ takes the largest unused prime $p\equiv1\bmod 2N$ below $2^b$, for a total of $211.99$ bits.
This set also scales its output by a constant $s$, $1.035$ at $\gamma=12$, chosen so that the error of $s\cdot\frac{\gamma}{2\pi}\sin(2\pi v/\gamma)$ equioscillates over $v\in[-1,1]$.
The constant is folded into the encodings of \accmul, so it costs no operation, and it raises the floor set by the sine from $4.5$ to $6.4$ bits.

\paragraph{Single-slot sets.}
A single message needs no \stc, which saves a prime, so it can stay at a smaller degree than a full-slot refresh.
$\mathtt{S13}$ and $\mathtt{S13m2}$ use $N=2^{13}$ and $h=32$.
The attacks of Section~\ref{s:sparse-keys} allow this: the row $(2^{13},32)$ of Table~\ref{tab:sparse-block} caps the base prime at $27$ bits, and Table~\ref{tab:may-spru} costs the combinatorial attacks there, whereas the estimate $T=S^{0.25}$ in \spru\ \cite[Section~4.5]{SPRU25} forced $N=2^{15}$ even for $n\le 128$ slots.
These values are close to the limit.
A Hamming weight near $16$ makes the cap on $q_0$ too small for the computation, and $N=2^{12}$ with $h=64$ no longer fits the bootstrapping into the $106$-bit budget of that degree.
Keeping $N=2^{13}$ with $h=64$ would be possible, but the product would then have depth $6$ and $63$ ciphertext multiplications instead of $31$.

Both sets use the circuit of Section~\ref{s:single-slot-impl}, with $G=2$ groups and the second harmonic.
$\mathtt{S13}$ spends its budget on larger Boot primes and a larger $P$, for $\log PQ=208.4$ out of $214$ bits: Boot primes of $27$ instead of $25$ bits alone gain one bit.
$\mathtt{S13m2}$ fits a second multiplicative level in the same ring instead, at $\log PQ=213.6$ bits, with the same precision of $6.53$ bits and $0.5$\,ms more on $32$ threads.

\begin{table}[t]
\centering
\caption{Parameter sets of \tutti.
$R$ is the subring reduction of the sparse key, dnum the number of digits of a GHS key switch at the top level, $D$ (base $2^\omega$) the number of digits of the BV key switch at $q_0$, and $\gamma=q_0/\Delta_{\rm in}$ the ratio of Section~\ref{s:spru-boot} for the base prime and the scale of the bootstrapped message.
$\mathtt{LL13}$ uses a dense key of Hamming weight $512$; its primes are the largest primes $p\equiv1\bmod 2N$ below $2^b$ for fractional widths $b$, here rounded to bits.}
\label{tab:lola-params}
\setlength{\tabcolsep}{2.3pt}
\small
\begin{tabular}{l*{13}{c}}
\toprule
\multirow{2}{*}{Set} & \multirow{2}{*}{Slots} & \multirow{2}{*}{$N$} & \multirow{2}{*}{$h$} & \multirow{2}{*}{$R$} & \multirow{2}{*}{$\log PQ$} & \multirow{2}{*}{dnum} & \multicolumn{5}{c}{(prime bits)\,$\times$\,(\# primes)} & \multirow{2}{*}{$D$ ($\omega$)} & \multirow{2}{*}{$\gamma$} \\
\cmidrule(lr){8-12}
 & & & & & & & Base & \stc & Mult & Boot & Aux & & \\
\midrule
$\mathtt{S13}$  & $1$   & $2^{13}$ & $32$ & $1$ & $208.4$ & $7$  & $24\times1$ & ---         & $20\times1$ & $27\times5$ & $30\times1$ & $8$ ($3$)  & $2^{4.5}$ \\
$\mathtt{S13m2}$ & $1$  & $2^{13}$ & $32$ & $1$ & $213.6$ & $8$  & $23\times1$ & ---         & $20\times2$ & $25\times5$ & $27\times1$ & $8$ ($3$)  & $2^{3.7}$ \\
\midrule
$\mathtt{LL13}$ & $N/2$ & $2^{13}$ & $64$ & $4$ & $212.0$ & $9$  & $24\times1$ & $17\times1$ & $20\times1$ & $19,20,21,22{\times}3$ & $25\times1$ & $24$ ($1$) & $2^{3.6}$ \\
$\mathtt{LL14}$ & $N/2$ & $2^{14}$ & $64$ & $4$ & $430.0$ & $9$  & $48\times1$ & $37\times1$ & $37\times1$ & $43\times6$ & $50\times1$ & $48$ ($1$) & $2^{11}$ \\
$\mathtt{HT14}$ & $N/2$ & $2^{14}$ & $64$ & $8$ & $429.5$ & $17$ & $27\times1$ & $18\times1$ & $23\times9$ & $25{\times}4,26{\times}2$ & $30\times1$ & $27$ ($1$) & $2^{5}$ \\
$\mathtt{HT15}$ & $N/2$ & $2^{15}$ & $64$ & $8$ & $769.0$ & $17$ & $50\times1$ & $39\times1$ & $39\times9$ & $46\times6$ & $53\times1$ & $50$ ($1$) & $2^{11}$ \\
\bottomrule
\end{tabular}
\end{table}

\subsection{Main results}

Table~\ref{tab:ship-compare} compares \tutti with \ship on the parameter sets of Table~\ref{tab:lola-params} and their \ship counterparts \cite{SHIP25}, and lists the single-slot set $\mathtt{S13}$.
Our implementation of \ship also uses the BV key switch at $q_0$ (Section~\ref{s:bv-ks}).
We note that conventional bootstrapping parallelizes poorly: \ship reports a latency on $32$ cores $2.5\times$ below the smallest it observed with HEaaN, on $48$ cores, and we refer to \cite{SHIP25} for that comparison.

\paragraph{Precision.}
The precision in Table~\ref{tab:ship-compare} is measured, not derived, but the construction of Section~\ref{s:more-slots} explains why we expect \tutti\ to match the precision of \spru.
\tutti\ changes where the factors of a message are stored, not how each factor is computed.
In both schemes, a factor is a sum of $B/R$ products of a key ciphertext with a public root, which in \spru\ the trace completes, and the $h$ factors of a message are multiplied at depth $\log h$.
$\tutti_n$ computes more of these products because it refreshes more messages, but they go to different slots, so each slot still receives a sum of $B/R$ products.
It also replaces rotations of the product operator by products of distinct accumulators, which removes key switches.
The precision of a set therefore depends on its primes and its ratio $\gamma$, chosen as in Section~\ref{s:params}, as it does for \spru.

\begin{table}[t]
\centering
\caption{Bootstrapping latency of \tutti and \ship, in ms, by number of threads.
Lev.\ is the number of multiplicative levels left to the application after bootstrapping.
Bold marks the faster scheme on sets with both, and the lowest latency of the row on the single-slot sets.
For \ship, precision and parameters are the ones reported in the original paper \cite{SHIP25}, except on its $\mathtt{LL13}$, which, like \tutti's $\mathtt{LL13}$, runs with a dense key of Hamming weight $512$ (Section~\ref{s:params}); its precision is that of our implementation.}
\label{tab:ship-compare}
\setlength{\tabcolsep}{3.2pt}
\small
\begin{tabular}{llrccrrrrrrcr}
\toprule
\multirow{2}{*}{Set} & \multirow{2}{*}{Scheme} & \multirow{2}{*}{Slots} & \multirow{2}{*}{Lev.} & Sec. & \multicolumn{6}{c}{Latency (ms) at $T$ threads} & Prec. & Mem. \\
\cmidrule(lr){6-11}
 & & & & (bits) & $1$ & $8$ & $32$ & $48$ & $192$ & $384$ & (bits) & (GB) \\
\midrule
$\mathtt{S13}$ & \tutti & $1$ & $1$ & $\ge128$ & $25.5$ & $5.97$ & $\mathbf{4.55}$ & $4.68$ & $6.27$ & $6.28$ & $6.53$ & $0.3$ \\
$\mathtt{S13m2}$ & \tutti & $1$ & $2$ & $\ge128$ & $30.1$ & $6.89$ & $5.04$ & $\mathbf{4.92}$ & $6.93$ & $5.88$ & $6.53$ & $0.3$ \\
\midrule
$\mathtt{LL13}$ & \tutti & $4096$ & $1$ & $\ge128$ & $\mathbf{1345}$ & $\mathbf{214}$ & $\mathbf{72.5}$ & $\mathbf{47.2}$ & $\mathbf{45.8}$ & $\mathbf{49.6}$ & $4.89$ & $2.6$ \\
$\mathtt{LL13}$ & \ship & $4096$ & $1$ & $125.5$ & $2942$ & $412$ & $132$ & $135$ & $84.0$ & $91.0$ & $4.96$ & $6.8$ \\
\midrule
$\mathtt{LL14}$ & \tutti & $8192$ & $1$ & $\ge128$ & $5424$ & $774$ & $260$ & $\mathbf{183.5}$ & $\mathbf{113}$ & $\mathbf{115}$ & $17.0$ & $17.5$ \\
$\mathtt{LL14}$ & \ship & $8192$ & $1$ & $126.3$ & $\mathbf{3717}$ & $\mathbf{511}$ & $\mathbf{186}$ & $183.5$ & $163$ & $169$ & $16.9$ & $22.0$ \\
\midrule
$\mathtt{HT14}$ & \tutti & $8192$ & $9$ & $\ge128$ & $\mathbf{5251}$ & $\mathbf{789}$ & $\mathbf{257}$ & $\mathbf{155}$ & $\mathbf{110}$ & $\mathbf{113}$ & $5.16$ & $9.5$ \\
$\mathtt{HT14}$ & \ship & $8192$ & $9$ & $125.7$ & $12393$ & $1743$ & $491$ & $467$ & $211$ & $202$ & $4.55$ & $29.5$ \\
\midrule
$\mathtt{HT15}$ & \tutti & $16384$ & $9$ & $\ge128$ & $23107$ & $3143$ & $1152$ & $\mathbf{839}$ & $\mathbf{424}$ & $\mathbf{427}$ & $18.44$ & $66.1$ \\
$\mathtt{HT15}$ & \ship & $16384$ & $9$ & $125.9$ & $\mathbf{22120}$ & $\mathbf{2897}$ & $\mathbf{959}$ & $919$ & $651$ & $690$ & $18.1$ & $103.5$ \\
\bottomrule
\end{tabular}
\end{table}

\paragraph{\tutti and \ship.}
\label{s:comparison}
Both schemes evaluate the decryption equation \(m_a=\sum_i s_iC_{i,a}\) on the unit circle with a sparse key at \(q_0\), and both end in a product of \(h\) factors.
They differ in which side of this matrix--vector product is encrypted.
\tutti computes \(\mathbf{C}\cdot s\): public roots built from \(\mathbf{C}\) multiply encryptions of the secret's block indicators, as plaintext--ciphertext products.
\ship computes, in effect, \(\mathbf{S}\cdot c_1\), with \(\mathbf{S}\) the negacyclic matrix of \(s\), so that \(c_1\cdot s=\mathbf{S}\cdot c_1\): an encryption of each secret position rotates the public roots blindly, so the secret acts as an encrypted permutation.
Both share everything around that product: \stc, the encapsulation, the product of the \(h\) factors, and the final conjugation.
The comparison is therefore between \accmul\ and blind rotation, and Table~\ref{tab:lola-vs-ship} lists what follows.

\begin{table}[t]
\centering
\caption{How \tutti and \ship produce one of the \(h\) factors, for the full-slot sets of Table~\ref{tab:lola-params}.
\stc, the encapsulation, the product of the factors, and the final conjugation are common to both.}
\label{tab:lola-vs-ship}
\setlength{\tabcolsep}{4pt}
\small
\begin{tabular}{p{0.25\textwidth}p{0.34\textwidth}p{0.34\textwidth}}
\toprule
 & \ship & \tutti \\
\midrule
Matrix--vector product & encrypted secret, public roots rotated: \(\mathbf{S}\cdot c_1\) & public roots, encrypted indicators: \(\mathbf{C}\cdot s\) \\
Position within \(\theta\) & column step: \(4\theta\) independent ${\sf ExtMult}$s & --- \\
Remaining position & \(3\) base-\(4\) digits and one alignment & \(B/R\) independent ${\sf ExtMult}$s \\
Key switches in sequence & \(4\), each over up to \(4\) keys & none \\
Unit of parallelism & \(h\) branches & \(t\cdot B/R\) single \accmul s \\
Fewer slots & only \stc\ gets lighter & \accmul\ load and key shrink with \(t\) \\
\bottomrule
\end{tabular}
\end{table}

\tutti trades total work for parallelism and targets latency on many cores; neither scheme dominates.
Single-threaded, \ship is faster on $\mathtt{LL14}$ and $\mathtt{HT15}$.
On $\mathtt{LL13}$ and $\mathtt{HT14}$, \tutti is faster at every thread count.
From $48$ threads on, \tutti is faster on $\mathtt{LL14}$, $\mathtt{HT14}$ and $\mathtt{HT15}$, on $\mathtt{LL14}$ at $48$ threads only by $0.02\%$, a tie.
On $\mathtt{LL13}$, where both schemes refresh $4096$ messages at about the same precision, \tutti is $1.8$ to $2.9\times$ faster and needs $2.6\times$ less memory.
Both schemes stream keys from memory: the \accmul\ load reads its key once, and each blind-rotation branch reads rotation keys through a sequential chain of digits.
\ship's blind rotation key-switches at the largest non-auxiliary modulus, so its larger sets need several auxiliary primes to control that error, four on $\mathtt{HT15}$.
\tutti's \accmul phase runs once and in parallel, so one auxiliary prime suffices, and $\mathtt{HT15}$ stays $99$ bits below the $868$-bit cap, room for more than two more multiplicative levels than \ship.
Concurrent work builds on \ship and reduces its latency further \cite{OLG+26}; it appeared too recently for a comparison.

\begin{figure}[t]
\centering
\begin{adjustbox}{max width=\textwidth}
\definecolor{stageStc}{HTML}{2A78D6}%
\definecolor{stageMain}{HTML}{EB6834}%
\definecolor{stageTrace}{HTML}{1BAF7A}%
\definecolor{stageProduct}{HTML}{8A5CC7}%
\definecolor{stageOther}{HTML}{B7B5AE}%
\definecolor{stageInk}{HTML}{52514E}%
\begin{tikzpicture}
\begin{axis}[
  scale only axis, width=262pt, height=170pt,
  font=\small, xbar stacked, bar width=9pt, y dir=reverse,
  xmin=0, xmax=100, ymin=-0.7, ymax=14.95,
  xtick={0,25,50,75,100}, xticklabel={$\pgfmathprintnumber{\tick}\,\%$},
  ytick={0,1,2,3.75,4.75,5.75,7.5,8.5,9.5,11.25,12.25,13.25}, yticklabels={$1$,$32$,$192$,$1$,$32$,$192$,$1$,$32$,$192$,$1$,$32$,$192$},
  axis lines*=left, axis line style={stageInk, line width=0.4pt},
  tick style={draw=none}, major tick length=0pt,
  xmajorgrids, grid style={black!12, line width=0.3pt},
  tick label style={text=stageInk, font=\footnotesize}, label style={text=stageInk},
  xlabel={Share of the latency, with the time in ms},
  legend style={at={(0.5,1.03)}, anchor=south, draw=none, fill=none, legend columns=-1,
                font=\footnotesize, /tikz/every even column/.append style={column sep=6pt}},
  area legend,
  every axis plot/.append style={draw=white, line width=0.5pt},
  clip=false,
]
\addplot[fill=stageStc] coordinates {(0.00,0) (0.00,1) (0.00,2) (11.23,3.75) (18.78,4.75) (29.71,5.75) (5.54,7.5) (10.21,8.5) (15.71,9.5) (12.76,11.25) (17.94,12.25) (30.71,13.25)};
\addplot[fill=stageMain] coordinates {(51.83,0) (30.68,1) (23.92,2) (75.56,3.75) (61.73,4.75) (48.06,5.75) (91.89,7.5) (77.93,8.5) (65.56,9.5) (63.66,11.25) (55.65,12.25) (35.71,13.25)};
\addplot[fill=stageTrace] coordinates {(6.24,0) (16.34,1) (21.03,2) (0.00,3.75) (0.00,4.75) (0.00,5.75) (0.00,7.5) (0.00,8.5) (0.00,9.5) (0.00,11.25) (0.00,12.25) (0.00,13.25)};
\addplot[fill=stageOther] coordinates {(3.22,0) (11.70,1) (11.08,2) (0.12,3.75) (1.39,4.75) (2.16,5.75) (0.03,7.5) (0.66,8.5) (1.12,9.5) (0.31,11.25) (0.97,12.25) (2.69,13.25)};
\addplot[fill=stageProduct] coordinates {(38.71,0) (41.28,1) (43.98,2) (13.09,3.75) (18.09,4.75) (20.07,5.75) (2.54,7.5) (11.20,8.5) (17.61,9.5) (23.27,11.25) (25.45,12.25) (30.89,13.25)};
\legend{{\stc},{\accmul\ or blind rotation},Trace,Other,Product}
  \node[anchor=west, inner sep=1pt, text=stageInk, font=\footnotesize] at (axis cs:101,0) {$25.5$\,ms};
  \node[anchor=west, inner sep=1pt, text=stageInk, font=\footnotesize] at (axis cs:101,1) {$4.55$\,ms};
  \node[anchor=west, inner sep=1pt, text=stageInk, font=\footnotesize] at (axis cs:101,2) {$6.27$\,ms};
  \node[anchor=west, inner sep=1pt, text=stageInk, font=\footnotesize] at (axis cs:101,3.75) {$1345$\,ms};
  \node[anchor=west, inner sep=1pt, text=stageInk, font=\footnotesize] at (axis cs:101,4.75) {$72.5$\,ms};
  \node[anchor=west, inner sep=1pt, text=stageInk, font=\footnotesize] at (axis cs:101,5.75) {$45.8$\,ms};
  \node[anchor=west, inner sep=1pt, text=stageInk, font=\footnotesize] at (axis cs:101,7.5) {$2942$\,ms};
  \node[anchor=west, inner sep=1pt, text=stageInk, font=\footnotesize] at (axis cs:101,8.5) {$132$\,ms};
  \node[anchor=west, inner sep=1pt, text=stageInk, font=\footnotesize] at (axis cs:101,9.5) {$84$\,ms};
  \node[anchor=west, inner sep=1pt, text=stageInk, font=\footnotesize] at (axis cs:101,11.25) {$5251$\,ms};
  \node[anchor=west, inner sep=1pt, text=stageInk, font=\footnotesize] at (axis cs:101,12.25) {$257$\,ms};
  \node[anchor=west, inner sep=1pt, text=stageInk, font=\footnotesize] at (axis cs:101,13.25) {$110$\,ms};
  \node[anchor=east, inner sep=0pt, text=stageInk, font=\footnotesize] at (axis cs:-7,1) {$\mathtt{S13}$, \tutti};
  \node[anchor=east, inner sep=0pt, text=stageInk, font=\footnotesize] at (axis cs:-7,4.75) {$\mathtt{LL13}$, \tutti};
  \node[anchor=east, inner sep=0pt, text=stageInk, font=\footnotesize] at (axis cs:-7,8.5) {$\mathtt{LL13}$, \ship};
  \node[anchor=east, inner sep=0pt, text=stageInk, font=\footnotesize] at (axis cs:-7,12.25) {$\mathtt{HT14}$, \tutti};
\end{axis}
\node[text=white, font=\scriptsize, inner sep=0pt] at (67.9pt,162.4pt) {$13.2$};
\node[text=white, font=\scriptsize, inner sep=0pt] at (211.3pt,162.4pt) {$9.86$};
\node[text=white, font=\scriptsize, inner sep=0pt] at (40.2pt,151.5pt) {$1.39$};
\node[text=black, font=\scriptsize, inner sep=0pt] at (101.8pt,151.5pt) {$0.74$};
\node[text=white, font=\scriptsize, inner sep=0pt] at (207.9pt,151.5pt) {$1.87$};
\node[text=white, font=\scriptsize, inner sep=0pt] at (31.3pt,140.7pt) {$1.49$};
\node[text=black, font=\scriptsize, inner sep=0pt] at (90.2pt,140.7pt) {$1.31$};
\node[text=white, font=\scriptsize, inner sep=0pt] at (204.4pt,140.7pt) {$2.74$};
\node[text=white, font=\scriptsize, inner sep=0pt] at (14.7pt,121.7pt) {$151$};
\node[text=white, font=\scriptsize, inner sep=0pt] at (128.4pt,121.7pt) {$1016$};
\node[text=white, font=\scriptsize, inner sep=0pt] at (244.9pt,121.7pt) {$176$};
\node[text=white, font=\scriptsize, inner sep=0pt] at (24.6pt,110.8pt) {$13.6$};
\node[text=white, font=\scriptsize, inner sep=0pt] at (130.1pt,110.8pt) {$44.7$};
\node[text=white, font=\scriptsize, inner sep=0pt] at (238.3pt,110.8pt) {$13.1$};
\node[text=white, font=\scriptsize, inner sep=0pt] at (38.9pt,99.9pt) {$13.6$};
\node[text=white, font=\scriptsize, inner sep=0pt] at (140.8pt,99.9pt) {$22$};
\node[text=white, font=\scriptsize, inner sep=0pt] at (235.7pt,99.9pt) {$9.19$};
\node[text=white, font=\scriptsize, inner sep=0pt] at (134.9pt,80.9pt) {$2703$};
\node[text=white, font=\scriptsize, inner sep=0pt] at (13.4pt,70.1pt) {$13.5$};
\node[text=white, font=\scriptsize, inner sep=0pt] at (128.9pt,70.1pt) {$103$};
\node[text=white, font=\scriptsize, inner sep=0pt] at (247.3pt,70.1pt) {$14.8$};
\node[text=white, font=\scriptsize, inner sep=0pt] at (20.6pt,59.2pt) {$13.2$};
\node[text=white, font=\scriptsize, inner sep=0pt] at (127.0pt,59.2pt) {$55.1$};
\node[text=white, font=\scriptsize, inner sep=0pt] at (238.9pt,59.2pt) {$14.8$};
\node[text=white, font=\scriptsize, inner sep=0pt] at (16.7pt,40.2pt) {$670$};
\node[text=white, font=\scriptsize, inner sep=0pt] at (116.8pt,40.2pt) {$3343$};
\node[text=white, font=\scriptsize, inner sep=0pt] at (231.5pt,40.2pt) {$1222$};
\node[text=white, font=\scriptsize, inner sep=0pt] at (23.5pt,29.3pt) {$46.1$};
\node[text=white, font=\scriptsize, inner sep=0pt] at (119.9pt,29.3pt) {$143$};
\node[text=white, font=\scriptsize, inner sep=0pt] at (228.7pt,29.3pt) {$65.4$};
\node[text=white, font=\scriptsize, inner sep=0pt] at (40.2pt,18.5pt) {$33.8$};
\node[text=white, font=\scriptsize, inner sep=0pt] at (127.2pt,18.5pt) {$39.3$};
\node[text=white, font=\scriptsize, inner sep=0pt] at (221.5pt,18.5pt) {$34$};
\end{tikzpicture}%
\end{adjustbox}
\caption{Share of each stage in the latency, with its time in ms where the segment is wide enough, by number of threads, and the total on the right.
Other is the switch into the sparse key together with ${\sf Im2}$, or with the conjugation on \ship.}
\label{fig:stages}
\end{figure}
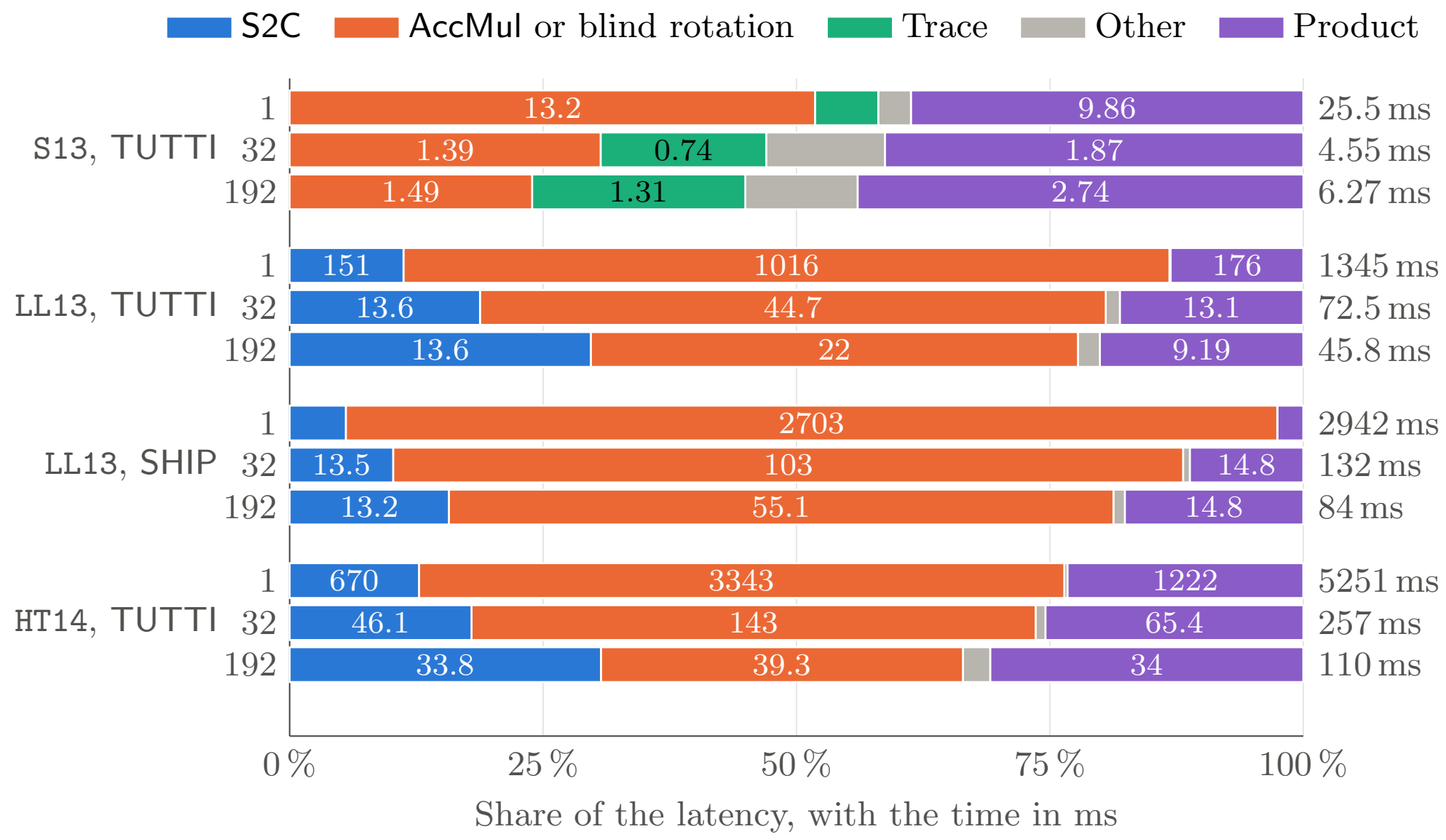

\paragraph{Where the time goes.}
\label{s:where-time}
Figure~\ref{fig:stages} splits the latency by stage.
\accmul\ dominates on one thread, with $76\%$ of the time on $\mathtt{LL13}$, and it scales well, by $46\times$ on $192$ threads.
\stc\ stops scaling at about $32$ threads, and the product operator gains little beyond it, so on $192$ threads the three stages are of similar size, on $\mathtt{HT14}$ $34$, $39$ and $34$\,ms.
Blind rotation scales as well as \accmul\ on $\mathtt{LL13}$, by $49\times$, but starts $2.7\times$ higher, and it takes most of \ship's time at every thread count.

\paragraph{Thread allocation.}
A budget of $T$ threads means three things in our implementation.
The library, and with it \stc, may use $T$ threads.
\accmul\ uses at most one pinned worker per physical core, so at most $192$, because workers on SMT siblings were slower; the single-slot sets use at most $32$.
The trace, the product and ${\sf Im2}$ run on a pool of at most $96$ workers, one socket's physical cores, because all $192$ cores were slower on the product.
Going from $192$ to $384$ threads therefore only changes \stc, and it does not help.
On $\mathtt{S13}$, the latency is lowest at $32$ threads: a key switch of its short chain has at most about $56$ parallel tasks, so further threads only add synchronization, and the trace and the product get slower from $32$ to $192$ threads, from $0.74$ to $1.31$ and from $1.87$ to $2.74$\,ms.
The machine has $6$ NUMA nodes.
We interleave the memory of the process over them, each \accmul\ worker allocates its buffers on its own node, and we move the key so that each worker's data sit on its node.
This also opts the process out of automatic NUMA balancing, whose page faults cost about $60$\,ms of \accmul\ per bootstrap on the $36$\,GiB key of $\mathtt{HT15}$.

\begin{figure}[t]
\centering
\begin{adjustbox}{max width=\textwidth}
\definecolor{slotsPanelLine}{HTML}{2A78D6}%
\definecolor{slotsPanelInk}{HTML}{52514E}%
\begin{tikzpicture}
\begin{axis}[
  scale only axis, width=150pt, height=86pt, font=\small,
  xmin=-0.4, xmax=13.4, xtick={0,2,...,12}, xticklabel={$2^{\pgfmathprintnumber{\tick}}$},
  xlabel={Messages $n$}, axis lines*=left, axis line style={slotsPanelInk, line width=0.4pt},
  tick style={slotsPanelInk, line width=0.4pt}, major tick length=2pt,
  ymajorgrids, grid style={black!12, line width=0.3pt},
  tick label style={text=slotsPanelInk, font=\footnotesize}, label style={text=slotsPanelInk},
  title style={text=slotsPanelInk, font=\small, at={(0,1)}, anchor=south west, inner xsep=0pt},
  every axis plot/.append style={color=slotsPanelLine, line width=1pt, mark=*, mark size=1pt}, clip=false,
  title={Precision (bits)}, ymin=4.5, ymax=9, ytick={5,6,7,8,9}]
\addplot coordinates {(0,8.37) (1,7.57) (2,7.3) (3,6.82) (4,6.69) (5,6.47) (6,6.36) (7,6.2) (8,6.11) (9,5.96) (10,5.81) (11,5.68) (12,5.42) (13,5.12)};
\node[anchor=south west, font=\footnotesize, text=slotsPanelInk] at (axis cs:0,8.37) {$8.37$};
\node[anchor=north east, font=\footnotesize, text=slotsPanelInk] at (axis cs:13,5.12) {$5.12$};
\end{axis}
\begin{axis}[xshift=186pt,
  scale only axis, width=150pt, height=86pt, font=\small,
  xmin=-0.4, xmax=13.4, xtick={0,2,...,12}, xticklabel={$2^{\pgfmathprintnumber{\tick}}$},
  xlabel={Messages $n$}, axis lines*=left, axis line style={slotsPanelInk, line width=0.4pt},
  tick style={slotsPanelInk, line width=0.4pt}, major tick length=2pt,
  ymajorgrids, grid style={black!12, line width=0.3pt},
  tick label style={text=slotsPanelInk, font=\footnotesize}, label style={text=slotsPanelInk},
  title style={text=slotsPanelInk, font=\small, at={(0,1)}, anchor=south west, inner xsep=0pt},
  every axis plot/.append style={color=slotsPanelLine, line width=1pt, mark=*, mark size=1pt}, clip=false,
  title={Peak memory (GB)}, ymin=0, ymax=10, ytick={0,2,4,6,8,10}]
\addplot coordinates {(0,1.3) (1,1.3) (2,1.4) (3,1.5) (4,1.5) (5,1.5) (6,1.6) (7,1.6) (8,1.9) (9,2.6) (10,3.5) (11,4.4) (12,6.1) (13,9.5)};
\node[anchor=south west, font=\footnotesize, text=slotsPanelInk] at (axis cs:0,1.3) {$1.3$};
\node[anchor=south east, font=\footnotesize, text=slotsPanelInk] at (axis cs:13,9.5) {$9.5$};
\end{axis}
\end{tikzpicture}%
\end{adjustbox}
\caption{Precision and peak memory of $\tutti_n$ on $\mathtt{HT14}$, for the runs of Figure~\ref{fig:slots}, with uniform random messages.}
\label{fig:slots-panels}
\end{figure}
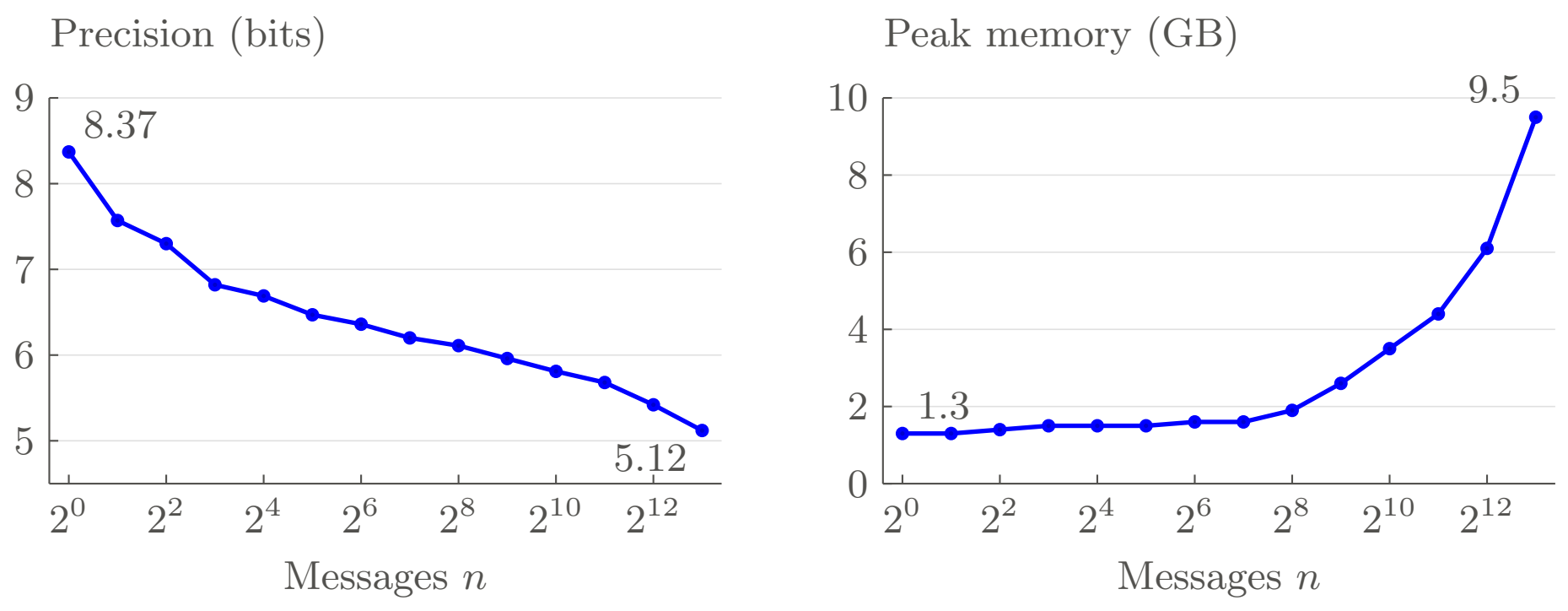

\paragraph{Varying the number of messages.}
Figure~\ref{fig:slots}, in the introduction, shows the latency of $\tutti_n$ on $\mathtt{HT14}$ for $n$ from $1$ to $N/2$, together with \ship on all slots and \paco at $128$ slots.
Dimension reduction cuts the \accmul load, so for $n\le2^{8}$ the latency barely changes with $n$ at any thread count: most of the time goes to the product operator, which is sequential and independent of $n$.
Above $2^{8}$, the \accmul load grows linearly in $n$ and the \stc\ load as $\sqrt{n}$, and parallelization absorbs most of that growth.
\ship, in contrast, gains at most $15\%$ from bootstrapping fewer messages, since only its \stc\ becomes lighter.
Figure~\ref{fig:slots-panels} adds the precision and the memory.
Precision rises as $n$ falls because the maximum runs over fewer slots, and memory falls with the key.

\paragraph{Repeated bootstrapping.}
We also ran three cycles of a bootstrap followed by $x\mapsto2x^2-1$, which maps $[-1,1]$ to itself.
With the real diagonals of Section~\ref{s:stc-placement}, each bootstrap keeps its precision against its own input: $4.88$, $4.85$ and $4.89$ bits on $\mathtt{LL13}$.

\paragraph{\paco.}
\paco~\cite{PaCo25} targets a medium number of slots: one instance bootstraps $C\le N/(4h)$ coefficients, and its parallel variant covers all $N/2$ slots with $N/(2C)$ instances side by side, which favours throughput over latency.
We run it within the modulus budget of $\mathtt{HT14}$, with $N=2^{14}$, $h=32$ and $C=N/(4h)=128$: its circuit takes $8$ levels ($2$ for its partial \cts, $5$ for the product and $1$ for its final transform) and leaves $7$ to the application, against $9$ for \tutti.
One instance takes $1307$, $149$ and $118$\,ms on $1$, $32$ and $192$ threads, at $6.1$ bits of precision, whereas \tutti refreshes the same $128$ slots in $245$, $35$ and $33$\,ms (Figure~\ref{fig:slots}); covering all $N/2$ slots with $2h=64$ instances takes \paco\ $7.1$\,s on $192$ threads, against $110$\,ms for \tutti.
Our choice of ring favours \paco: its original parameters use $N=2^{15}$ and $2^{16}$ with $h=64$ \cite{PaCo25}, but an instance that bootstraps only $N/(4h)$ coefficients is faster in the smaller ring.

\paragraph{TFHE-rs.}
We ran TFHE-rs~1.8.1 \cite{TFHErs} on our machine with the parameters of Zama's benchmark \cite{TFHErsBench}: a keyswitch followed by a programmable bootstrapping (PBS) on $4$-bit messages, with failure probability $2^{-128}$.
The PBS is the \textit{``multi-bit''} variant based on \cite{ZYLZL18}, which processes $4$ secret coefficients per step and so exposes parallelism within one bootstrapping.
Its fastest configuration takes $7.3$\,ms, on $32$ threads.
On the same number of threads, single-slot \tutti takes $4.55$\,ms, $1.6\times$ less, and $\tutti_n$ takes $25$ to $35$\,ms for up to $256$ slots on $\mathtt{HT14}$ and $15$\,ms for $256$ slots on $\mathtt{LL13}$.
Zama reports $4.57$\,ms on a different machine, with two AMD EPYC 9R45 processors (192 cores) \cite{TFHErsBench}, which single-slot \tutti matches on ours.
The nominal precisions are similar: $6.5$ bits for single-slot \tutti, and $5.1$ to $8.4$ bits for $\tutti_n$ on $\mathtt{HT14}$, rising as $n$ falls since the maximum runs over fewer slots.
The guarantees differ, however: the $4$ bits of TFHE-rs are exact up to a failure probability of $2^{-128}$, whereas our precision is a median worst-slot error, for which we did not measure a failure probability.
TFHE-rs also evaluates a lookup table during bootstrapping \cite{DM15,BR15,BGGJ19}, and the two schemes also differ in key sizes and security assumptions \cite{BV14}.
We therefore see single-slot \tutti and TFHE as complementary: at matching precision, an application may use \tutti for fixed-point multiplications and switch to TFHE~\cite{BGGJ20,LHHMQ21} to evaluate arbitrary functions.

\paragraph{On a laptop.}
We also ran single-slot \tutti and TFHE-rs on a MacBook Pro with an Apple M5 Pro ($15$ cores), with an AArch64 implementation of \tutti whose outputs are bit-identical to those of the implementation above, so that its precision is unchanged.
On one thread, \tutti takes $12.0$\,ms on $\mathtt{S13}$, and TFHE-rs~1.8.1, on the same $4$-bit parameters, $13.1$\,ms with its single-threaded PBS.
On $14$ threads, the fastest configuration of both, \tutti takes $2.4$\,ms and TFHE-rs $4.5$\,ms with its multi-bit PBS and a parallel keyswitch, $1.9\times$ more.
For \tutti these are medians over three processes of $500$ bootstraps each.

\paragraph{Cosets.}
\label{s:coset-results}
We implemented the coset variant of Section~\ref{s:cosets} with $h_t=8$ on $\mathtt{HT15}$, at the same precision.
It shrinks the bootstrapping key by almost $h_t$, from $36$ to $4.75$\,GiB, since its mask needs one more prime, but is faster than full-slot \tutti only on few threads, by $6\%$ on one thread, where its \accmul, reading each key row once for all cosets, is $23\%$ faster with the same arithmetic.
Table~\ref{tab:cosets} gives the full comparison.
From $48$ threads on, cosets lose: every coset repeats its product operator, and the cores spent on cosets are lost inside each $\tutti_n$.

\begin{table}[t]
\centering
\caption{Cosets on $\mathtt{HT15}$ with $h_t=8$, against full-slot \tutti: latency in ms at $T$ threads, precision in bits, key in GiB and peak memory in GB.}
\label{tab:cosets}
\setlength{\tabcolsep}{3.5pt}
\small
\begin{tabular}{lrrrrrrccr}
\toprule
 & $1$ & $8$ & $32$ & $48$ & $192$ & $384$ & Prec. & Key & Mem. \\
\midrule
Full slots & $23107$ & $\mathbf{3143}$ & $\mathbf{1152}$ & $\mathbf{839}$ & $\mathbf{424}$ & $\mathbf{427}$ & $18.44$ & $36$ & $66.1$ \\
Cosets & $\mathbf{21779}$ & $3483$ & $1205$ & $992$ & $972$ & $971$ & $18.63$ & $4.75$ & $48.9$ \\
\bottomrule
\end{tabular}
\end{table}

\section*{Acknowledgements}
We thank Jean-S\'ebastien Coron and Tim Seur\'e for helpful discussions.
Large language models assisted in writing and editing the text and in developing the implementation and the security-estimation scripts; the authors verified all results, proofs and claims.
Robin K\"ostler was supported by the ERC Advanced Grant Cloudmap
no. 787390. Antonio Guimar\~aes was supported by grants JDC2024-055789-I and CEX2024-001471-M, funded by
MICIU/AEI/10.13039/501100011033 and ESF+, and by the ERC Proof of Concept grant VERIFHE (GA: 101287502).

\bibliographystyle{alpha}
\bibliography{paper}

\pagebreak

\appendix
\renewcommand{\theHsection}{app.\Alph{section}}

\section{Details on \spru, \tutti\ and \ship}
\label{app:spru}

\paragraph{\spru\ key generation.}
Algorithm~\ref{alg:spru-keygen} generates the bootstrapping key of Algorithm~\ref{alg:spru-boot}.
For each sub-block $u$, it packs the secret coefficients that~\eqref{eq:spru-index} selects and encrypts their encoding under the public key.

\begin{algorithm}[H]
\caption{\spru\ key generation \cite{SPRU25}}
\label{alg:spru-keygen}
\begin{algorithmic}[1]
  \Require Block-binary $s$, weight $h$, and $n\le B/2$
  \Ensure Key $({\sf cs}_u)_{0\le u<2n}$
  \For{$u=0$ to $2n-1$}
    \ForAll{$0\le k<B/(2n)$, $0\le b<h$, $0\le a<n$}
      \State $\tilde{s}^{u}_{\iota(k,b,a)}\leftarrow s_{bB+j(u,k)}$
    \EndFor
    \State $S_u\leftarrow{\sf Ecd}(\tilde{\vec s}^{u})$
    \State ${\sf cs}_u\leftarrow{\sf Enc}_{\pk}(S_u)$
  \EndFor
  \State \Return $({\sf cs}_u)_{0\le u<2n}$
\end{algorithmic}
\end{algorithm}

\paragraph{\tutti\ key generation.}
Algorithm~\ref{alg:lola-keygen} generates the key of Algorithm~\ref{alg:lola-boot}: for each group $g$ and offset $u$, it packs the secret coefficients that~\eqref{eq:2k-index} selects and encrypts their encoding under the public key.

\begin{algorithm}[h]
\caption{$\tutti_n$ key generation}
\label{alg:lola-keygen}
\begin{algorithmic}[1]
  \Require Block-binary $s$, weight $h$, and $1\le k\le\log h$
  \Require $t=2^k$, $h_t=h/t$, $\nmax=B/2$, $n=t\cdot\nmax$
  \Ensure Key $({\sf cs}_u^{(g)})_{g\in\intv{t},\,u\in\intv{B}}$
  \For{$g=0$ to $t-1$}
    \For{$u=0$ to $B-1$}
      \ForAll{$r\in\intv{h_t}$, $a\in\intv{n}$}
        \State $\tilde{s}^{u,(g)}_{rn+a}\leftarrow s_{(gh_t+r)B+u}$
      \EndFor
      \State $S_u^{(g)}\leftarrow{\sf Ecd}(\tilde{\vec s}^{u,(g)})$
      \State ${\sf cs}_u^{(g)}\leftarrow{\sf Enc}_{\pk}(S_u^{(g)})$
    \EndFor
  \EndFor
  \State \Return $({\sf cs}_u^{(g)})_{g\in\intv{t},\,u\in\intv{B}}$
\end{algorithmic}
\end{algorithm}

\paragraph{\ship\ parameter sets.}
\label{app:ship-params}
Table~\ref{tab:ship-params} lists the \ship sets of Table~\ref{tab:ship-compare}, from \cite[Table~2]{SHIP25}, as our benchmark builds them.
The chain is listed from the bottom, as in Table~\ref{tab:lola-params}: the base prime, the \stc\ prime, the data levels, five primes for the product of depth $5$, and the auxiliary primes.
Every set has $h=31$, multiplexer base $\nu=4$, and the published half-width $w=\max(175,\lfloor N/(2h)\rfloor)$.
Every set exceeds the dense-key cap of Table~\ref{tab:dense-key} at its degree, by $4.5$ to $14$ bits: $216.5$ bits against the $212$ of the dense key of Hamming weight $512$ that $\mathtt{LL13}$ runs with, $437.0$ and $438.9$ against $430$, and $882.0$ against $868$.
We keep the published chains, which favours \ship in the comparison.

\begin{table}[t]
\centering
\caption{The \ship parameter sets of our comparison \cite[Table~2]{SHIP25}.
A prime of $b$ bits is a prime $p<2^b$ with $p\equiv 1\bmod 2N$ chosen by the prime generator of our implementation, and $\log PQ$ is computed on the actual primes; on $\mathtt{LL13}$ the \stc\ and Mult primes have $19.5$ and $19.3$ bits.
dnum is the number of digits of a GHS key switch at the top level, $\theta$ the column step, and $\lceil\log_4\lceil 2w/\theta\rceil\rceil$ the number of multiplexer digits; $\gamma=q_0/\Delta_{\rm in}$ as in Table~\ref{tab:lola-params}.}
\label{tab:ship-params}
\setlength{\tabcolsep}{3pt}
\small
\begin{tabular}{l*{12}{c}}
\toprule
\multirow{2}{*}{Set} & \multirow{2}{*}{$N$} & \multirow{2}{*}{$\log PQ$} & \multirow{2}{*}{dnum} & \multicolumn{5}{c}{(prime bits)\,$\times$\,(\# primes)} & \multirow{2}{*}{$\theta$} & \multirow{2}{*}{$w$} & \multirow{2}{*}{digits} & \multirow{2}{*}{$\gamma$} \\
\cmidrule(lr){5-9}
 & & & & Base & \stc & Mult & Boot & Aux & & & & \\
\midrule
$\mathtt{LL13}$ & $2^{13}$ & $216.5$ & $8$ & $25\times1$ & $20\times1$ & $20\times1$ & $25\times5$ & $28\times1$ & $6$  & $175$ & $3$ & $2^{5}$ \\
$\mathtt{LL14}$ & $2^{14}$ & $437.0$ & $4$ & $48\times1$ & $37\times1$ & $37\times1$ & $43\times5$ & $50\times2$ & $9$  & $264$ & $3$ & $2^{11}$ \\
$\mathtt{HT14}$ & $2^{14}$ & $438.9$ & $8$ & $28\times1$ & $23\times1$ & $23\times9$ & $25\times5$ & $29\times2$ & $9$  & $264$ & $3$ & $2^{5}$ \\
$\mathtt{HT15}$ & $2^{15}$ & $882.0$ & $4$ & $50\times1$ & $39\times1$ & $39\times9$ & $46\times5$ & $53\times4$ & $17$ & $528$ & $3$ & $2^{11}$ \\
\bottomrule
\end{tabular}
\end{table}

\section{Single-slot bootstrapping with fewer sequential steps}
\label{app:single-slot}

Algorithms~\ref{alg:single-keygen} and~\ref{alg:single-boot} give the single-slot circuit of Section~\ref{s:single-slot-impl}, with \(G\) groups and the second harmonic, in the notation of that section.

\begin{algorithm}[H]
\caption{Single-slot key generation with \(G\) groups and the second harmonic}
\label{alg:single-keygen}
\begin{algorithmic}[1]
  \Require Block-binary \(s\), weight \(h\), \(B=N/h\), a power of two \(G\mid B\)
  \Ensure Key \(({\sf cs}_j)_{0\le j<B/G}\)
  \State \(a_0\leftarrow(4/3)^{1/h}\), \(a_1\leftarrow(1/6)^{1/h}e^{\mathrm{I}\pi/h}\) \Comment{\(h\)-th roots of the weights \(\frac43\) and \(-\frac16\)}
  \For{\(j=0\) to \(B/G-1\)}
    \For{\(\ell=0\) to \(N/2-1\)}
      \State \(\tilde{s}^{j}_{\ell}\leftarrow a_{\eta(\ell)}\cdot s_{i_j(\ell)}\)
    \EndFor
    \State \(S_j\leftarrow{\sf Ecd}(\tilde{\vec s}^{j})\)
    \State \({\sf cs}_j\leftarrow{\sf Enc}_{\pk}(S_j)\) \Comment{at the top modulus times \(P\) (Section~\ref{s:extprod})}
  \EndFor
  \State \Return \(({\sf cs}_j)_{0\le j<B/G}\)
\end{algorithmic}
\end{algorithm}

\begin{algorithm}[H]
\caption{Single-slot bootstrapping with \(G\) groups and the second harmonic}
\label{alg:single-boot}
\begin{algorithmic}[1]
  \Require \(q\), key \(({\sf cs}_j)_{j<B/G}\), ciphertext \(\ct\) under the sparse key, \(\delta=(q/(4\pi\Delta))^{1/h}\)
  \Ensure Refreshed ciphertext \(\ct'\)
  \For{\(i=0\) to \(N-1\)} \Comment{\(2N\) exponentiations}
    \State \(r_{i,0}\leftarrow\delta\cdot\exp(2\pi\mathrm{I}\,C_i/q)\), \(r_{i,1}\leftarrow\delta\cdot\exp(4\pi\mathrm{I}\,C_i/q)\)
  \EndFor
  \State \({\sf acc}\leftarrow 0\)
  \For{\(j=0\) to \(B/G-1\)} \Comment{independent rows}
    \For{\(\ell=0\) to \(N/2-1\)}
      \State \(\rho^{j}_{\ell}\leftarrow r_{i_j(\ell),\,\eta(\ell)}\)
    \EndFor
    \State \(E_j\leftarrow{\sf Ecd}(\vec\rho^{j})\) \Comment{\(4Gh\) coefficients, transforms of \(4Gh\) points}
    \State \({\sf acc}\leftarrow{\sf Add}({\sf acc},{\sf ExtMultR}({\sf cs}_j,E_j))\)
  \EndFor
  \State \({\sf acc}\leftarrow\tr{2Gh}{2h}({\sf acc})\) \Comment{adds the groups; \(\log G\) automorphisms}
  \State \({\sf acc}\leftarrow\pr{2h}{2}({\sf acc})\) \Comment{multiplies the blocks; \(\log h\) automorphisms and levels}
  \State \({\sf acc}\leftarrow{\sf Add}({\sf acc},{\sf Rot}({\sf acc},1))\) \Comment{adds the two harmonics}
  \State \Return \(\ct'\leftarrow{\sf Im2}({\sf acc})\)
\end{algorithmic}
\end{algorithm}

\section{BV and GHS key switching}
\label{app:ks}

This appendix recalls the two key switches compared in Section~\ref{s:sparse-keys} and counts their cost.
Throughout, $\ct=(c_0,c_1)\in{\cal R}_{q_\ell}^2$ encrypts a message under $\overline{\sk}=(1,\bar s)$, so $c_0+c_1\cdot\bar s=m+e\pmod{q_\ell}$, and the output should encrypt the same message under $\sk=(1,s)$.
In the sparse-secret encapsulation, $\bar s$ is the dense key, $s$ the sparse key, and $q_\ell=q_0$.
When an automorphism produced $(c_0,c_1)$, the permutation of coefficients is not part of the counts below, which cover the switch itself.

\subsection{The BV decomposition}
\label{app:ks-bv}

The decomposition of Brakerski and Vaikuntanathan \cite{BV11} works entirely modulo $q_\ell$.
It writes each coefficient of $c_1$, taken in $\{0,\ldots,q_\ell-1\}$, in base $W=2^\omega$ with
\[
  D=\ceil{\log q_\ell/\omega}
\]
digits.
This produces digit polynomials $c_{1,0},\ldots,c_{1,D-1}$ with coefficients in $\{0,\ldots,W-1\}$ such that
\begin{equation}
\label{eq:bv-digits}
  c_1=\sum_{i=0}^{D-1} c_{1,i}\, W^i \pmod{q_\ell}.
\end{equation}
The key consists of the samples ${\sf swk}_i=(b_i,a_i)\in{\cal R}_{q_\ell}^2$ under $\sk$ of the multiples $W^i\bar s$,
\begin{equation}
\label{eq:bv-key}
  b_i+a_i\cdot s=W^i \bar s+e_i\pmod{q_\ell},
\end{equation}
with $e_i$ the error of the $i$-th sample.
The switched ciphertext is
\begin{equation}
\label{eq:bv-switch}
  (c'_0,c'_1)
  =(c_0,0)+\sum_{i=0}^{D-1} c_{1,i}\cdot{\sf swk}_i
  \pmod{q_\ell}.
\end{equation}
Substituting \eqref{eq:bv-key} and then \eqref{eq:bv-digits} gives
\begin{align*}
  c'_0+c'_1\cdot s
  &=c_0+\sum_{i=0}^{D-1} c_{1,i}(W^i \bar s+e_i)\\
  &=c_0+c_1\cdot \bar s+\sum_{i=0}^{D-1} c_{1,i}e_i
  \pmod{q_\ell},
\end{align*}
so the message is preserved and $\sum_i c_{1,i}e_i$ is the switching error.
A coefficient of a ring product collects at most $N$ terms, and the digits are at most $W-1$, so
\begin{equation}
\label{eq:bv-err}
  \Bigl\|\sum_{i=0}^{D-1} c_{1,i}e_i\Bigr\|_\infty
  \le
  D\cdot N\cdot(W-1)\cdot\max_i\|e_i\|_\infty.
\end{equation}
Narrower digits lower this bound and add further pairs of multiplications.
The published key is the tuple $({\sf swk}_i)_{i<D}$, all of it in ${\cal R}_{q_\ell}^2$.

\begin{algorithm}[H]
\caption{BV key switch with $\omega$-bit digits}
\label{alg:bv-ks}
\begin{algorithmic}[1]
\Require $\ct=(c_0,c_1)\in{\cal R}_{q_\ell}^2$ under $\overline{\sk}=(1,\bar s)$, digit width $\omega$, keys ${\sf swk}_i$ as in \eqref{eq:bv-key}
\Ensure A ciphertext under $\sk=(1,s)$
\State $W\gets 2^\omega$, $D\gets\ceil{\log q_\ell/\omega}$
\State Decompose $c_1=\sum_{i=0}^{D-1} c_{1,i} W^i$ with coefficients of $c_{1,i}$ in $\{0,\ldots,W-1\}$
\State $(c'_0,c'_1)\gets(c_0,0)$
\For{$i=0$ to $D-1$}
  \State $(c'_0,c'_1)\gets(c'_0,c'_1)+c_{1,i}\cdot{\sf swk}_i\pmod{q_\ell}$
\EndFor
\State \Return $(c'_0,c'_1)$
\end{algorithmic}
\end{algorithm}

\subsection{Operation count}
\label{app:ks-cost}

Equation~\eqref{eq:bv-switch} multiplies each of the $D$ digits by both polynomials of ${\sf swk}_i$, which is $2D$ multiplications in ${\cal R}_{q_\ell}$.
The first output component is $c_0$ plus $D$ products, which takes $D$ additions, and the second is the sum of $D$ products, which takes $D-1$.
The switch therefore performs
\[
  2D\text{ multiplications and }2D-1\text{ additions in }{\cal R}_{q_\ell}.
\]
Line~2 of Algorithm~\ref{alg:bv-ks} only reads the base-$W$ digits of the $N$ coefficients of $c_1$.
A digit width $\omega\ge\log q_\ell$ gives $D=1$: the single digit is $c_1$ itself, and \eqref{eq:bv-err} multiplies the key error by a polynomial of coefficient size $q_\ell$.
Every useful decomposition therefore has $D\ge 2$.

\subsection{The GHS switch}
\label{app:ks-ghs}

Gentry, Halevi, and Smart \cite{GHS12} avoid the digit sum by enlarging the modulus by an auxiliary prime $P$.
The key is one sample ${\sf swk}=(b_{\sf swk},a_{\sf swk})\in{\cal R}_{Pq_\ell}^2$ under $\sk$ of $P\cdot\bar s$,
\[
  b_{\sf swk}+a_{\sf swk}\cdot s=P\cdot\bar s+e_{\sf swk}\pmod{Pq_\ell}.
\]
Lift $c_1$ to ${\cal R}_{Pq_\ell}$ by its representative in $\{0,\ldots,q_\ell-1\}$ and set
\[
  (d_0,d_1)=c_1\cdot{\sf swk}\pmod{Pq_\ell},
  \qquad
  (c'_0,c'_1)
  =(c_0,0)+\near{P^{-1}(d_0,d_1)}\pmod{q_\ell}.
\]
Then $d_0+d_1\cdot s=c_1\cdot P\cdot\bar s+c_1\cdot e_{\sf swk}\pmod{Pq_\ell}$, and the division by $P$ recovers $c_1\cdot\bar s$ with the scaled key error $c_1\cdot e_{\sf swk}/P$ and a rounding error of at most $1/2$ per coefficient.
With $P\ge q_\ell$ this error is small without any digits.
The switch costs two multiplications modulo $Pq_\ell$, about four of the size of one modulo $q_\ell$ when $\log P=\log q_\ell$, against the $2D$ of the BV switch.

The price is the modulus of the published sample: one sample under $s$ modulo $Pq_\ell$, against $D$ samples modulo $q_\ell$.
For the sparse key at $q_0$, any auxiliary prime lifts that sample above the cap of Table~\ref{tab:sparse-block} once $q_0$ sits at it, to at least $54$ bits for a $27$-bit $q_0$ with $\log P=\log q_0$.
This is why the encapsulation uses the decomposition, in \tutti as in our implementation of \ship.

\section{Combinatorial attacks on block-binary and windowed secrets}
\label{app:may}

This appendix details the combinatorial estimates of Section~\ref{s:may}.
Sections~\ref{app:may-instance} to~\ref{app:may-search} cost Odlyzko's split \(T_{\mathrm{Odl}}\) (Table~\ref{tab:may-spru}) and May's Meet-LWE with Gaussian error guessing, \(T_{\mathrm{ML}}\) (Table~\ref{tab:may-params}), on the anchored block-binary secret of \spru, following May's paper \cite{May21}.
Odlyzko's meet-in-the-middle is his Section~3, applied to a product of blocks; Meet-LWE is his Algorithm~3, with the representation counts of his Section~7 rewritten for a binary secret and the list-cost recurrence of his Section~7.4.
Section~\ref{app:may-free} records the tree behind the guessing-free columns \(T_{\mathrm{free}}\) and \(T_{\mathrm{ops}}\), Section~\ref{app:may-rep2} checks that digits \(\pm 2\) (his Section~8) leave the rounded costs unchanged, Section~\ref{app:may-ship} treats the windowed secret of \ship, and Section~\ref{app:may-out} lists what the estimate leaves out.

The procedure prices the attack without building the lists.
Which term dominates is read off his list-merging recurrence, and the polynomial factors of sorting, hashing, and the number of lists are omitted, as in his tables (his Section~2.3).
Where his Lemma~1 replaces multinomial coefficients by binary entropy, the numbers below use the multinomials themselves,
\[
  \binom{n}{c_1,\ldots,c_k}
  =
  \frac{n!}{c_1!\cdots c_k!\,(n-c_1-\cdots-c_k)!},
\]
evaluated as \(\log(n!)=\ln\Gamma(n+1)/\ln 2\); this is the exact count behind his non-asymptotic tables.

\subsection{The anchored instance}
\label{app:may-instance}

A block-binary secret lies in \(\{0,1\}^N\) and has exactly one coefficient equal to \(1\) in each of \(h\) blocks of length \(B=N/h\); \spru\ fixes \(s_0=1\), so the first block is known.
Deleting its \(B\) columns leaves a secret of dimension \(N'=N-B\) and weight \(h'=h-1\), uniform on the \(S=B^{h'}\) placements of the remaining ones.

One \rlwe sample is read in the coefficient embedding.
May's Definition~1 asks for a uniform matrix, and his Section~2.1 notes that Ring-LWE samples are dependent and that the attack does not use the dependence.
We follow that choice: the \(N'\) equations are charged as uniform \lwe samples modulo \(q\), where \(q\) is the base prime \(q_0\) of the row of Table~\ref{tab:sparse-block}.

The error is the discrete Gaussian of standard deviation \(3.2\) (Section~\ref{s:sparse-keys}), whereas May's secrets and errors are ternary (his Definition~1).
Odlyzko's collision only needs the two sides of a genuine pair to differ by a short vector (his Section~3.1), which holds for this Gaussian at every modulus in the table; the guessing cost inside Meet-LWE depends on the alphabet and is computed in Section~\ref{app:may-tree}.

Two rows of Table~\ref{tab:may-spru} serve as running examples: \(N=2^{13}\), \(h=32\), \(\log q=27\), with \(B=256\), \(N'=7936\), and \(h'=31\); and \(N=2^{11}\), \(h=64\), \(\log q=27\), with \(B=32\), \(N'=2016\), and \(h'=63\), the row of the sparse key of \(\mathtt{LL13}\) and \(\mathtt{HT14}\).

\subsection{Odlyzko's meet-in-the-middle on the blocks}
\label{app:may-odl}

May's Algorithm~1 splits a ternary secret by coordinates, with lists of size \(\sqrt{S}\) after a permutation of the columns (his Section~3.2).
The block-binary secret is a product of \(h'\) independent symbols from an alphabet of size \(B\), so the same meet-in-the-middle splits the blocks instead, and no permutation is needed (Algorithm~\ref{alg:odl-blocks}).

\begin{algorithm}[H]
\caption{Odlyzko's meet-in-the-middle on the blocks, following Algorithm~1 of~\cite{May21}}
\label{alg:odl-blocks}
\begin{algorithmic}[1]
\Require Matrix \(\mathbf{A}\) with \(N'\) columns, target \(\mathbf{b}\), block length \(B\), weight \(h'\)
\Ensure A block-binary \(\mathbf{s}\) of weight \(h'\) with \(\mathbf{A}\mathbf{s}-\mathbf{b}\) short
\State \(a\gets\lceil h'/2\rceil\). Fix a partition of the \(h'\) blocks into a left group of size \(a\) and a right group of size \(h'-a\).
\ForAll{placements \(\mathbf{s}_L\) of a single \(1\) in each left block}
  \State Store \(\bigl(\mathbf{s}_L,\,\ell(\mathbf{A}_L\mathbf{s}_L)\bigr)\).
\EndFor
\ForAll{placements \(\mathbf{s}_R\) of a single \(1\) in each right block}
  \State Store \(\bigl(\mathbf{s}_R,\,\ell(\mathbf{b}-\mathbf{A}_R\mathbf{s}_R)\bigr)\).
\EndFor
\ForAll{pairs whose hash labels agree}
  \State \(\mathbf{s}\gets\mathbf{s}_L+\mathbf{s}_R\)
  \If{\(\mathbf{s}\) has one \(1\) in every block and \(\mathbf{A}\mathbf{s}-\mathbf{b}\) is short}
    \State \Return \(\mathbf{s}\)
  \EndIf
\EndFor
\end{algorithmic}
\end{algorithm}

The hash \(\ell\) is May's Equation~(2): the most significant bit of each coordinate, with both labels stored on a residue where a short error can flip the bit.
A coordinate of a genuine pair fails to collide only when it lies within \(\lvert e_i\rvert\) of \(0\) or of \(q/2\).
At \(\log q\ge 27\) and standard deviation \(3.2\), that fraction of coordinates is negligible beside the polynomial factors his Section~3 already drops.
The algorithm enumerates secret halves only, and the larger list has size \(B^{a}\), so time and memory are
\[
  T_{\mathrm{Odl}}
  =
  \lceil h'/2\rceil\cdot\log B
\]
bits.
For an odd weight the larger half has one block more; fixing \(s_0=1\) turns an even \(h\) into \(h'=h-1\) and leaves \(T_{\mathrm{Odl}}\) unchanged, since \(\lceil (h-1)/2\rceil=h/2\).
On the first running example, \(a=16\) and \(T_{\mathrm{Odl}}=16\cdot\log 256=128\); the coordinate split of his Section~3.2, which ignores the blocks, costs \(147.0\) bits there.

\subsection{Representation 1 for a binary secret}
\label{app:may-rep1}

Meet-LWE writes the secret as a sum \(\mathbf{s}=\mathbf{s}_1+\mathbf{s}_2\) in many ways and merges lists of the summands (May's Algorithm~3 and Figure~2).
A binary vector has, up to order, one writing of each one, \(1=1+0=0+1\), and one of each zero, \(0=0+0\): this is his Rep-0 (Section~6).
Rep-1 (his Section~7) also writes
\[
  0=1+(-1)=(-1)+1,
\]
and each such pair is an extra representation.
The summands become ternary, so the enumerated set grows, but each secret is hit more often, so the merge keeps a smaller fraction of it; the list size at a level is the size of the set divided by the number of representations.

May's Section~7.1 starts from a balanced ternary secret, whose two signs each contribute a binomial coefficient, so the coefficients appear squared.
A block-binary secret has no \(-1\): the first level splits only the \(h'\) ones, and the cancellations are the only source of minus signs.
From the second level on, the parent is an arbitrary ternary vector with \(p\) entries \(+1\) and \(m\) entries \(-1\), and his Section~7.2, with the rounding he allows for concrete parameters (his Section~2.1), applies to each sign separately.
The block length \(B\) enters only \(T_{\mathrm{Odl}}\) and the search space \(S=B^{h'}\) against which Section~\ref{s:may} measures exponents; the representation counts depend on \(h'\) and \(N'\) alone, as for the ambient set his Section~7 enumerates.

\begin{algorithm}[H]
\caption{Representation counts for a binary secret of weight \(h'\) (May, Section~7, unbalanced)}
\label{alg:rep-counts}
\begin{algorithmic}[1]
\Require Dimension \(N'\), weight \(h'\), and level parameters \(\varepsilon^{(1)},\ldots,\varepsilon^{(k)}\ge 0\)
\Ensure Search-space sizes \(S^{(j)}\) and representation counts \(R^{(j)}\) for \(j=1,\ldots,k\), or rejection
\State \(a\gets\lceil h'/2\rceil\), \(\;p\gets a+\varepsilon^{(1)}\), \(\;m\gets\varepsilon^{(1)}\)
\If{\(2\varepsilon^{(1)}>N'-h'\)} \Return rejection \EndIf
\State \(S^{(1)}\gets\binom{N'}{p,\,m}\)
\State \(R^{(1)}\gets\binom{h'}{a}\binom{N'-h'}{\varepsilon^{(1)},\,\varepsilon^{(1)}}\)
\For{\(j=2\) to \(k\)}
  \State \(z\gets N'-p-m\)
  \If{\(2\varepsilon^{(j)}>z\)} \Return rejection \EndIf
  \State \(p'\gets\lceil p/2\rceil+\varepsilon^{(j)}\), \(\;m'\gets\lceil m/2\rceil+\varepsilon^{(j)}\)
  \State \(R^{(j)}\gets\binom{p}{\lceil p/2\rceil}\binom{m}{\lceil m/2\rceil}\binom{z}{\varepsilon^{(j)},\,\varepsilon^{(j)}}\)
  \State \(S^{(j)}\gets\binom{N'}{p',\,m'}\)
  \State \(p\gets p'\), \(\;m\gets m'\)
\EndFor
\State \Return \((S^{(j)}, R^{(j)})_{j=1}^{k}\)
\end{algorithmic}
\end{algorithm}

At the first level, \(\mathbf{s}_1\) has \(a+\varepsilon^{(1)}\) entries \(+1\) and \(\varepsilon^{(1)}\) entries \(-1\); the factor \(\binom{h'}{a}\) chooses which of the ones lie in \(\mathbf{s}_1\), and the multinomial in \(N'-h'\) chooses the zero coordinates that carry the two kinds of cancellation.
When the parent is balanced and both sign counts are even, the later levels reduce to the squared binomial of his Section~7.2.

On the first running example, one level with \(\varepsilon^{(1)}=4\) gives \(a=16\), so \(\mathbf{s}_1\) has \(20\) entries \(+1\) and \(4\) entries \(-1\), and
\[
  \log S^{(1)}=245.19,
  \qquad
  \log\binom{31}{16}=28.16,
  \qquad
  \log\binom{7905}{4,\,4}=94.41,
\]
so \(\log R^{(1)}=122.58\) and the level-1 list \(S^{(1)}/R^{(1)}\) has \(122.61\) bits.
With \(\varepsilon^{(1)}=0\), May's Rep-0 on this binary vector, the list has \(134.8\) bits, larger than both the list for \(\varepsilon^{(1)}=4\) and \(T_{\mathrm{Odl}}=128\).

\subsection{The cost of one tree}
\label{app:may-tree}

Algorithm~\ref{alg:rep-counts} fixes the shape of the lists.
May's Algorithm~3 then builds them as a binary tree (his Figure~4) and guesses part of the error so that one representation of the true secret survives an exact filter.

His Section~5 takes that filter to be \(r=\lfloor\log_q R^{(1)}\rfloor\) whole coordinates, because the range \(q^{r}\) must be at most the number of representations.
Inner levels match exactly on a nonincreasing sequence of coordinate counts \(r^{(j)}=\lfloor\log_q R^{(j)}\rfloor\) (his Sections~7.2 and~7.3).
Only the top merge is approximate, by the hash \(\ell\), on the remaining \(N'-r\) coordinates (his Section~7.4).
On the moduli of Table~\ref{tab:sparse-block} this coordinate count is often zero at an inner level: \(R^{(j)}<q\) leaves that level with no filter, and the list stays at size \(S^{(j)}\).

The costs in Table~\ref{tab:may-params} use a filter whose range is the representation count itself.
Any hash of at most \(\log R^{(j)}\) bits has that range, as does reduction modulo a divisor of \(q\) of the same bit length.
May's argument in Section~5 still applies: the range is at most \(R^{(j)}\), so some representation is expected to hit the target.
Inner levels consume a nonincreasing number of those bits, which is his requirement that each level fix a subset of the bits fixed above it.
A tree whose representation counts increase is rejected.
The error is guessed only at the top, and only on whole coordinates.
If \(\log R^{(1)}\) is positive, the number of coordinates is the smallest \(r\) with \(r\cdot\log q\ge\log R^{(1)}\),
\[
  r=\bigl\lceil\log R^{(1)}/\log q\bigr\rceil.
\]
Bits of the last coordinate beyond \(\log R^{(1)}\) are left unused, so the filter range stays equal to \(R^{(1)}\).
Each of the \(r\) coordinates is nevertheless drawn from the full guessing alphabet.

That alphabet is where the Gaussian error enters.
May's Section~5.2 guesses a ternary error and charges \(3^{r/2}\), the meet-in-the-middle over \(r/2\) coordinates on each side of his Algorithm~3.
Here each guessed coordinate lies in
\[
  \mathcal{E}=\{-8,\ldots,8\},
  \qquad
  \lvert\mathcal{E}\rvert=17,
\]
about \(\pm 2.5\) standard deviations.
The guessing cost is \(\lvert\mathcal{E}\rvert^{r/2}\), that is \(T_g=(r/2)\log\lvert\mathcal{E}\rvert\) bits.
Under the discrete Gaussian proportional to \(\exp(-t^2/(2\cdot 3.2^2))\) on the integers, a coordinate falls outside \(\mathcal{E}\) with probability \(0.77\%\).
Over the \(r\le 7\) coordinates that occur on these rows, redrawing such a guess costs under a tenth of a bit and does not move a cost rounded to the nearest bit.

The list-merging recurrence is his Section~7.4, with \(q^{r^{(j)}-r^{(j+1)}}\) replaced by the bit budget \(R^{(j)}/R^{(j+1)}\).
The leaves are an ordinary meet-in-the-middle of size \(\sqrt{S^{(k)}}\), as in his depth-\(d\) tree.
Match-and-filter (his Section~7.3) drops sums whose weight is wrong.
The time of a level is therefore the larger of the stored list and the number of pairs that survive the filter; in bits, the level cost is \(\max(\log L^{(j)},\,2\log L^{(j+1)}-\log(R^{(j)}/R^{(j+1)}))\), which is how Algorithm~\ref{alg:tree-cost} is evaluated.

\begin{algorithm}[H]
\caption{Bit cost of one representation tree (May, Section~7.4, bit-granular filter)}
\label{alg:tree-cost}
\begin{algorithmic}[1]
\Require Dimension \(N'\), modulus \(q\), alphabet \(\mathcal{E}\), and counts \((S^{(j)},R^{(j)})_{j=1}^{k}\) from Algorithm~\ref{alg:rep-counts}
\Ensure Bit cost \(T=T_\ell+T_g\), or rejection
\If{\(R^{(1)}\ge\cdots\ge R^{(k)}>0\) does not hold} \Return rejection \EndIf
\State \(L^{(j)}\gets S^{(j)}/R^{(j)}\) for \(j=1,\ldots,k\). Reject if any \(L^{(j)}<1\).
\State \(L^{(k+1)}\gets\sqrt{S^{(k)}}\) and \(R^{(k+1)}\gets 1\)
\State \(T^{(k+1)}\gets L^{(k+1)}\)
\For{\(j=k\) down to \(1\)}
  \State \(T^{(j)}\gets\max\Bigl(L^{(j)},\;\bigl(L^{(j+1)}\bigr)^{2}\cdot R^{(j+1)}/R^{(j)}\Bigr)\)
\EndFor
\State \(r\gets\bigl\lceil\log R^{(1)}/\log q\bigr\rceil\)
\State \(T^{(0)}\gets\max\Bigl(L^{(1)},\;\bigl(L^{(1)}\bigr)^{2}/2^{N'-r}\Bigr)\)
\State \(T_\ell\gets\log\bigl(\max_{0\le j\le k+1}T^{(j)}\bigr)\) and \(T_g\gets(r/2)\log\lvert\mathcal{E}\rvert\)
\State \Return \(T_\ell+T_g\)
\end{algorithmic}
\end{algorithm}

The same recurrence, fed with May's balanced ternary counts and with his coordinate filter \(r=\lfloor\log_q R\rfloor\), is what reproduces his NTRU-HPS tables to within \(1\)--\(5\) bits.

On the first running example the tree has one level, \(\log L^{(1)}=122.61\), and the leaves have the same size because \(\log R^{(1)}\) accounts for the whole gap between \(S^{(1)}\) and the meet-in-the-middle.
Here \(r=\lceil 122.58/27\rceil=5\), so
\[
  T_g=\log 17^{5/2}=10.22
  \qquad\text{and}\qquad
  T=122.61+10.22=132.83.
\]
The approximate top merge costs \(T^{(0)}=L^{(1)}\): a hash of length \(N'-r\) eliminates the false pairs, and writing the list is the expensive step.
Rounded, this is \(133\) bits, above \(T_{\mathrm{Odl}}=128\), so with error guessing Odlyzko's split is cheaper on this row; the guessing-free tree (Section~\ref{app:may-free}) brings it to \(125\) list elements.

The second running example is a depth-\(3\) tree in May's numbering: two representation levels, with \((\varepsilon^{(1)},\varepsilon^{(2)})=(6,0)\), and a meet-in-the-middle underneath.
Algorithm~\ref{alg:rep-counts} returns
\[
\begin{aligned}
  \log S^{(1)}&=324.27, & \log R^{(1)}&=171.81, & \log L^{(1)}&=152.46,\\
  \log S^{(2)}&=181.99, & \log R^{(2)}&=39.36,  & \log L^{(2)}&=142.63,
\end{aligned}
\]
and the leaves have size \(\tfrac12\log S^{(2)}=91.00\) bits.
The inner join has range \(R^{(2)}\) and costs \(\max(142.63,\,181.99-39.36)=142.63\).
The join of the two level-\(2\) lists has range \(R^{(1)}/R^{(2)}\) and costs
\[
  \max\bigl(152.46,\;2\cdot 142.63-(171.81-39.36)\bigr)=152.81.
\]
That join is the most expensive level, and with \(r=\lceil 171.81/27\rceil=7\) one has \(T_g=\log 17^{7/2}=14.31\), so the tree costs \(167.12\) bits, which rounds to the \(167\) of Table~\ref{tab:may-params}, above Odlyzko's \(160\).
May's own coordinate counts \(r^{(1)}=\lfloor 171.81/27\rfloor=6\) and \(r^{(2)}=\lfloor 39.36/27\rfloor=1\) give \(167.3\) bits after re-optimizing \(\varepsilon\).
The bit-granular filter matters when an inner level receives no whole coordinate: on the row \(N=2^{13}\), \(h=64\), \(\log q=102\), May's rule gives \(r^{(2)}=\lfloor 43.13/102\rfloor=0\), and the cost rises from \(218.0\) to \(230.7\) bits, above Odlyzko's \(224\).
The table reports the bit-granular filter.

List construction is independent of \(q\), while a smaller modulus increases \(r\) and hence \(T_g\); this is the correction of May's ternary guessing term for a Gaussian error.
With error guessing, it is why Meet-LWE loses to Odlyzko's split (Algorithm~\ref{alg:odl-blocks}) on the rows with \(\log q\in\{27,59\}\); the guessing-free lists (Section~\ref{app:may-free}) drop \(T_g\) and are the cheaper attack on those rows.

\subsection{The search}
\label{app:may-search}

A tree is a depth \(d\) in May's sense, that is \(k=d-1\) representation levels and a final meet-in-the-middle, together with a tuple \((\varepsilon^{(1)},\ldots,\varepsilon^{(k)})\).
His Section~7.5 found depth \(5\) no better than depth \(4\) on ternary secrets of much larger relative weight, so the search covers depths \(2\), \(3\), and \(4\), with \(\varepsilon^{(1)}\in\{0,\ldots,40\}\), \(\varepsilon^{(2)}\le\min(\varepsilon^{(1)},24)\), and \(\varepsilon^{(3)}\le\min(\varepsilon^{(2)},12)\).
Tuples rejected by Algorithm~\ref{alg:rep-counts} or Algorithm~\ref{alg:tree-cost} are discarded, and the Meet-LWE cost of the instance is the minimum of \(T\) over the survivors.

\begin{table}[t]
\centering
\caption{Trees attaining the Meet-LWE cost \(T_{\mathrm{ML}}=T_\ell+T_g\) with Gaussian error guessing, against Odlyzko's split \(T_{\mathrm{Odl}}\) of Table~\ref{tab:may-spru}.
\(T_\ell\) and \(T_g\) are in bits before rounding; \(T_{\mathrm{ML}}\) and \(T_{\mathrm{Odl}}\) are rounded to the nearest bit.
The depth is May's tree depth.
On every row the parameters below the first level are zero, and depth \(4\) is never chosen.}
\label{tab:may-params}
\setlength{\tabcolsep}{4pt}
\small
\begin{tabular}{cccrcrrrrr}
\toprule
\(N\) & \(h\) & \(\log q_0\) & depth & \(\varepsilon\) & \(r\) & \(T_\ell\) & \(T_g\) & \(T_{\mathrm{ML}}\) & \(T_{\mathrm{Odl}}\) \\
\midrule
\(2^{11}\) & \(64\) & \(27\)  & \(3\) & \((6,0)\) & \(7\) & \(152.81\) & \(14.31\) & \(167\) & \(160\) \\
\(2^{12}\) & \(64\) & \(51\)  & \(3\) & \((8,0)\) & \(5\) & \(179.95\) & \(10.22\) & \(190\) & \(192\) \\
\(2^{13}\) & \(64\) & \(102\) & \(3\) & \((8,0)\) & \(3\) & \(211.85\) & \(6.13\) & \(218\) & \(224\) \\
\(2^{14}\) & \(64\) & \(209\) & \(3\) & \((8,0)\) & \(2\) & \(243.80\) & \(4.09\) & \(248\) & \(256\) \\
\midrule
\(2^{13}\) & \(32\) & \(27\) & \(2\) & \((4)\)   & \(5\) & \(122.61\) & \(10.22\) & \(133\) & \(128\) \\
\(2^{14}\) & \(32\) & \(59\) & \(2\) & \((4)\)   & \(3\) & \(138.61\) & \(6.13\) & \(145\) & \(144\) \\
\bottomrule
\end{tabular}
\end{table}

The winning first-level counts are \(4\), \(6\), and \(8\), so the cap at \(40\) does not bind.
With error guessing, the cost of May's original attack is the cheaper of \(T_{\mathrm{ML}}\) and \(T_{\mathrm{Odl}}\); the guessing-free cost is the separate column \(T_{\mathrm{free}}\) of Table~\ref{tab:may-spru}.

\subsection{The guessing-free tree}
\label{app:may-free}

Kirshanova and May remove the error guess by locality-sensitive hashing \cite{KM21}; the success-probability analysis of that attempt is flawed \cite[Appendix~A]{HKLS22}.
The column \(T_{\mathrm{free}}\) of Table~\ref{tab:may-spru} instead replaces the exact match by a box, as in \cite{HKLS22}, with the cost model of \cite[Section~3.3]{BLLLW26}, on the representation counts of Algorithm~\ref{alg:rep-counts} at depth \(2\).
On the first running example the tree uses \(\varepsilon=3\), seven coordinates in boxes of half-width \(3072\) and one of half-width \(4096\).
The top list has \(2^{125.1}\) elements and is the bottleneck; the leaves have size \(2^{113}\), with \(2^{11.8}\) hash repetitions.
The column \(T_{\mathrm{ops}}\) adds \(\log(\log L)\) bits for hashing each element of the largest list \(L\) on about \(\log L\) coordinates.
With the ten hash repetitions per expected hit that Bi et al.\ charge \cite{BLLLW26}, where May charges one, the row \(N=2^{13}\), \(h=32\) costs \(128\) list elements and \(135\) word operations instead of \(125\) and \(132\).
Its top list stores \(2^{125}\) elements and Odlyzko's \(2^{128}\), far above the memory of a lattice attack at the \(128\)-bit line, so a memory-aware cost would put both well above \(128\).

\subsection{Digits \texorpdfstring{\(\pm 2\)}{±2}}
\label{app:may-rep2}

May's Rep-2 (his Section~8) enlarges the digit set by \(\pm 2\).
Appendix~A.1 of his paper introduces three parameters, collected in his Table~11.
The parameter \(\varepsilon_{10}\) counts extra pairs \((1,-1)\) on zeros, which is \(\varepsilon\) from Rep-1.
The parameter \(\varepsilon_{21}\) counts pairs that realize a parent \(\pm 1\) by \((2,-1)\) or \((-1,2)\).
The parameter \(\varepsilon_{20}\) counts pairs \((2,-2)\) on zeros.
His formulas square a multinomial because the parent secret has both signs.
The block-binary parent has only the sign \(+1\), so the square becomes one multinomial on the \(h'\) ones.
With \(a=\lceil h'/2\rceil\) and \(b=\lfloor h'/2\rfloor\), the ones are partitioned into \(\varepsilon_{21}\) coordinates written \((2,-1)\), \(\varepsilon_{21}\) coordinates written \((-1,2)\), \(a-\varepsilon_{21}\) coordinates written \((1,0)\), and \(b-\varepsilon_{21}\) coordinates written \((0,1)\).
The zeros are partitioned into \(\varepsilon_{20}\) coordinates of each of \((2,-2)\) and \((-2,2)\), and \(\varepsilon_{10}\) coordinates of each of \((1,-1)\) and \((-1,1)\).
The first summand then has
\begin{align*}
  n_{2+}&=\varepsilon_{21}+\varepsilon_{20}, &
  n_{2-}&=\varepsilon_{20}, \\
  n_{1+}&=a-\varepsilon_{21}+\varepsilon_{10}, &
  n_{1-}&=\varepsilon_{21}+\varepsilon_{10}
\end{align*}
entries of \(+2\), \(-2\), \(+1\), and \(-1\) respectively, and
\begin{align*}
  S^{(1)}
  &=
  \binom{N'}{n_{2+},\,n_{2-},\,n_{1+},\,n_{1-}},
  \\
  R^{(1)}
  &=
  \binom{h'}{\varepsilon_{21},\,\varepsilon_{21},\,a-\varepsilon_{21},\,b-\varepsilon_{21}}
  \binom{N'-h'}{\varepsilon_{20},\,\varepsilon_{20},\,\varepsilon_{10},\,\varepsilon_{10}}.
\end{align*}
These are the level-1 formulas of his Appendix~A.1 for a one-signed parent.
The tree is then priced by Algorithm~\ref{alg:tree-cost}.

Deeper levels in his Appendix~A.2 keep the digits \(\pm 2\).
The search here folds each \(+2\) into a \(+1\) and each \(-2\) into a \(-1\), and continues with Algorithm~\ref{alg:rep-counts}.
The folded tail is at least as large as a tail that keeps \(\pm 2\).
The first level is searched for \(\varepsilon_{10}\in\{0,\ldots,27\}\), \(\varepsilon_{21}\in\{0,\ldots,7\}\), and \(\varepsilon_{20}\in\{0,\ldots,5\}\).
The tail is either empty, a single further parameter in \(\{0,\ldots,15\}\), or a pair \((\varepsilon',\varepsilon'')\) with \(\varepsilon'\le 11\) and \(\varepsilon''\le\min(\varepsilon',8)\), under the same monotonicity constraint as Algorithm~\ref{alg:tree-cost}.
On the six rows of Table~\ref{tab:may-params} this search stays within \(0.08\) bits of Representation~1.
The rounded column \(T_{\mathrm{ML}}\) of Table~\ref{tab:may-params} does not change.
On the two rows with \(h=32\) the minimum uses \(\varepsilon_{21}=\varepsilon_{20}=0\), so the digits \(\pm 2\) are absent and the two trees coincide.

\subsection{May's attack on the windowed secret of \ship}
\label{app:may-ship}

Table~\ref{tab:may-ship} applies the same attacks to \ship's windowed secret, with Hamming weight \(h=31\) and dimension \(N\).
\ship fixes no coefficient.
Our estimate fixes one window, which only removes search space, so the costs are conservative for \ship's secret.
The modulus is the base prime \(q_0\) of the BV switch: \(25\), \(48\), \(28\), and \(50\) bits for \(\mathtt{LL13}\), \(\mathtt{LL14}\), \(\mathtt{HT14}\), and \(\mathtt{HT15}\) \cite{SHIP25}.
Odlyzko splits the remaining windows, each with \(4w\) signed positions.
On every set the best tree uses no extra cancellation, so the bold cost is the window split.
That cost is at least \(142\) bits at the published half-width \(w=\max\bigl(175,\lfloor N/(2h)\rfloor\bigr)\), and it stays at least \(128\) bits down to the smallest integer half-width at least \(N/(2h)\).
For \(\mathtt{LL14}\) and \(\mathtt{HT14}\), \(\lfloor N/(2h)\rfloor=264\) sits just below \(N/(2h)\), so that minimum is \(265\).
On every set the smaller window leaves the number \(\lceil\log_4\lceil 2w/\theta\rceil\rceil=3\) of multiplexer digits unchanged (Table~\ref{tab:ship-params}), and with it the sequential key switches of a blind rotation, so our implementation of \ship keeps the published windows.

\begin{table}[t]
\centering
\caption{May's attack on \ship's windowed secret (\(h=31\), dimension \(N\)).
One window is fixed in our estimate; \ship fixes none.
\(\log q_0\) is the base prime of the set.
\ship's lattice check used \(P\cdot q_0\), not this prime.
\(w=\max(175,\lfloor N/(2h)\rfloor)\) is the published half-width, and \(w^\star\) is the smallest integer at least \(N/(2h)\) whose cost is still at least \(128\) bits.
Green columns are Odlyzko's split, which is cheaper than Meet-LWE on every row.
Bold entries are that cost.
Costs are rounded to the nearest bit.}
\label{tab:may-ship}
\setlength{\tabcolsep}{4pt}
\small
\begin{tabular}{l|cc|c>{\columncolor{maygreen}}c c>{\columncolor{maygreen}}c}
\toprule
Set & \(N\) & \(\log q_0\) & \(w\) & May & \(w^\star\) & May\(^\star\) \\
\midrule
LL13  & \(2^{13}\) & \(25\) & \(175\) & \(\mathbf{142}\) & \(133\) & \(\mathbf{136}\) \\
LL14  & \(2^{14}\) & \(48\) & \(264\) & \(\mathbf{151}\) & \(265\) & \(\mathbf{151}\) \\
HT14  & \(2^{14}\) & \(28\) & \(264\) & \(\mathbf{151}\) & \(265\) & \(\mathbf{151}\) \\
HT15  & \(2^{15}\) & \(50\) & \(528\) & \(\mathbf{166}\) & \(529\) & \(\mathbf{166}\) \\
\bottomrule
\end{tabular}
\end{table}

\subsection{What is left out}
\label{app:may-out}

Three parts of May's paper are outside the estimate.
His Section~9 combines the same representations with the collision search of van Oorschot and Wiener, which removes the lists at the price of a higher time exponent; the estimate uses the list version of his Section~7.4, whose memory is the largest list, \(T_\ell\) on the rows above.
His Section~10 discusses a hybrid of the representation tree with lattice reduction: the tree keeps the vectors that satisfy the \lwe equation on the filtered coordinates, and the hybrid projection discards those coordinates.
He leaves open whether the combination improves on either attack alone; for unstructured secrets the primal and dual hybrids of Section~\ref{s:may} have since answered it \cite{HKLS22,BLLW22,BLLLW26}.
Finally, the estimate omits the ring structure he declines to use (his Section~2.1).

The lattice estimate of Section~\ref{s:sparse-budget} leaves out the blocks in turn.
A block with a single \(1\) takes \(B\) values, \(\log B\) bits, whereas \(B\) free coordinates of density \(1/B\) carry \(B\cdot H(1/B)\approx\log(e\cdot B)\) bits, with \(H\) the binary entropy.
Guessing whole blocks in the hybrid's meet-in-the-middle stage would thus save about \(\log e\approx 1.44\) bits of entropy per guessed block; we have not computed its effect on the cost of the attack.

The polynomial factors omitted throughout are those he omits: sorting and hashing the lists, the duplicate labels of \(\ell\) near the two boundaries, the match-and-filter arithmetic, and the number of lists in a tree of depth at most \(4\).

\end{document}